\documentclass{article}
\usepackage{graphicx}  
\usepackage{amsmath}   
\usepackage[compress]{cite}
\usepackage{amssymb}   
\usepackage{bm} 
\usepackage{dcolumn}
\usepackage{color}
\usepackage{mathrsfs}
\usepackage{amsfonts}
\usepackage{varioref}
\usepackage{textcomp}
\usepackage[parfill]{parskip}
\RequirePackage[colorlinks,citecolor=blue,urlcolor=magenta,linkcolor=blue]{hyperref}
\allowdisplaybreaks
\def\gr{general relativity}
\def\RN{Reissner-Nordstr\"{o}m}

\def\dS{de Sitter}
\def\SCC{strong cosmic censorship}
\def\wrt{with respect to }
\labelformat{section}{Section #1} 
\labelformat{subsection}{Section #1} 
\labelformat{subsubsection}{Section #1}
\labelformat{subsubsubsection}{Section #1}
\labelformat{equation}{Eq.~(#1)} 
\labelformat{figure}{Fig.~#1} 
\labelformat{subfigure}{Fig.~\thefigure#1} 
\labelformat{table}{Table.~#1} 
\labelformat{appendix}{Appendix #1}
\title{Fate of Strong Cosmic Censorship Conjecture in Presence of Higher Spacetime Dimensions}
\author{Mostafizur Rahman\footnote{ mostafizur@ctp-jamia.res.in}$~^{1}$, Sumanta Chakraborty\footnote{sumantac.physics@gmail.com}$~^{2}$ , Soumitra SenGupta\footnote{tpssg@iacs.res.in}$~^{2}$ and 
Anjan A. Sen\footnote{aasen@jmi.ac.in}$~^{1}$\\
{$~^{1}$\small{Center for Theoretical Physics, Jamia Millia Islamia, New Delhi-110025, India}}\\
{$^{2}$\small{School of Physical Sciences, Indian Association for the Cultivation of Science, Kolkata-700032, India}}}
\begin{document}

\maketitle
\begin{abstract}
Strong cosmic censorship conjecture has been one of the most important leap of faith in the context of general relativity, providing assurance in the deterministic nature of the associated field equations. Though it holds well for asymptotically flat spacetimes, a potential failure of the strong cosmic censorship conjecture might arise for spacetimes inheriting Cauchy horizon along with a positive cosmological constant. We have explicitly demonstrated that violation of the censorship conjecture holds true in the presence of a Maxwell field even when higher spacetime dimensions are invoked. In particular, for a higher dimensional Reissner-Nordstr\"{o}m-de Sitter black hole the violation of cosmic censorship conjecture is at a larger scale compared to the four dimensional one, for certain choices of the cosmological constant. On the other hand, for a brane world black hole, the effect of extra dimension is to make the violation of cosmic censorship conjecture weaker. For rotating black holes, intriguingly, the cosmic censorship conjecture is always respected even in presence of higher dimensions. A similar scenario is also observed for a rotating black hole on the brane.
\end{abstract}
\section{Introduction and Motivation}\label{SCC_Intro}

The strong cosmic censorship conjecture, in its simplest form, can be stated as follows: \textit{for a generic initial data}, all the physically reasonable solutions of Einstein's Field equations are globally hyperbolic and apart from a possible initial singularity, no other space-time singularity will ever be visible to any observer \cite{PhysRevLett.14.57,Wald:106274,nla.cat-vn3002454}. This conjecture was originally formulated in order to ensure the deterministic nature of general relativity, governing the dynamics of gravity. For a given initial data, it is possible to construct a maximal global hyperbolic extension of the Lorentzian manifold governed by the Einstein's field equations \cite{Costa:2017tjc,Costa:2014yha}. In several situations, such a maximal global hyperbolic extension forms a subset of a larger manifold, whose boundary is referred to as the Cauchy horizon. Hence the existence of the Cauchy horizon indicates a possible breakdown of determinism (or, equivalently predictability) of general relativity. As evident, the strong cosmic censorship conjecture exactly restricts this possibility. The Einstein's field equations being of second order, a mathematically precise formulation of the censorship conjecture demands that for a generic initial data, \textit{it is not possible to extend the  spacetime across the Cauchy horizon}, such that the spacetime metric is still twice differentiable \cite{Costa:2014yha}. A challenge is posed to the cosmic censorship conjecture by certain parameter space of several exact solutions to the Einstein's Field equations, known to possess Cauchy horizon inside the black hole region, e.g., Reissner-Nordstr\"{o}m or Kerr black holes \cite{Chandrasekhar:579245,poisson_2004}. In these black hole spacetimes, the metric itself is perfectly regular at the Cauchy horizon and hence an observer can safely cross the Cauchy horizon in a finite proper time \cite{poisson_2004}. This situation is troublesome, since after crossing the Cauchy horizon, the future of the observer is no longer determined by the initial data and Einstein's Field equations, pointing towards a possible violation of the censorship conjecture. A possible resolution to this problem, as advocated by Penrose has to do with fact that the Cauchy horizon is unstable under external perturbations, i.e., even a small perturbation can turn it into a curvature singularity \cite{1973IJTP....7..183S}. Later on the reason for such instability, leading to divergence of curvature at the Cauchy horizon was attributed to the blow up of the mass function at the Cauchy horizon. This phenomenon, known as \textit{mass inflation} is applicable for asymptotically flat black holes alone \cite{PhysRevD.41.1796,Dafermos:2003wr,Dafermos2014}.

Recently, several interesting alternative views have been proposed to understand the phenomenon of mass inflation in a better way. These results possibly stem from the fact that an infinite tidal force due to the mass inflation singularity at the Cauchy horizon, does not necessarily spell doom on an observer who attempt to cross it \cite{PhysRevLett.67.789}. In fact, there exist several possible solutions to the Einstein's field equations with regularity lower than $C^{2}$, i.e., the metric need not be twice differentiable \cite{0264-9381-16-12A-302}. This urges one to take a more modern look into the cosmic censorship conjecture, which essentially demands the Christoffel symbols $\Gamma ^{\mu}_{\alpha \beta}$, constructed out of the first derivatives of the metric to belong to $L^{2}_{\rm loc}$ space, i.e., they are locally square integrable functions \cite{Cardoso:2017soq}. Using this one can construct a more refined version of the strong cosmic censorship conjecture as formulated by Christodoulou, which reads: \textit{it is impossible to extend the spacetime across the Cauchy horizon with Christoffel symbols being square integrable locally, i.e., with $\Gamma ^{\mu}_{\alpha \beta} \in L^{2}_{\rm loc}$} \cite{Christodoulou:2008nj,Dafermos2014}.

It turns out that this version of the strong cosmic censorship conjecture holds true for asymptotically flat black holes \cite{Dafermos2014}. But it is very important to understand the fate of the same in the presence of a positive cosmological constant. In particular, it was demonstrated that one may not be able to conclude the same in presence of a positive cosmological constant  \cite{Chambers:1997ef}. The fate of Cauchy horizon, essential in understanding the strong cosmic censorship conjecture, under a small perturbation (may be due to an external scalar field $\Phi$) depends on two factors --- (a) its growth at the Cauchy horizon and (b) the rate of its decay along the event horizon \cite{PhysRevD.41.1796}. In the case of asymptotically flat black holes, the power law decay of the perturbation along the event horizon \cite{PhysRevD.5.2419,Dafermos:2014cua,Angelopoulos:2016wcv} is overwhelmed by its exponential growth at the Cauchy horizon \cite{PhysRevD.19.2821,PhysRevLett.67.789,HISCOCK1981110,PhysRevLett.80.3432}. The growth at the Cauchy horizon, governed by the surface gravity $\kappa_{-}$ of the Cauchy horizon, subsequently turns the Cauchy horizon into a curvature singularity. But in presence of a positive cosmological constant, massless scalar field perturbations decay exponentially along the event horizon, rather than power law as in the case of asymptotically flat spacetime \cite{Brady:1996za}. Such an exponential decay of the perturbing scalar field $\Phi$ in presence of a positive cosmological constant takes the following form \cite{Dyatlov:2013hba,Bony2008,Dyatlov2012,Cardoso:2017soq}
\begin{equation}\label{3}
|\Phi-\Phi_{0}|\leq C e^{-\alpha t}~,
\end{equation}
where $C$ and $\alpha$ are both positive constants along with $\Phi_{0}$. In particular, the constant $\alpha$ is called the \textit{spectral gap}, corresponding to the longest-lived quasi-normal modes of the black hole, which is simply given by $-\{\textrm{Im}(\omega)\}_{\rm min}$ \cite{Cardoso:2017soq}. Thus this exponential decay can nullify the effect of the exponential growth of perturbation at the Cauchy horizon and hence may lead to a possible violation of strong cosmic censorship conjecture \cite{Christodoulou:2008nj,PhysRevD.61.064016,Costa:2014aia}. In fact, it has been explicitly demonstrated that in presence of a positive cosmological constant, an Einstein-Maxwell-scalar field system will violate the strong cosmic censorship conjecture for a finite parameter space of the model under consideration \cite{Cardoso:2017soq}. In particular, it turns out that if a dimensionless quantity $\beta$, constructed out of $\alpha$ and $\kappa_{-}$, such that, 
\begin{equation}\label{SCC violation}
\beta \equiv \frac{\alpha}{\kappa_{-}}=-\frac{\{\textrm{Im}(\omega)\}_{\rm min}}{\kappa_{-}}>\frac{1}{2}~,
\end{equation}  
then cosmic censorship conjecture would be violated \cite{Dafermos2014,Cardoso:2017soq,Hintz:2015jkj,Dafermos:2017dbw}. The computation for the spectral gap $\alpha$ in the eikonal limit is straightforward and is performed by computing the Lyapunov exponent associated with the stability of the photon circular orbit. Besides, there are two additional quasi-normal modes, which are also of importance, namely the de Sitter modes and the near extremal modes. Using these three modes, as well as numerical methods, e.g., continued fraction method, the violation of strong cosmic censorship conjecture for \RN-\dS\ black holes has been demonstrated in \cite{Cardoso:2017soq}. Subsequently these results have been generalized for Kerr-de Sitter black holes in \cite{PhysRevD.97.104060}. Surprisingly, it turned out that the condition $\beta>(1/2)$ is never satisfied in the context of rotating black holes and further the numerical analysis of the quasi-normal modes shows that the error in using \emph{only} the photon sphere modes is negligible. Thus one need not worry about the de Sitter or near extremal modes in the context of rotating black hole spacetime. This suggests that for astrophysical scenarios there will be no violation of cosmic censorship conjecture. However, all these analysis are in the context of four dimensional spacetime and thus it seems legitimate to understand the validity of strong cosmic censorship conjecture in presence of higher dimensions. For this purpose we will mainly use two possibilities --- (a) The black hole itself could live in a higher dimensional spacetime (for a incomplete set of references, see  \cite{Emparan:2008eg,Reall:2015esa,Emparan:2001wn,Emparan:2007wm,Arcioni:2004ww,Gibbons:2002av,Gregory:1993vy,Horowitz:2012nnc,Myers:1986un}) or, (b) The black hole is living on the four dimensional spacetime (which we will call brane), while the spacetime itself is higher dimensional \cite{Shiromizu:1999wj,Maartens:2001jx,Dadhich:2000am,Germani:2001du,Casadio:2012pu,Harko:2004ui,Chakraborty:2014xla,Chakraborty:2015bja,Chakraborty:2015taq}. In the first context the gravitational field equations will remain the same, but the effect of higher dimension will change the metric elements from the four dimensional one. While in the second, presence of higher dimensions will modify the field equations non-trivially leading to departure from four dimensional solution. Both of these result into non-trivial departures from general relativistic solution \cite{Chakraborty:2017qve,Mukherjee:2017fqz,Banerjee:2017hzw,Chakraborty:2016lxo} and it will be important to understand the consequences as far as cosmic censorship conjecture is concerned. This is what we will explore in this work.  

The paper is organized as follows: In \ref{lyaintro}, we start by introducing the Lyapunov exponent for photon circular orbit, its relation to the effective potential for radial motion on the equatorial plane and finally how the quantity $\beta$ is related to it. Then in \ref{SCC_Static} we have computed this quantity $\beta$ for a general static and spherically symmetric spacetime, which was applied in \ref{SCC_Static_App} for static and spherically symmetric black holes in higher dimensions or in four dimensions inheriting effects from higher dimensions. Subsequently, we have demonstrated a computation of $\beta$ for an arbitrary rotating black hole in \ref{SCC_Rotating}, which was applied in \ref{SCC_Rot_App} for a rotating black hole in higher dimension as well as on the four dimensional brane. Finally we conclude with discussions on the results obtained. 
\section{Violation of Strong Cosmic Censorship Conjecture in Higher Dimensions}\label{SCC_HD}

In this paper we will be working with spacetimes inheriting extra spatial dimensions and hence it is legitimate to ask whether the above condition on $\beta$, namely \ref{SCC violation}, still results into violation of strong cosmic censorship conjecture, even for higher dimensional black holes. To see the same one may consider a perturbing scalar field living on a higher dimensional static and spherically symmetric spacetime, satisfying the equation $\square \Phi=0$ (this can be trivially generalized to a conformally coupled scalar field as well). Due to existence of angular and timelike Killing vectors in the spacetime, the scalar field can be decomposed as $\Phi(t,r,\Omega)=e^{-i\omega t}R(r)h(\Omega)$, where $h(\Omega)$ corresponds to the spherical harmonics associated with the $(d-2)$ dimensional sphere and $R(r)$ satisfies a second order differential equation, which resembles time independent Schr\"{o}dinger equation with a potential (see e.g., \cite{Du:2004jt, Berti:2009kk}). Near the Cauchy horizon (assuming it exists), the second order differential equation for $R(r)$ has two linearly independent solutions, namely,
\begin{align}\label{radial_solution}
\Phi^{(1)}&=e^{-i\omega u}\mathcal{R}^{(1)}(r)Y_{\ell m}(\theta,\phi)
\\
\Phi^{(2)}&=e^{-i\omega u}\mathcal{R}^{(2)}(r)Y_{\ell m}(\theta,\phi)\left(r-r_{-}\right)^{i\omega/\kappa_{-}}
\end{align}
where $u$ is the retarded time coordinate and $r_{-}$ is the location of the Cauchy horizon. Note that $\mathcal{R}^{(1)}(r)$ and $\mathcal{R}^{(2)}(r)$ are two linearly independent radial functions, having smooth limit to the Cauchy surface. Given the above solutions, one can determine the integral of $(\partial_{\mu}\Phi\partial ^{\mu}\Phi)$ over the Cauchy surface, which boils down to the integral of $(r-r_{-})^{2(i\omega/\kappa_{-}-1)}$. Thus if the quantity $\beta$, defined in \ref{SCC violation}, is greater than $(1/2)$, the perturbing scalar field $\Phi(t,r,\Omega)$ is regular at the Cauchy horizon and can be extended beyond. This is sufficient to ensure the violation of strong cosmic censorship conjecture. Thus even for higher dimensional black holes, if the relation $\beta>(1/2)$ holds true, one can safely argue about violation of strong cosmic censorship conjecture.

The above argument continues to hold true for rotating black holes in higher dimensions as well, with minimal modifications. To see this explicitly, one can again consider a scalar field living on a rotating black hole spacetime in higher dimension (see, e.g., \cite{Cardoso:2004cj,Ida:2002zk}). Following the same analogy as the spherically symmetric spacetime, for rotating higher dimensional black hole spacetime as well, the scalar field can be decomposed into individual parts depending on time and the angular coordinates, while the radial part satisfies a Schr\"{o}dinger-like second order differential equation with certain potential \cite{Cardoso:2004cj,Ida:2002zk}. In this case as well the differential equation for the radial part can be solved in the near Cauchy horizon limit, yielding two independent solutions as in \ref{radial_solution}. One of which is certainly regular at the Cauchy horizon, while the other has non-smooth radial dependance at the Cauchy horizon, such that, $\Phi ^{\rm non-smooth}\sim (r-r_{-})^{p}$, where $p=i(\omega-\sum _{i}m_{i} \Omega _{-}^{i})/\kappa _{-}$. Here $\Omega_{-}^{i}$ corresponds to the angular velocity of the Cauchy horizon along the $i$th direction. Thus the problematic part in the integral of $\partial _{\mu}\Phi\partial ^{\mu}\Phi$ corresponds to $(r-r_{-})^{2(\beta-1)}$ in the integrand, where $\beta$ is defined as in \ref{SCC violation}. Thus for $\beta>(1/2)$, the perturbation can be continued across the Cauchy horizon leading to possible violation of strong cosmic censorship conjecture.  Thus even in presence of higher dimensions, for both spherically symmetric and rotating black hole spacetimes, $\beta>(1/2)$ signals possible violation of strong cosmic censorship conjecture. We will use this input in our subsequent sections. Finally, note that the existence of the Cauchy horizon is absolutely essential for the above argument to work and thus we need to work with higher dimensional black hole solutions inheriting Cauchy horizon. Based on the above discussion, in the next section we provide an estimation of the parameter $\beta$ in the eikonal approximation, using the Lyapunov exponent associated with circular null geodesics. 
\section{Lyapunov Exponent for a Black Hole and Cosmic Censorship Conjecture}\label{lyaintro}

Computation of black hole quasi-normal modes is of utmost importance since it enables one to understand and possibly differentiate between various black holes in gravity theories beyond \gr. Apart from few simple scenarios, the computation of quasi-normal modes associated with perturbation of black holes, in general involves numerical techniques. However, under certain circumstances it is indeed possible to determine analytical methods to compute the quasi-normal modes. One such method was developed in \cite{PhysRevD.31.290}, where the computation of the quasi-normal modes follow from geometrical-optics approximation, where the null geodesics trapped at the unstable photon orbit plays an important role. In particular, the real part of the quasi-normal mode frequency is given by the angular frequency of rotation of a photon in the photon circular orbit, while the imaginary part is related to the largest Lyapunov exponent measuring the growth of the perturbation around photon circular orbit \cite{Cornish:2003ig,Cardoso:2008bp,0264-9381-9-12-004,PhysRevD.31.290,PhysRevLett.52.1361} (see also \cite{Konoplya:2017wot}). Later on, this approach has received further attention since one could demonstrate that the results derived using the above analytical technique are in accord with numerical methods in the eikonal limit \cite{Hod:2009td}. To see, how the derivation of the Lyapunov exponent associated with the growth of perturbation around photon circular orbit on the equatorial plane goes, we write down the equation for radial null geodesics in the the equatorial plane $\theta=\pi/2$ as 
\begin{equation}\label{rdotVr}
\dot{r}^{2}=V_{\rm eff}(r)~.
\end{equation}
Here, `dot' denotes derivative \wrt the affine parameter associated with the null geodesics and $V_{\rm eff}$ is the effective potential associated with the radial null geodesics. On the other hand, the circular photon orbit is a solution of the following equation
\begin{equation}\label{rph}
V_{\rm eff}(r_{\rm ph})=V_{\rm eff}'(r_{\rm ph})=0~,
\end{equation}
where $r_{\rm ph}$ stands for the radius of the circular photon orbit. The above photon orbit is unstable, since it appears at the maxima of the effective potential. Therefore the Lyapunov exponent $\lambda$ is associated with infinitesimal fluctuations around the photon circular orbit, i.e., we will consider $r=r_{\rm ph}+\delta r$, where $\delta r$ is assumed to be a small perturbation. Substituting this expression and then expanding the right hand side of \ref{rdotVr} around the photon circular orbit, we obtain the following expression for the time evolution of the perturbed quantity $\delta r$,
\begin{equation}\label{prelya}
\begin{aligned}
\left(\dot{\delta r}\right)^{2}=\frac{1}{2}V_{\rm eff}''(r_{\rm ph})\delta r^{2}~.
\end{aligned}
\end{equation}
In the above expression we have kept terms upto quadratic order in the expansion of the right hand side of \ref{rdotVr} and have used \ref{rph}. As evident, there will be two solutions of the above differential equation, one will be growing in nature while the other will be decaying. To present the solutions without any reference to the affine parameter, we divide both sides of \ref{prelya} with $\dot{t}^{2}$ and hence we obtain,
\begin{equation}\label{lya}
\delta r=A \exp(\pm \lambda t)~,
\end{equation}
where $A$ is a constant of integration and $\lambda$ presents the Lyapunov exponent yielding the decay (growing) rate of the photon circular orbit. This is given by the following analytical expression \cite{Cardoso:2008bp},
\begin{equation}\label{lyapunov}
\lambda=\sqrt{\dfrac{V_{\rm eff}''}{2\dot{t}^{2}}}\bigg|_{r=r_{\rm ph}}~.
\end{equation} 
This enables one, following \cite{Cardoso:2008bp}, to relate the imaginary part of the quasi-normal mode frequencies with the Lyapunov exponent in the eikonal limit as
\begin{equation}\label{imomega}
\operatorname{Im}(\omega)=-\left(\nu+\frac{1}{2}\right)\lambda~,
\end{equation}
where $\nu=0,1,2...$ is the overtone number. Since the frequencies of the quasi-normal modes appear as $\exp(-i\omega t)$, it follows that larger the value of $\nu$, the faster that corresponding quasi-normal mode will decay. The longest lived mode corresponds to the one having smallest imaginary part, which must be the mode with $\nu=0$. Thus from \ref{imomega}, we can conclude that $\{\textrm{Im}(\omega)\}_{\rm min}=-\lambda/2$. Furthermore, if the spacetime admits a Cauchy horizon, we can certainly compute the surface gravity $\kappa_{-}$ associated with it and hence the parameter $\beta$, defined in \ref{SCC violation}, in the eikonal limit becomes 
\begin{equation}\label{sccviolation}
\beta_{\rm ph}=\frac{\lambda}{2\kappa_{-}}~.
\end{equation}
The suffix `ph' to $\beta$ is to remind us that this expression holds true in the eikonal approximation, namely for the photon sphere modes. As emphasized earlier, this parameter is a deterministic factor in understanding the validity (or, possible violation) of the strong cosmic censorship conjecture in various black hole spacetimes. In particular, in four spacetime dimensions, if the parameter $\beta$ becomes larger than half, we can conclude that strong cosmic censorship conjecture is violated, see e.g., \cite{Dafermos2014,Cardoso:2017soq,Hod:2018dpx,Cardoso:2018nvb,Ge:2018vjq,Mo:2018nnu}. In the subesquent sections we will compute the Lyapunov exponent and hence $\beta$ in the eikonal approximation for both static and spherically symmetric spacetime as well as for a rotating black hole spacetime as well. The results so derived can be used to understand the violation of cosmic censorship conjecture in the presence of higher dimensions. 
\section{Lyapunov Exponent and Strong Cosmic Censorship Conjecture for a General Static and Spherically Symmetric Spacetime}
\label{SCC_Static}

In this section, we would like to present the computation of the Lyapunov exponent in the context of a general static and spherically symmetric spacetime, so that the result derived here can further be used in various other contexts as well, whenever a static and spherically symmetric solution becomes available. Any such static and spherically symmetric spacetime in $d$ spacetime dimensions can be expressed through the following line element,
 \begin{equation}\label{staticspherically}
ds^{2}=-f(r)dt^{2}+g(r)^{-1}dr^{2}+r^{2}d\Omega^{2}_{d-2}~,
\end{equation} 
where the functions $f(r)$ and $g(r)$ are as of now arbitrary. These functions can be determined by solving the associated gravitational field equations, which could be the Einstein's field equations or field equations associated with gravity theories beyond \gr. Further, $d\Omega^{2}_{d-2}$ denotes the line element of the $(d-2)$-sphere. Since we are interested in the geodesic motion in four dimensional spacetime alone and the spacetime inhibits spherical symmetry we will set all the angular coordinates, except one, to $\pi/2$. This ensures that the Lagrangian associated with the motion of a particle on the equatorial plane takes the following form,
\begin{equation}\label{lagrangian}
\mathcal{L}=\frac{1}{2}\Big\{-f(r)\dot{t}^{2}+g(r)^{-1}\dot{r}^{2}+r^{2}\dot{\phi}^{2}\Big\}~.
\end{equation}
Here `dot' denotes derivative with respect to proper time or proper length in the context of timelike or spacelike trajectories, while it is the derivative with respect to the affine parameter in the context of null geodesics. Since the metric is independent of the co-ordinates $t$ and $\phi$, the Lagrangian $\mathcal{L}$ is cyclic with respect to these co-ordinates. Hence the corresponding conjugate momentums are constants of motion which we identify as the Energy $p_{t}=-E$ and angular momentum $p_{\phi}=L$ of the trajectory. Then the geodesic equation for $\phi$ and $t$ are trivial to solve for, while the geodesic equation for the radial coordinate becomes \cite{Cardoso:2008bp}, 
\begin{align}\label{geodesicnospin}
\dot{r}^{2}=\frac{g(r)}{f(r)}\left[E^{2}-f(r)\left(-\epsilon+\frac{L^{2}}{r^{2}}\right)\right]~,
\end{align}
where, $\epsilon=g_{\mu\nu}u^{\mu}u^{\nu}=(1,0,-1)$ for spacelike, null and timelike geodesics respectively. Since the determination of Lyapunov exponent depends explicitly on the photon circular orbit in this spacetime, we are interested in the null geodesics corresponding to $\epsilon=0$. Thus from \ref{geodesicnospin}, the radial null geodesics satisfy the following equation,
\begin{equation}\label{potentialnospin}
\dot{r}^{2}=\frac{g(r)}{f(r)}\left[E^{2}-f(r)\frac{L^{2}}{r^{2}}\right]\equiv V_{\rm eff}(r)~,
\end{equation}
where `dot' denotes derivative with respect to the affine parameter along the null geodesic. Given the above potential one can immediately determine the circular photon orbit $r_{\rm ph}$ starting from \ref{rph}, by setting both $V_{\rm eff}(r)$ and $V_{\rm eff}'(r)$ to zero. This result into the following equations
\begin{equation}\label{rphnospin}
\begin{aligned}
\frac{E^{2}}{L^{2}}&=\frac{f(r)}{r^{2}}~;
\\
2f(r)&=rf'(r)~,
\end{aligned}
\end{equation}
where ``prime" denotes denotes derivative \wrt $r$. Further, given the potential in \ref{potentialnospin}, one can immediately compute $V_{\rm eff}''(r)$ necessary to determine the Lyapunov exponent. Similarly from the fact that $p_{t}=-E$, it follows that $\dot{t}=\{E/f(r)\}$. Thus using \ref{lyapunov} and \ref{rphnospin} along with the expression for $V_{\rm eff}''(r)$ on the photon circular orbit, we can determine the Lyapunov exponent for any general static and spherically symmetric spacetime to be, 
\begin{equation}\label{lyanospin}
\lambda =\sqrt{\frac{g(r_{\rm ph})}{2}\left(\frac{2f(r_{\rm ph})}{r_{\rm ph}^{2}}-f''(r_{\rm ph})\right)}~,
\end{equation}
where all the quantities have been evaluated at $r=r_{\rm ph}$ and the subscript `S' stands for Static spacetime. The above provides the expression for the Lyapunov exponent in a general static and spherically symmetric spacetime with arbitrary choices of the functions $f(r)$ and $g(r)$ respectively. In order to have any possibility of violation of cosmic censorship conjecture, it is necessary that the solution presented in \ref{staticspherically} inherits a Cauchy horizon, which is a null surface and is the smallest root $r_{-}$ of the equation $g(r)=0$, the larger root $r_{+}$ presents the event horizon. The surface gravity associated with the Cauchy horizon corresponds to $\kappa_{-}=(1/2)g'(r_{-})$. Hence the parameter $\beta_{\rm ph}$, associated with quasi-normal modes in the large $\ell$-limit, defined in \ref{sccviolation}, takes the following form,
\begin{equation}\label{betastatic}
\beta _{\rm ph}=\sqrt{\frac{g(r_{\rm ph})}{2g'(r_{-})^{2}}\left(\frac{2f(r_{\rm ph})}{r_{\rm ph}^{2}}-f''(r_{\rm ph})\right)}~,
\end{equation}
Thus given any static and spherically symmetric spacetime in the presence of a positive cosmological constant, inheriting Cauchy horizon, one can explicitly compute the parameter $\beta_{\rm ph}$. If for any choices of the parameters in the spherically symmetric solution, which allows for a non-trivial Cauchy as well as event horizon, if $\beta _{\rm ph}$ turns out to be larger than half, then it will lead to a violation of the cosmic censorship conjecture. This provides a robust and quantitative way to understand the violation of the cosmic censorship conjecture in terms of the photon sphere modes. 

So far the above discussion is purely based on the quasi-normal modes in the large $\ell$ limit and originates from the perturbation of the photon sphere. The modes so obtained are referred to as the photon sphere modes and they are well described by the  WKB approximation. The imaginary parts of the photon sphere modes are given by \ref{imomega} and hence the lowest lying mode corresponds to the following choice, $\{\textrm{Im}(\omega)/\lambda\}=-(1/2)$. However in the context of asymptotically de Sitter black holes with electromagnetic charge, there are two additional modes which are also of importance. These are the de Sitter modes and the near extremal modes. The frequencies associated with the de Sitter modes essentially depend on the asymptotic structure of the spacetime itself and is mostly independent of the other hairs inherited by the spacetime. The pure de Sitter modes have been studied in the presence of higher dimensions in \cite{Du:2004jt,LopezOrtega:2012vi,Abdalla:2002hg} and the frequencies of the lowest lying quasi-normal modes, for spacetime dimensions $d>4$ correspond to,
\begin{equation}\label{qnmdS}
\frac{\omega_{0,\textrm{dS}}}{\kappa _{\rm c}}=-i\ell;\qquad 
\frac{\omega _{n\neq 0,\textrm{dS}}}{\kappa _{\rm c}}=-i\left(\ell+2n\right)~.
\end{equation}
Here $\kappa _{c}=\sqrt{2\Lambda/(d-1)(d-2)}$ is the surface gravity associated with the cosmological horizon for a d-dimensional asymptotically de Sitter black hole. Thus the dominant, lowest lying mode corresponds to $\ell=1$ and $n=0$, while the other modes are also present. Note that the imaginary part of the lowest lying de Sitter mode ($\ell=1$, $n=0$) in a d-dimensional spacetime has a structure identical to the de Sitter mode in a four dimensional spacetime \cite{Cardoso:2017soq} albeit with a modified surface gravity. However, the higher order quasi-normal modes differ in a higher dimensional spacetime from the four-dimensional result, see \cite{LopezOrtega:2012vi}. Thus one must take into account of these modes before making any conclusive statement regarding the strong cosmic censorship conjecture.

Finally, there is another mode of interest, these are the near extremal modes and comes into the picture as event and Cauchy horizon come closer to each other. In four dimensions, such near extremal modes for perturbing scalar field have been derived in \cite{Kim:2012mh, Chen:2012zn}. However, they are not known for higher dimensional black holes. Thus we have presented the computation of these near extremal modes in \ref{App_NE} and they become,
\begin{equation}\label{qnmNE}
\omega_{\textrm{NE}}=-i\left(n+\sigma+1\right)\kappa _{\rm +}=-i\left(n+\sigma+1\right)\kappa _{\rm -}~,
\end{equation}
where, $n$ is an integer and $\sigma$ is related to the angular momentum quantum number $\ell$ through the following relation, $\sigma(\sigma+1)=\ell(\ell+d-3)$. Note that for four dimensional spacetime $\sigma=\ell$ and thus \ref{qnmNE} matches with the result presented in \cite{Kim:2012mh,Cardoso:2017soq}. Moreover, even in presence of extra dimensions, the lowest lying near extremal mode correspond to $\sigma=0$ (or, equivalently $\ell=0$), which structurally coincide with the four dimensional result modulo the expression for surface gravity. However, near extremal modes with $\ell>0$ for higher dimensional black hole are distinct from the four dimensional counterpart, since integer values of $\ell$ do not imply integer $\sigma$, see \ref{App_NE} for details. As we are concerned about the lowest lying modes alone, the above difference for higher $\ell$ modes will not bother us in the subsequent discussion. Thus most of the considerations presented in \cite{Cardoso:2017soq} remains unaffected even with the introduction of higher dimensions. Detailed behaviour of these modes will certainly depend on the exact nature of the solutions, which we will discuss on a case by case basis.

At this point we should emphasize that all these analytical calculations must be backed up by appropriate numerical analysis as well. The fact that Lyapunov exponent yields the quasi-normal modes in an appropriate manner must be verified using appropriate numerical techniques, e.g., the continued fraction method. In particular, it must be ascertained that the errors in the determination of the quasi-normal modes using analytical techniques are small. Keeping this in mind, we have supplemented the analytical computations presented above by appropriate numerical techniques.

In the next section, we will apply the formalism devised above in the context of a higher dimensional \RN-\dS\ black hole, as well as an effective four dimensional black hole in the presence of higher dimensions. This will be the first step to investigate how the presence of higher dimensions can affect the fate of cosmic censorship conjecture.
\section{Application: Charged Black Hole in Higher Dimensions}\label{SCC_Static_App}
 
In this section, we will apply the formalism presented in the previous section in the context of higher dimensional static and spherically symmetric black hole. There exist two paths that one may follow --- (a) one may consider a black hole on the four dimensional brane, inheriting some higher dimensional properties, or, (b) it is possible to consider a truly higher dimensional black hole. In what follows, we will discuss both these scenarios, starting with the charged black hole solution on the brane.
\subsection{Charged Black Hole on the Brane}

Even a four dimensional black hole can encode signatures of higher dimensions, if the four dimensional hypersurface the black hole is living on (referred to as brane), is properly embedded into a higher dimensional spacetime manifold. In such a scenario the gravitational field equations on the brane inherits additional corrections due to presence of extra dimensions. These corrections can be broadly classified into two pieces, one originating from bulk Weyl tensor and another from orbifold symmetry and Israel junction condition on the brane. To keep the discussion along similar lines as in the previous section, we consider the Maxwell field on the brane. This in turn would require a non-zero bulk Weyl tensor as well, since the effective gravitational field equations on the brane involves quadratic terms involving brane energy momentum tensor \cite{Shiromizu:1999wj,Maartens:2001jx}. With both the effect of bulk Weyl tensor and quadratic terms of brane energy momentum tensor included, the following solution to the effective gravitational field equation is obtained \cite{Chamblin:2000ra},
\begin{equation}\label{RNdSeff}
f(r)=g(r)=1-\frac{2M}{r}+\frac{Q^{2}-q}{r^{2}}+\frac{\alpha Q^{4}}{r^{6}}-\frac{\Lambda}{3}r^{2}~.
\end{equation}
Here, $Q$ is the electric charge of the Maxwell field on the brane, $q$ is the charge inheriting from bulk Weyl tensor and $\Lambda$ is the brane cosmological constant. Further the constant $\alpha$ is also being inherited from higher dimensions, which can be written as, $\alpha=(3/80\pi)\lambda _{\rm b}^{-1}$, where $\lambda _{\rm b}$ is the brane tension. Note that in the general relativistic limit, we have $\lambda _{\rm b}^{-1}\rightarrow 0$ and $q\rightarrow 0$, such that \ref{RNdSeff} leads to four dimensional \RN-\dS\ black hole spacetime. 

Given this black hole solution to the effective four dimensional gravitational field equations inheriting Maxwell field on the brane, one can try to see whether cosmic censorship conjecture is violated in this case or not. The first and most important quantity associated with the computation of $\beta_{\rm ph}$ is the Lyapunov exponent $\lambda _{\rm RNbrane}$ and hence the photon circular orbit $r_{\rm ph}$. The location of the photon circular orbit can be easily determined following \ref{rphnospin}, leading to the following algebraic equation,
\begin{equation}\label{rpheff}
1=\frac{3M}{r}-\frac{2\left(Q^{2}-q\right)}{r^{2}}-\frac{4\alpha Q^{4}}{r^{6}}
\end{equation}
Thus having determined the algebraic equation for the photon circular orbit $r_{\rm ph}$, one can find out the expression for Lyapunov exponent, determining the stability of the photon circular orbit and hence the imaginary part of the smallest quasi-normal mode frequency. The determination of the Lyapunov exponent follows directly from \ref{lyanospin}, by substituting the metric elements written down in \ref{RNdSeff}, which can be presented as follows,
\begin{equation}\label{lyabrane}
\lambda _{\rm RNbrane}=\sqrt{\frac{f(r_{\rm ph})}{r_{\rm ph^{2}}}\left[1-2\frac{Q^{2}-q}{r_{\rm ph}^{2}}-20\frac{\alpha Q^{4}}{r_{\rm ph}^{6}}\right]}~.
\end{equation}
Note that in the general relativity limit with $\alpha$ and $q$ tending to vanish, the above relation reduces to \ref{lyaRNdSin4}, as it should.  The above completes one part of the journey in our quest to determine the parameter $\beta_{\rm ph}$ and hence possibility of violation of cosmic censorship conjecture.

The other ingredient necessary for the computation of $\beta_{\rm ph}$ requires determination of the location of the Cauchy horizon $r_{-}$. This can be achieved by solving for the lowest root of the algebraic equation $f(r)=0$. Larger roots of the above equation determines the location of the event horizon and the cosmological horizon respectively. Having determined the Cauchy horizon one needs to know the surface gravity at the Cauchy horizon as well, which reads,
\begin{equation}\label{kappaeff}
\kappa^{(-)}_{\rm RNbrane}=\frac{1}{2}f'(r_{-})=\frac{1}{2}\left[\frac{2M}{r_{-}^{2}}-\frac{2\left(Q^{2}-q\right)}{r_{-}^{3}}-\frac{6\alpha Q^{4}}{r_{-}^{7}}-\frac{2\Lambda}{3}r_{-} \right]~.
\end{equation}
Thus having determined the Lyapunov exponent and the surface gravity of the Cauchy horizon, it is straightforward to determine the quantity $\beta_{\rm ph}$. As evident from \ref{lyabrane} and \ref{kappaeff}, presence of higher dimension leads to significant changes to the expressions of both these quantities and hence to the expression for $\beta_{\rm ph}$, which reads,
\begin{equation}\label{betaeff}
\beta^{\rm RNbrane}_{\rm ph}=\frac{\sqrt{\frac{f(r_{\rm ph})}{r_{\rm ph^{2}}}\left[1-2\frac{Q^{2}-q}{r_{\rm ph}^{2}}-20\frac{\alpha Q^{4}}{r_{\rm ph}^{6}}\right]}}{\left[\frac{2M}{r_{-}^{2}}-\frac{2\left(Q^{2}-q^{2}\right)}{r_{-}^{3}}-\frac{6\alpha Q^{4}}{r_{-}^{7}}-\frac{2\Lambda}{3}r_{-} \right]}~.
\end{equation}
It is clear from the above expression that the quantity $\beta_{\rm ph}^{\rm RNbrane}$ depends explicitly on the Maxwell charge $Q$, brane tension $\lambda _{\rm b}$, bulk Weyl tensor through $q$ and finally the brane cosmological constant $\Lambda$.  

\begin{figure}
\minipage{0.33\textwidth}
  \includegraphics[width=\linewidth]{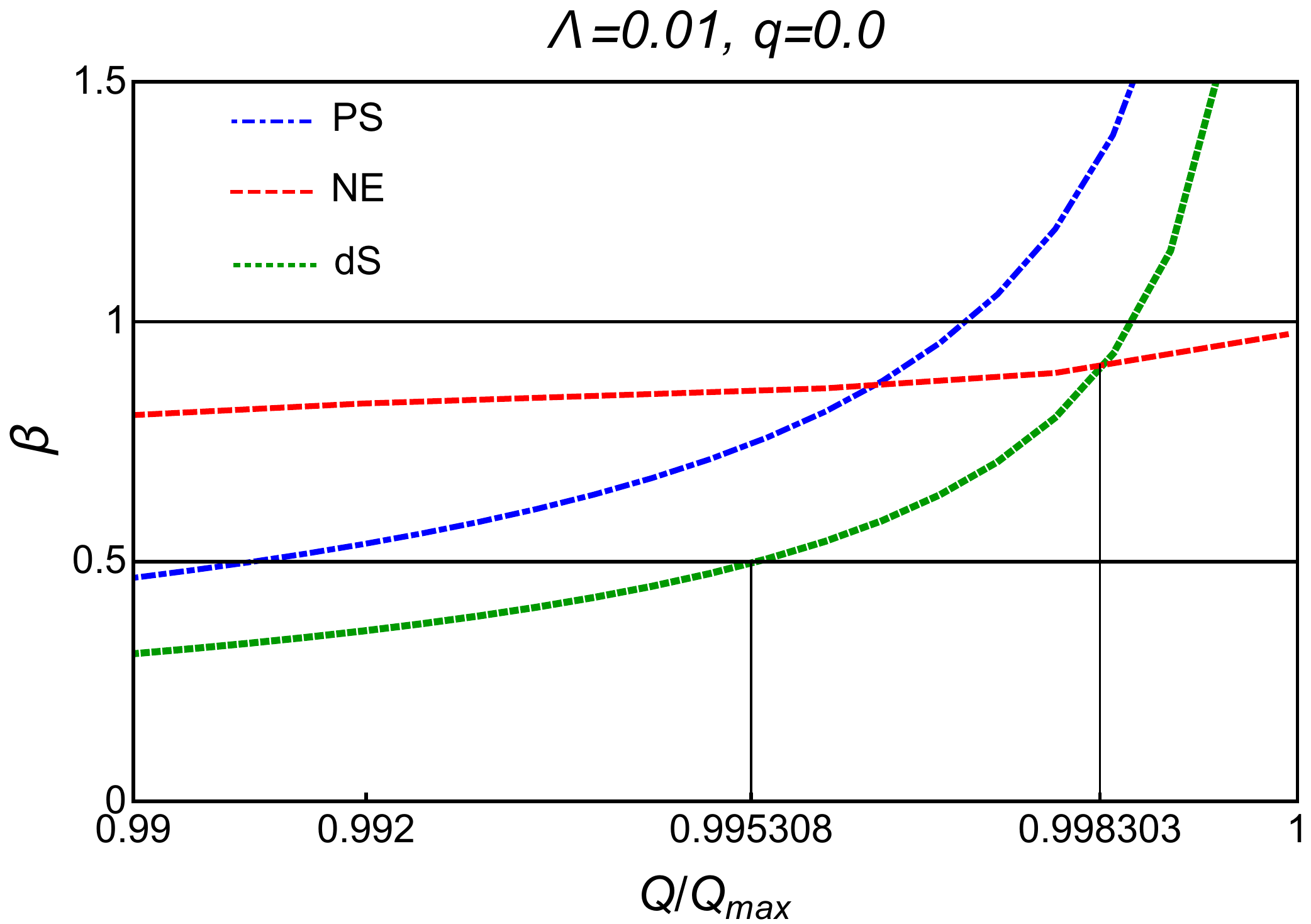}
\endminipage\hfill
\minipage{0.33\textwidth}
  \includegraphics[width=\linewidth]{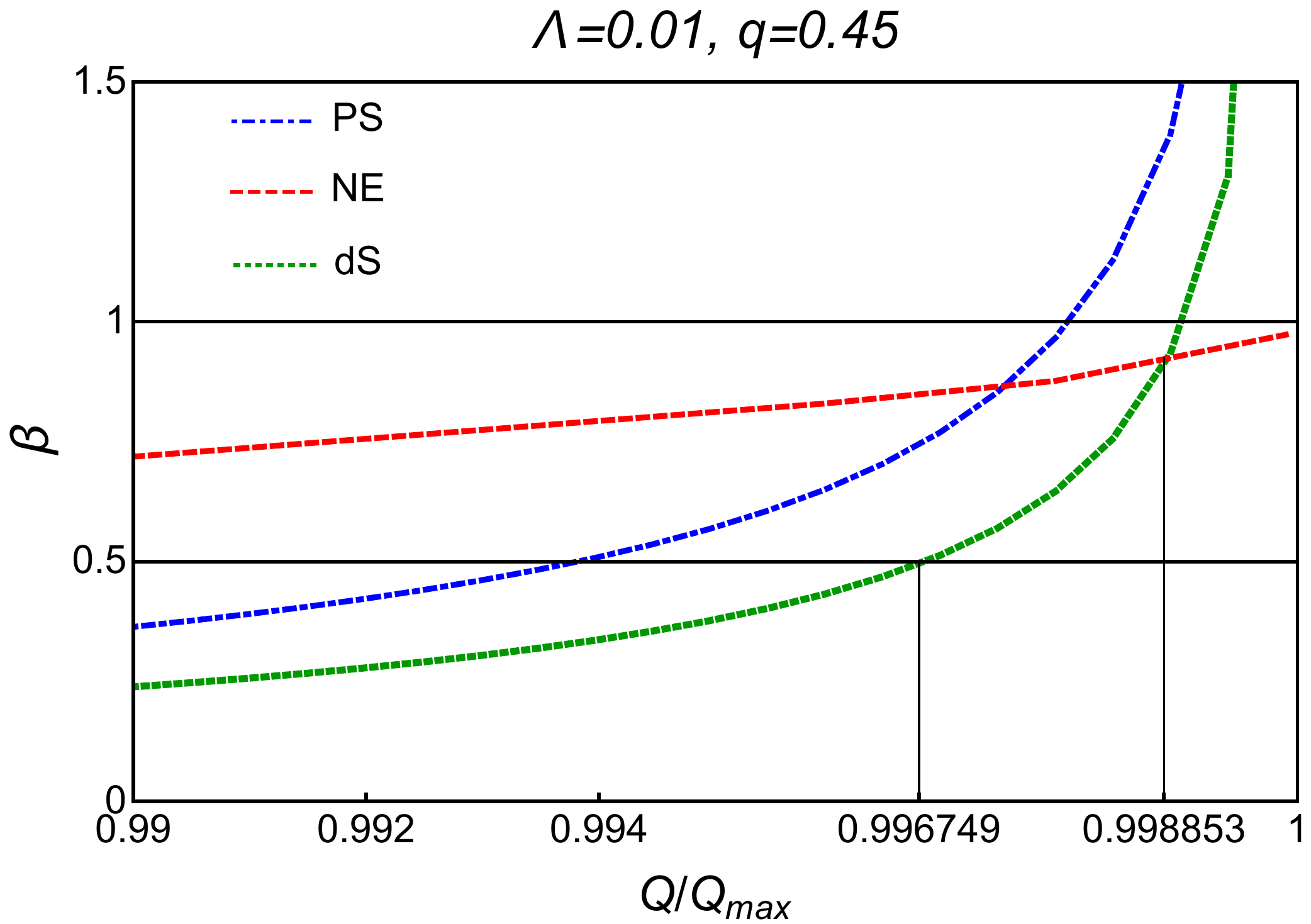}
\endminipage\hfill
\minipage{0.33\textwidth}%
  \includegraphics[width=\linewidth]{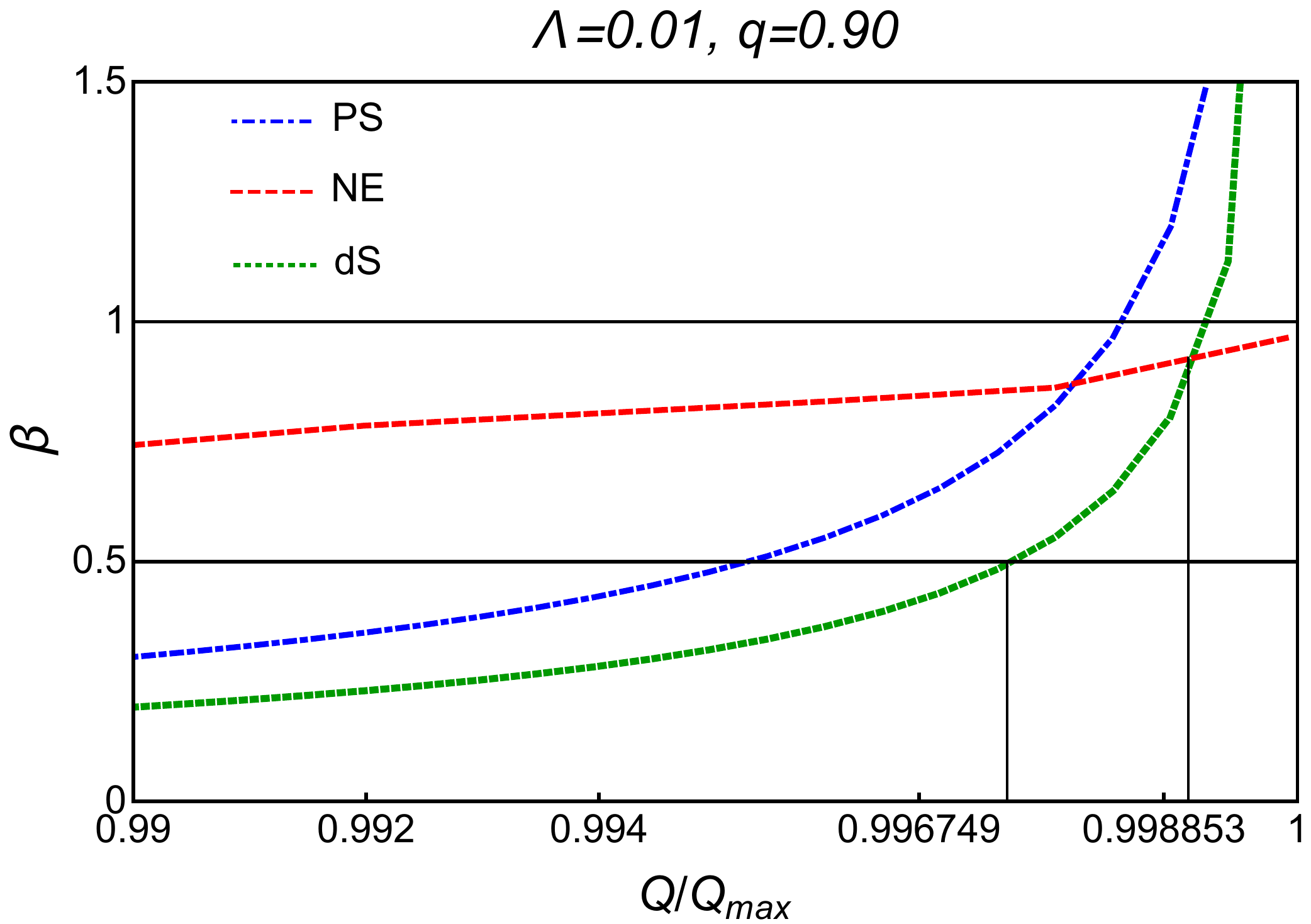}
\endminipage\hfill
\minipage{0.33\textwidth}
  \includegraphics[width=\linewidth]{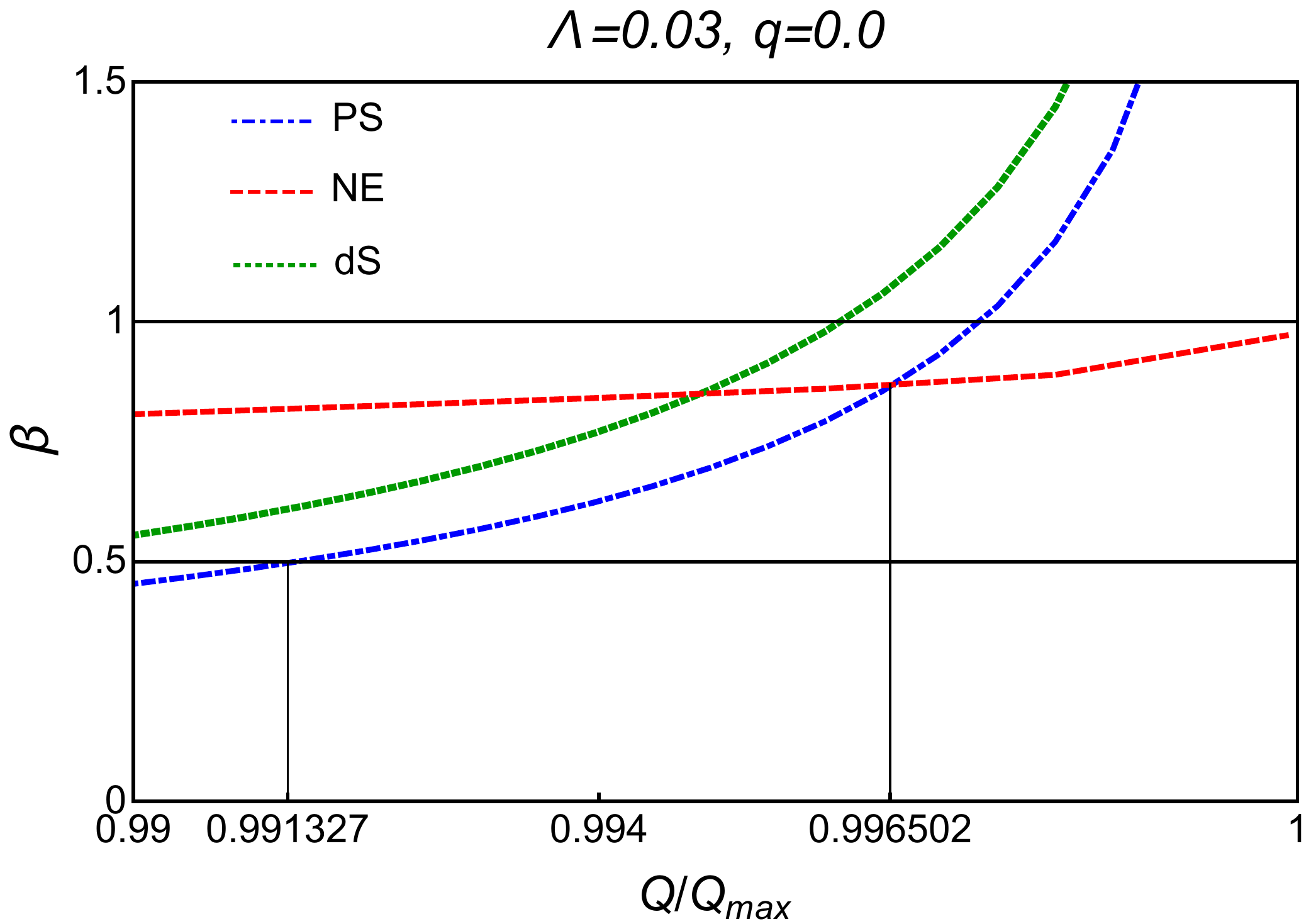}
\endminipage\hfill
\minipage{0.33\textwidth}
  \includegraphics[width=\linewidth]{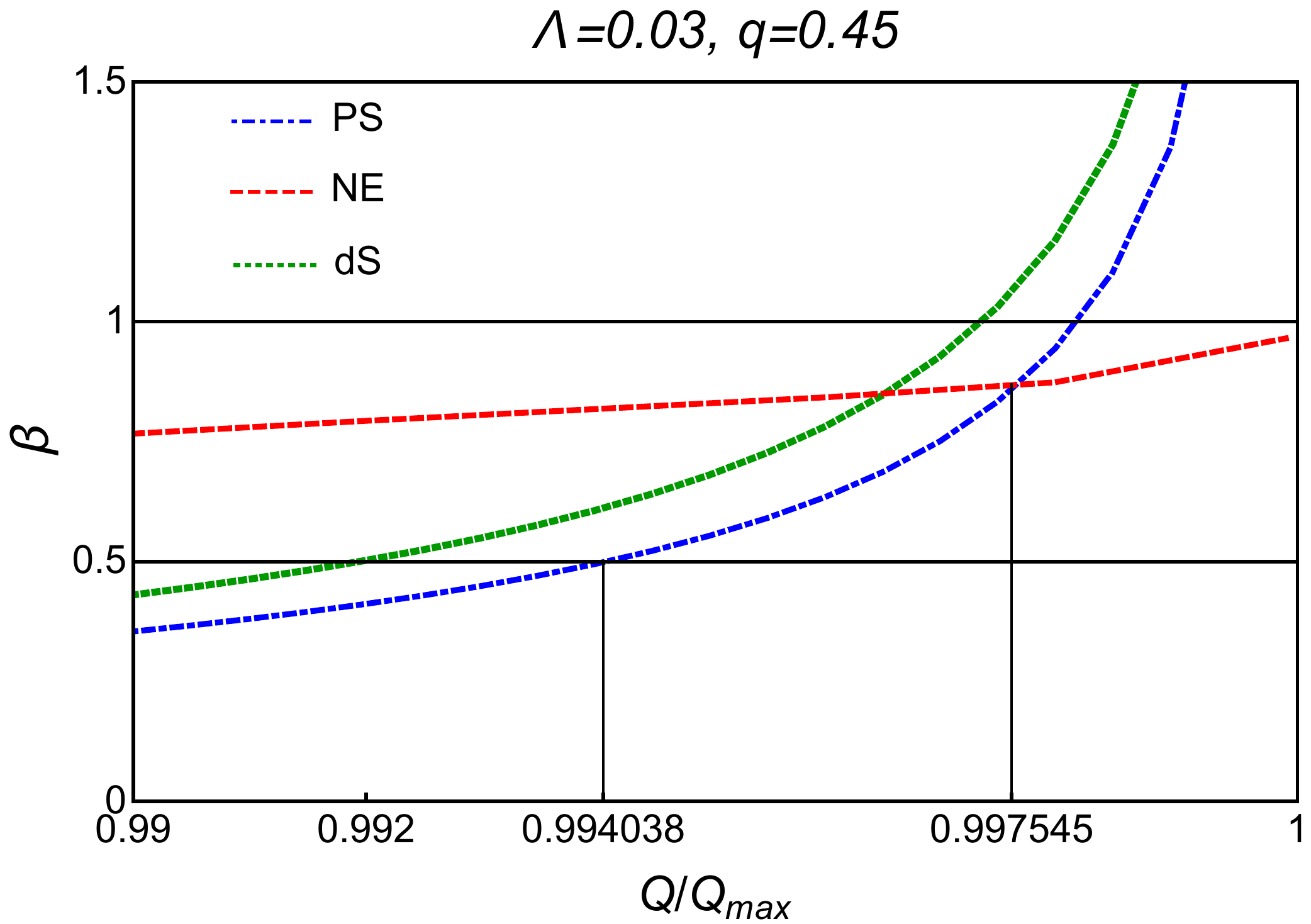}
\endminipage\hfill
\minipage{0.33\textwidth}%
  \includegraphics[width=\linewidth]{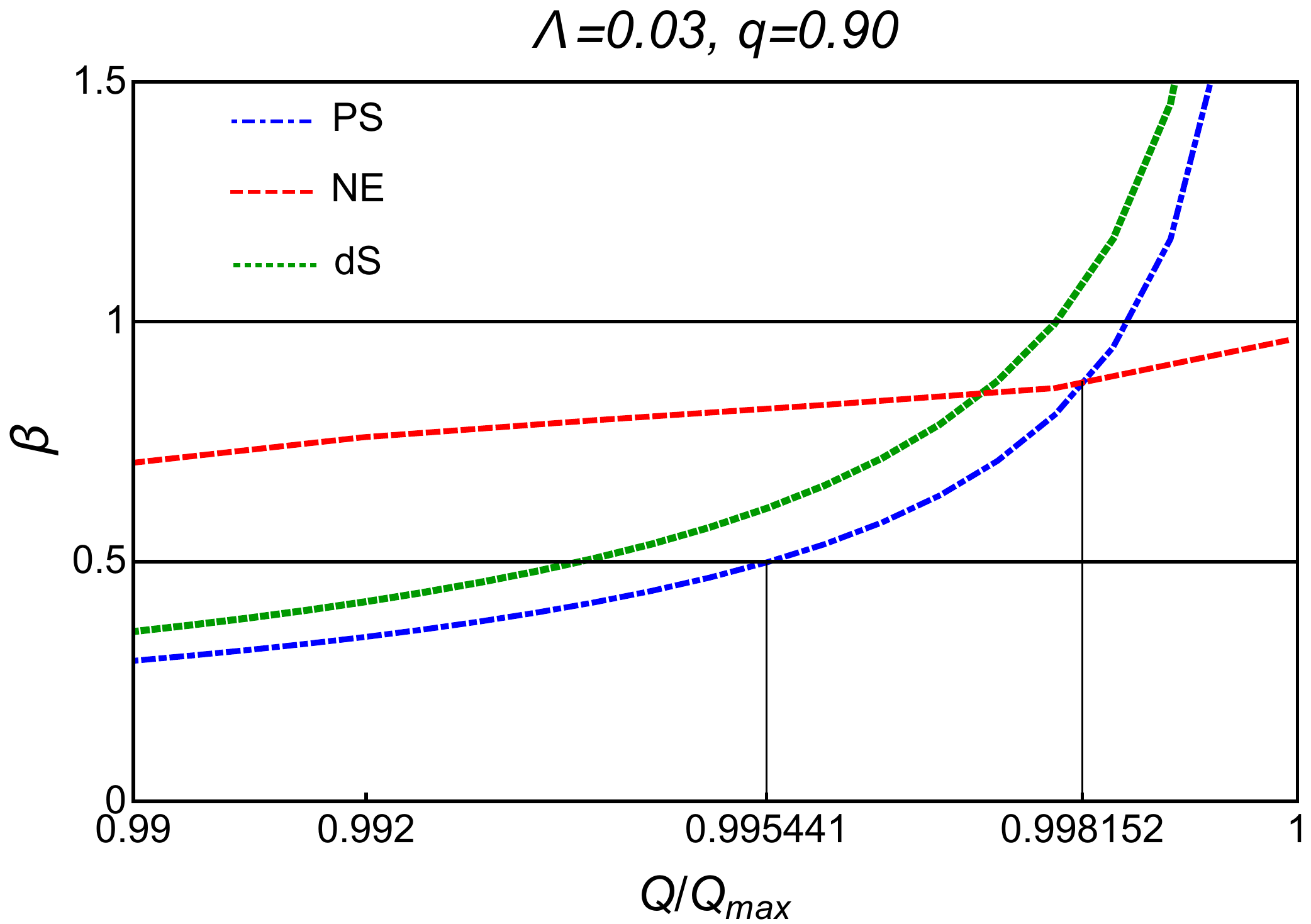}
\endminipage\hfill
\minipage{0.33\textwidth}
  \includegraphics[width=\linewidth]{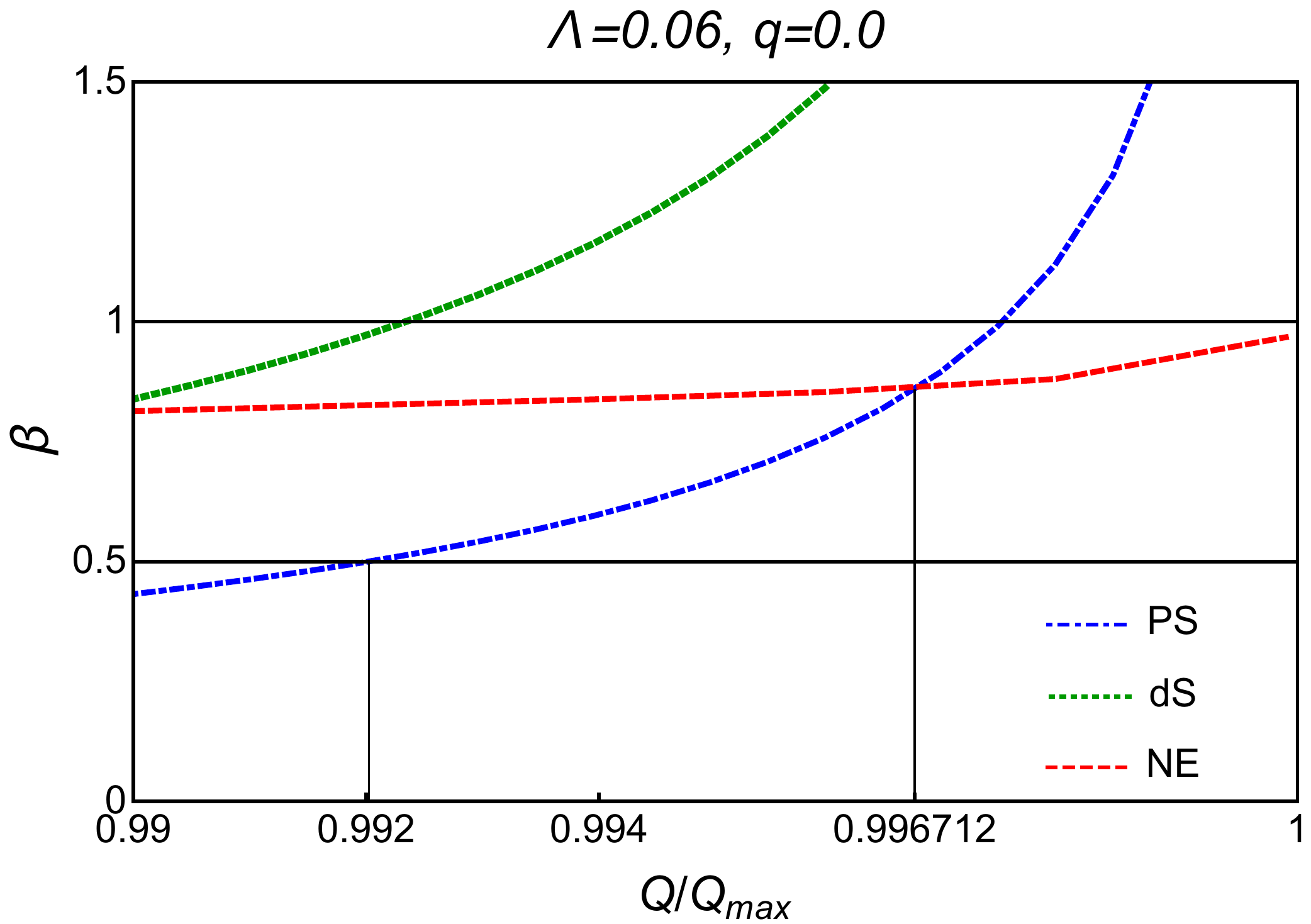}
\endminipage\hfill
\minipage{0.33\textwidth}
  \includegraphics[width=\linewidth]{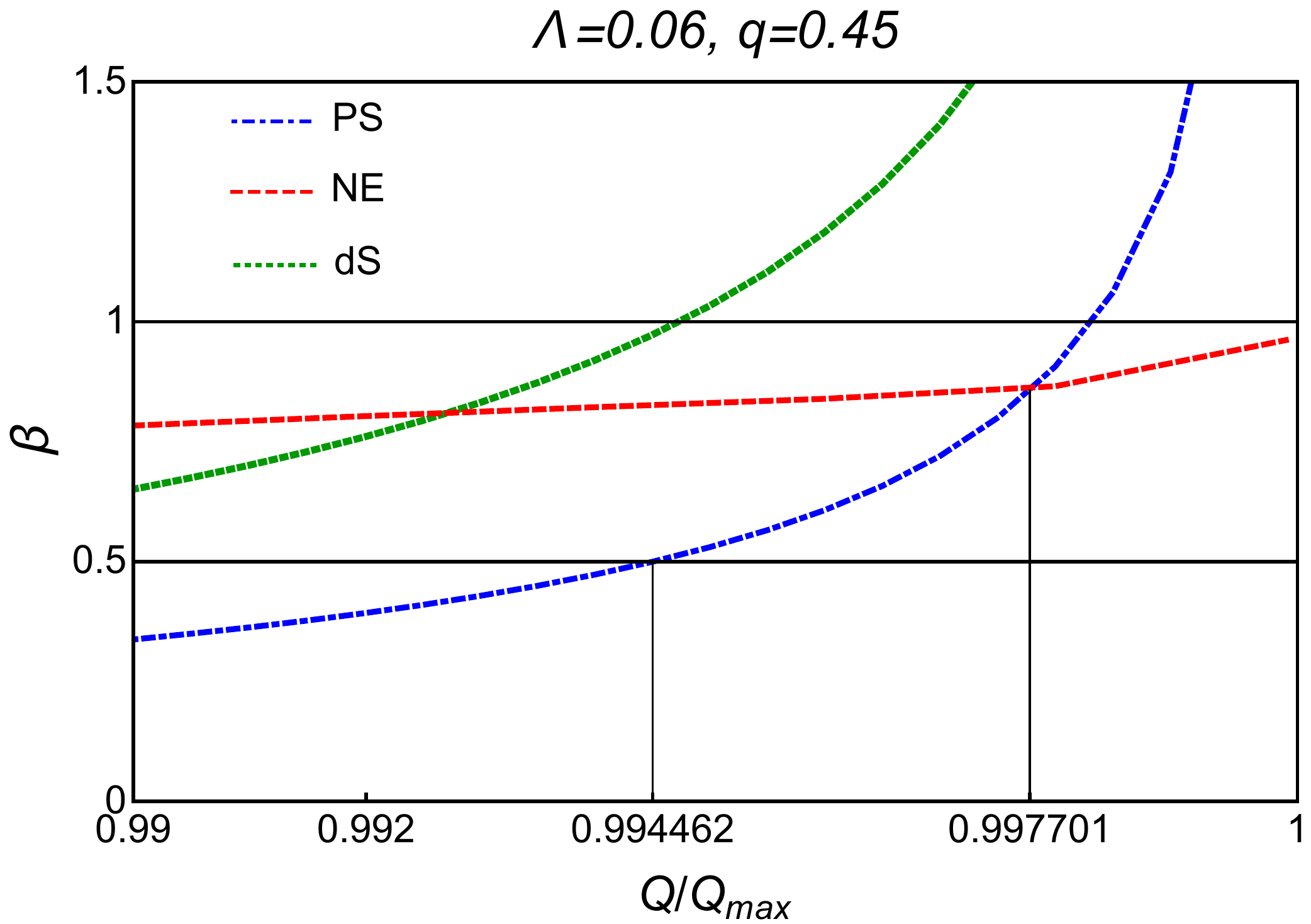}
\endminipage\hfill
\minipage{0.33\textwidth}%
  \includegraphics[width=\linewidth]{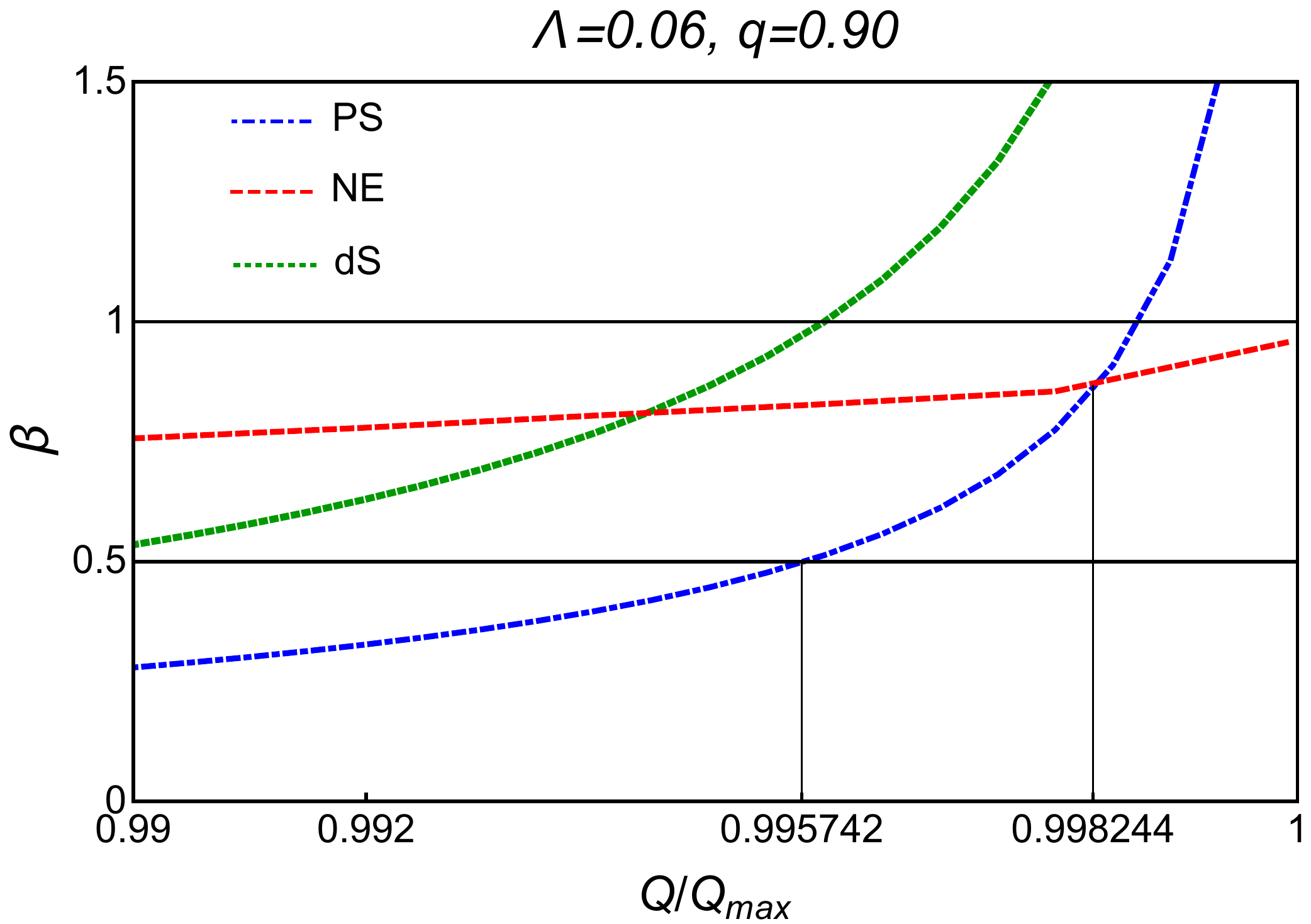}
\endminipage
\caption{The variation of $\beta$ constructed out of the imaginary part of the quasi-normal mode with $(Q/Q_{\rm max})$ has been presented for different values of the cosmological constant and tidal charge parameter $q$ inherited from higher dimensions (for a fixed $\alpha=10^{-5}$). The photon sphere modes (drawn for $\ell=10$) are presented by blue dashed line, while the de Sitter (drawn for $\ell=1$) and the near extremal modes (drawn for $\ell=0$) are depicted by green dashed and red dashed line respectively. The plots for $\beta\equiv -(\textrm{Im}~\omega/\kappa _{-})$ explicitly demonstrate that as the tidal charge parameter $q$ increases, for a fixed $\Lambda$, the violation of strong cosmic censorship conjecture happens at higher and higher values of $Q/Q_{\rm max}$. On the other hand, for a fixed $q$, the de Sitter modes dominate for small values of the cosmological constant, while the photon sphere modes dominate as value of cosmological constant increases. In each of these plots, the first black vertical line corresponds to the value of $(Q/Q_{\rm max})$ where strong cosmic censorship conjecture is violated and the second vertical line presents the location of $(Q/Q_{\rm max})$, where near extremal modes starts to dominate over and above the de Sitter/photon sphere modes. Note that always the near extremal modes starts to dominate before $\beta$ reaches unity.}
\label{dS_NE_PS_Modes}
\end{figure}

\begin{figure}
\centering
\begin{minipage}[b]{0.65\textwidth}
\includegraphics[width=\textwidth]{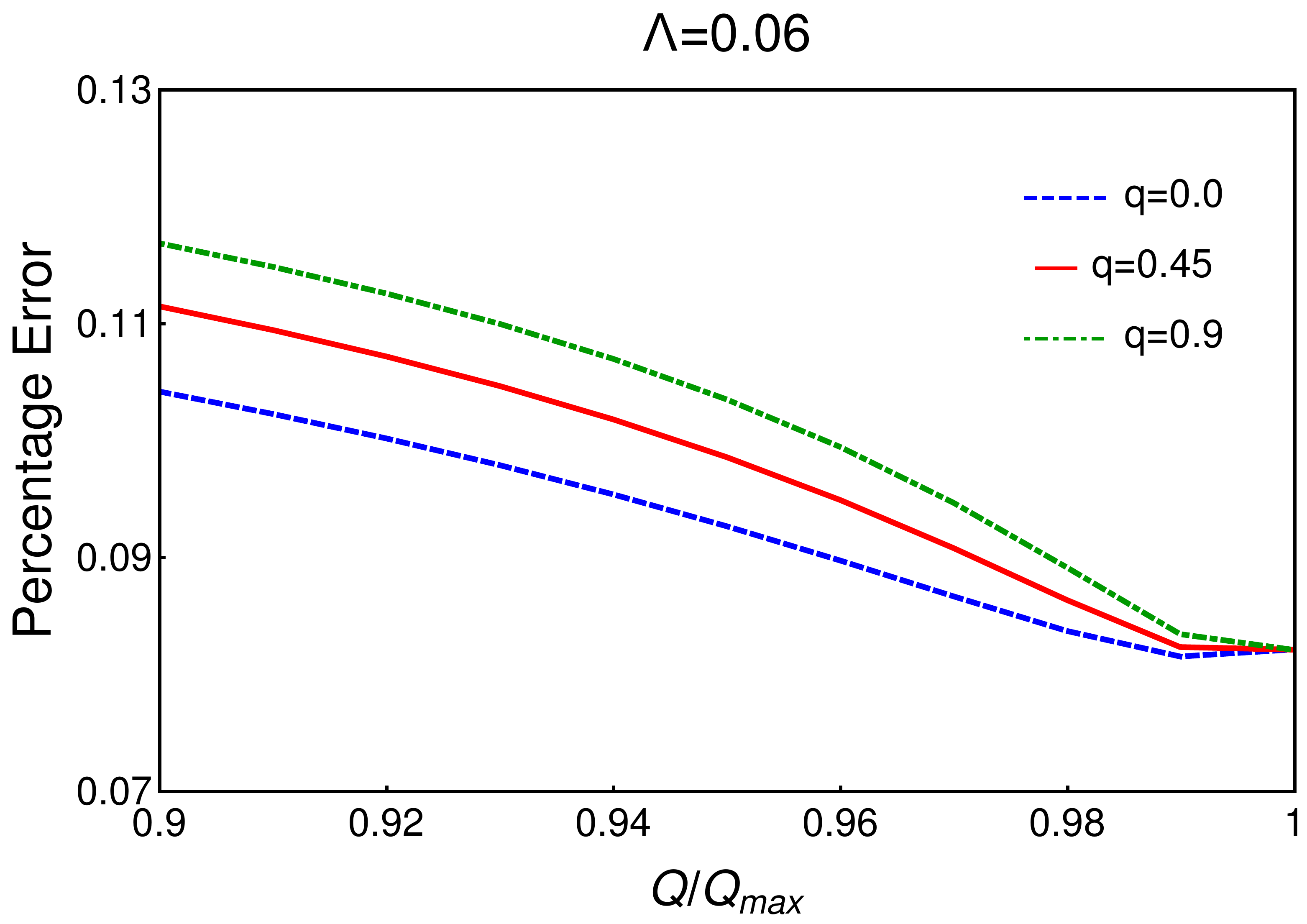}
\end{minipage}
\caption{Percentage error in the analytical expression of the parameter $\beta_{\rm ph}$ against the numerical estimates as a function of $(Q/Q_{\rm max})$ has been depicted for a fixed value of the cosmological constant ($\Lambda=0.06$) for three different choices of the tidal charge $q$. As evident, the error decreases with increase of $q$ and on the average the error stays at $\sim 0.1\%$. However with increase of $(Q/Q_{\rm max})$ the error decreases and reaches a value $\sim 0.08\%$ in the extremal limit.}
\label{Relative_Error}
\end{figure}

In the context of charged asymptotically de Sitter brane world black hole as well, besides photon sphere modes, we will have both de Sitter and near extremal modes. The photon sphere modes have already been discussed, we will now briefly comment on the de Sitter and the near extremal modes as well. Since we are interested in the perturbative modes associated with a test scalar field, both the de Sitter and near extremal modes will parallel the result presented in \cite{Cardoso:2017soq}. In particular, the lowest lying pure de Sitter mode has the following frequencies,
\begin{equation}\label{qnmdS_brane}
\frac{\omega_{0,\textrm{dS}}}{\kappa _{\rm c}}=-i\ell~.
\end{equation}
Here $\kappa _{c}=\sqrt{\Lambda/3}$, where $\Lambda$ is the brane cosmological constant, is the surface gravity associated with the cosmological horizon for a d-dimensional asymptotically de Sitter black hole. Thus the dominant, lowest lying mode corresponds to $\ell=1$ and $n=0$, while the other modes will decay down quickly. Finally for the near extremal modes in the context of four dimensional brane world black hole they become,
\begin{equation}\label{qnmNE_brane}
\omega_{\textrm{NE}}=-i\left(n+\ell+1\right)\kappa _{\rm -}~,
\end{equation}
where, $n$ is an integer and $\ell$ is the angular momentum quantum number. Thus most of the considerations presented in \cite{Cardoso:2017soq} remains unaffected on the brane as long as we consider scalar perturbations. In what follows we have taken into account all of these modes before making any conclusive statement regarding the strong cosmic censorship conjecture for a brane world black hole.

To understand the effect of extra dimensions on the cosmic censorship conjecture, we have plotted the quantity $\beta$ against the Maxwell charge $Q$, normalized to its extremal value $Q_{\rm max}$ for different choices of $q$ and brane cosmological constant $\Lambda$ in \ref{dS_NE_PS_Modes}. The figures in each row of \ref{dS_NE_PS_Modes} clearly demonstrates the effect of extra dimension on cosmic censorship conjecture for a fixed brane cosmological constant. As evident from \ref{dS_NE_PS_Modes} as the ``charge'' $q$ induced from bulk Weyl tensor is increased, the censorship conjecture is still violated, but the violation happens at larger and larger values of the Maxwell charge. This implies that as the effect from extra dimension is increased, the parameter space leading to violation of censorship conjecture becomes smaller. Hence the effect of extra dimension on a brane world black hole is to protect it from violation of cosmic censorship conjecture. This should also be evident from the fact that the influence of extra dimension is to change the charge term to $Q^{2}-q$, thereby reducing the effective charge. In particular, it should be emphasized that for large enough $q$, it is entirely possible to completely change the nature of the spacetime as Cauchy horizon may cease to exist. Further, for small values of brane cosmological constant, the de Sitter modes dominate the decay of perturbations, while for larger values of $\Lambda$, the photon sphere modes dominate as evident from \ref{dS_NE_PS_Modes}. The above figure also demonstrates that the near extremal modes dominate over either de Sitter or photon sphere modes as $\beta$ approaches unity, which is another desired property of any spacetime.

The analysis presented above computes the quasi-normal modes numerically, not only for photon sphere modes but also for de Sitter and near extremal modes as well. To provide confidence over our numerical estimation, the error/discrepancy between analytical expectation and numerical estimate of $\beta$ has been presented in \ref{Relative_Error}. It is clear from the figure that the percentage error decreases with increasing $q$ while remaining at an average level of $\sim 0.1\%$. While in the extremal limit the error decreases and reaches a consistent value of $\sim 0.08\%$. This is because the near extremal behaviour is governed by the near extremal modes and they are independent of the tidal charge $q$. Thus in the context of a four dimensional black hole, embedded in a higher dimensional spacetime, the cosmic censorship conjecture is only weakly violated or, not violated at all, depending on the ``charge'' $q$ inherited from the bulk Weyl tensor. 
\subsection{The Case of Higher Dimensional Reissner-Nordstr\"{o}m-de Sitter Black Hole}

As another application of the general formalism derived in the previous section to asses the validity of the cosmic censorship conjecture, we consider a higher dimensional \RN-\dS\ black hole in this section. This will provide another avenue to understand how the presence of higher dimensions influence the validity of the cosmic censorship conjecture. The metric element of a d-dimensional \RN-\dS\ black hole is given by the line element as in \ref{staticspherically} with the metric coefficients $f(r)$ and $g(r)$ taking the following form \cite{Cai:2001tv, refId0},
\begin{equation}\label{RNdS}
f(r)=g(r)=1-\frac{\varpi_{d-2}M}{r^{d-3}}+\frac{(d-2)\varpi_{d-2}^{2}}{8(d-3)}\frac{Q^{2}}{r^{2d-6}}-\frac{2\Lambda}{(d-1)(d-2)}r^{2}~.
\end{equation}
Here $M$ denotes the mass of the black hole, $Q$ represents the electric charge of the Maxwell field in the spacetime and $\Lambda$ is the cosmological constant. Further the constant $\varpi_{d-2}$ appearing in \ref{RNdS} is a purely dimension dependent factor and is given by
\begin{equation}\label{varpi}
\varpi_{d-2}=\frac{\Gamma(\frac{d-1}{2})}{(d-2)\pi^{\frac{d-3}{2}}}~.
\end{equation}
The positions of the horizon can be found by solving the equation $f(r)=0$ for any general $M$, $Q$ and $\Lambda$. It turns out that the corresponding equation has three real and positive roots, denoted by $r_{-}$, $r_{+}$ and $r_{c}$ respectively. Since these roots follow the inequality, namely $r_{-}\leq r_{+}\leq r_{c}$, they in turn define the position of the Cauchy horizon, the Event horizon and the Cosmological horizon, respectively. Given the above metric elements, one can immediately determine the surface gravity at the Cauchy horizon, which becomes 
\begin{equation}\label{surfacegravity}
\kappa^{(-)}_{\rm RNdS}=\frac{f'(r_{-})}{2}=\frac{1}{2}\left[\frac{(d-3)\varpi_{d-2}M}{r_{-}^{d-2}}-\frac{(d-2)\varpi_{d-2}^{2}}{4}\frac{Q^{2}}{r_{-}^{2d-5}}-\frac{4\Lambda}{(d-1)(d-2)}r_{-}\right]~,
\end{equation}
where $r_{-}$ is the location of the Cauchy horizon. This provides the first step in the computation of the parameter $\beta$ introduced in \ref{betastatic}. The second ingredient corresponds to the Lyapunov exponent $\lambda$, which in the context of \RN-\dS\ black hole can be determined using \ref{lyanospin} and \ref{RNdS}, yielding
\begin{equation}\label{lyaRNdSind}
\lambda_{\rm RNdS}=\sqrt{\frac{f(r_{\rm ph})}{r_{\rm ph}^{2}}\left[1+\frac{\varpi_{d-2}M}{r_{\rm ph}^{d-3}}\left\{\frac{(d-1)(d-4)}{2}\right\}-\frac{(d-2)\varpi_{d-2}^{2}}{8(d-3)}\frac{Q^{2}}{r_{\rm ph}^{2d-6}}\left\{(d-3)(2d-5)-1\right\}\right]}~.
\end{equation}
Here $r_{\rm ph}$ stands for the photon circular orbit, which is to be determined by the condition $2f(r)=rf'(r)$ written down in \ref{rphnospin} by using the expression for $f(r)$ presented in \ref{RNdS}. 
\begin{figure}[!htb]
\minipage{0.32\textwidth}
  \includegraphics[width=\linewidth]{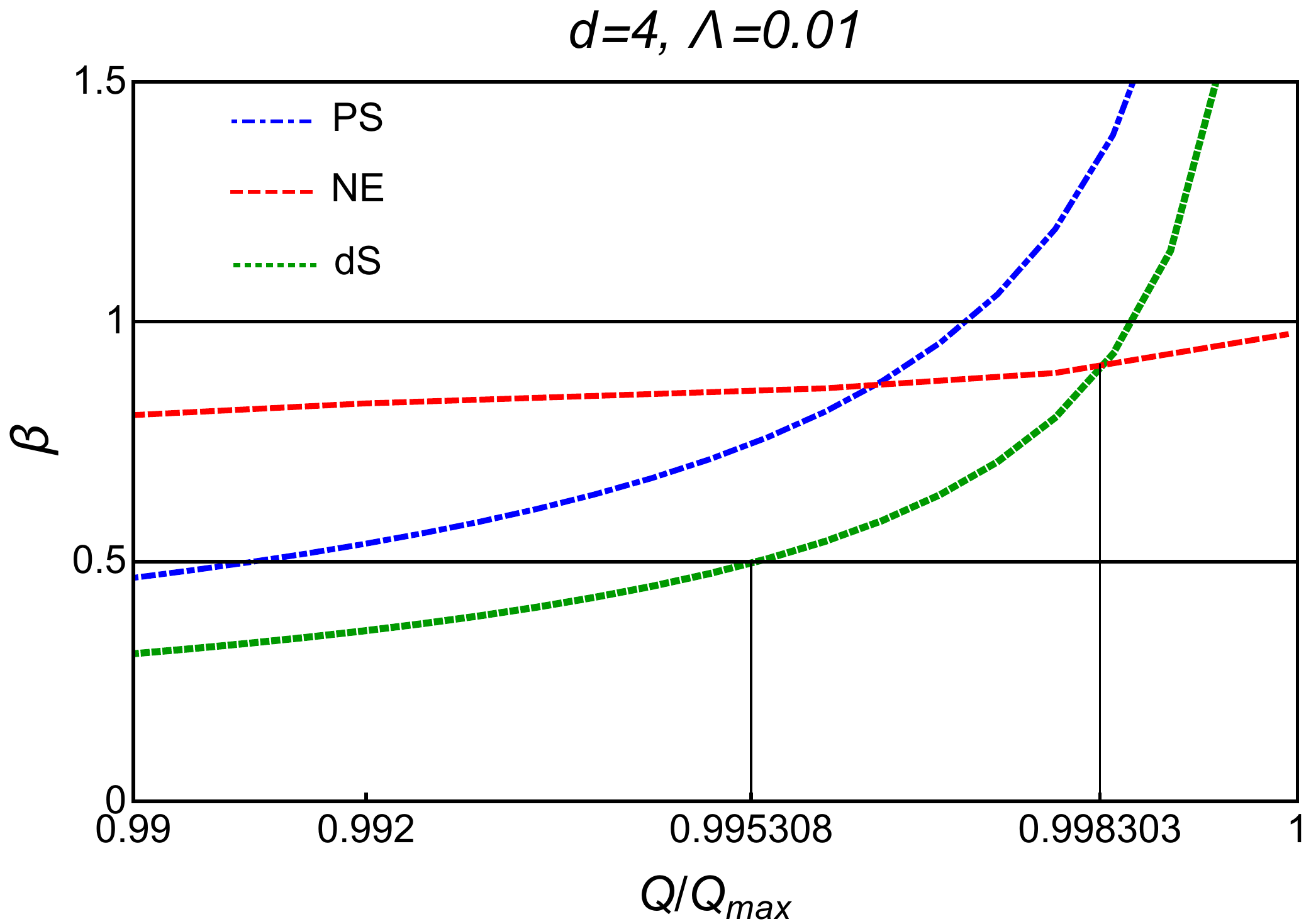}
\endminipage\hfill
\minipage{0.32\textwidth}
  \includegraphics[width=\linewidth]{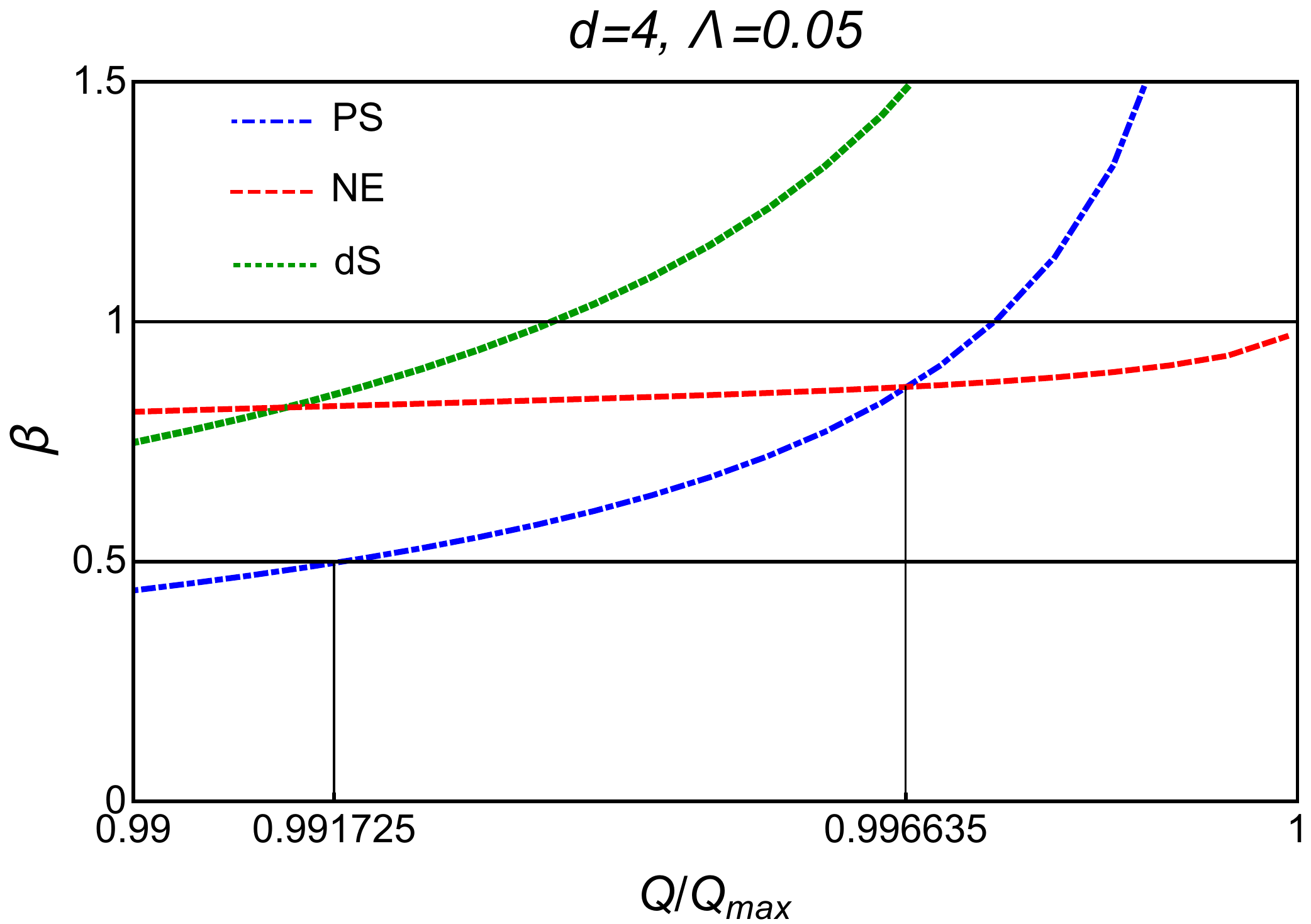}
\endminipage\hfill
\minipage{0.32\textwidth}%
  \includegraphics[width=\linewidth]{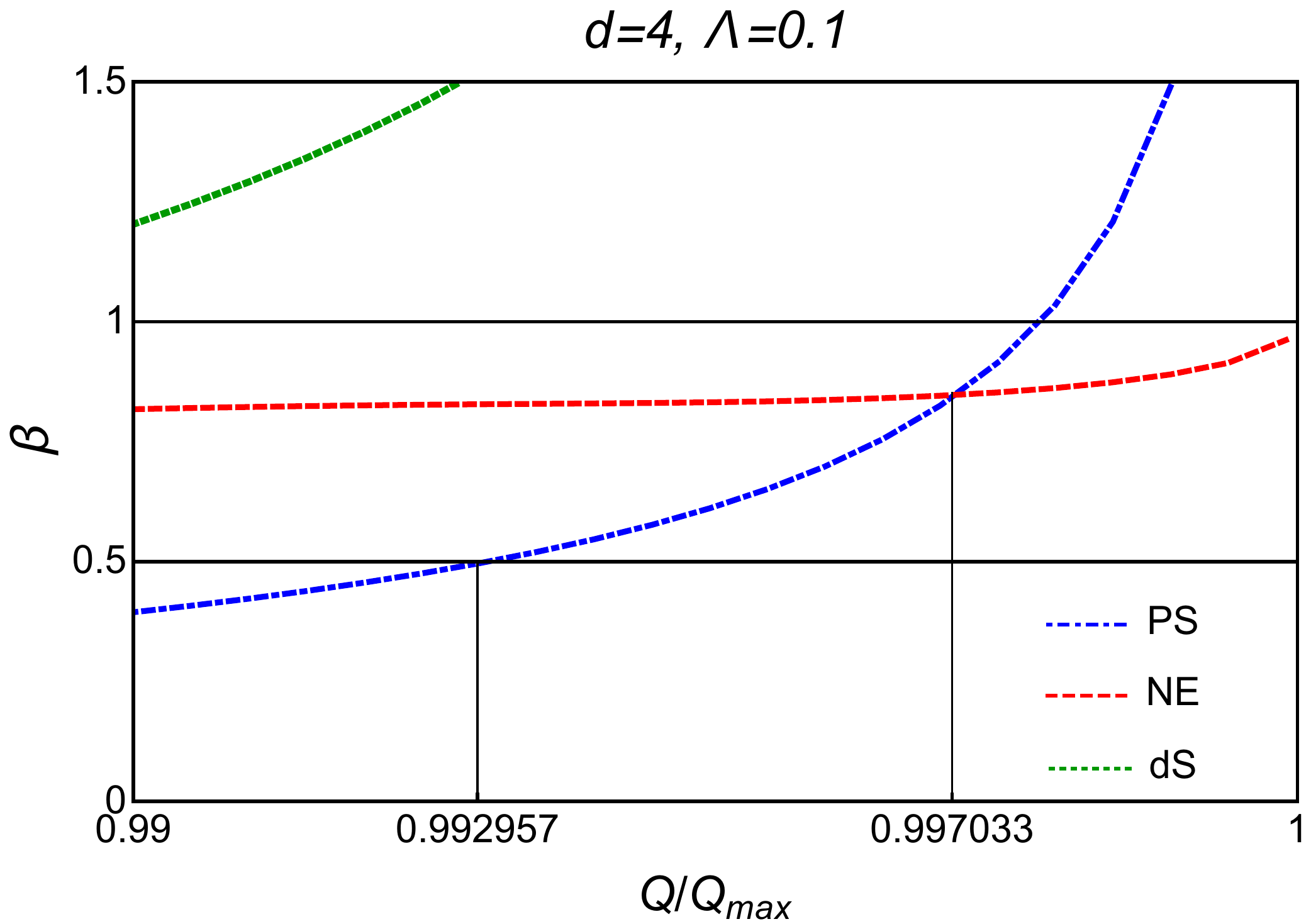}
\endminipage\hfill
\minipage{0.32\textwidth}
  \includegraphics[width=\linewidth]{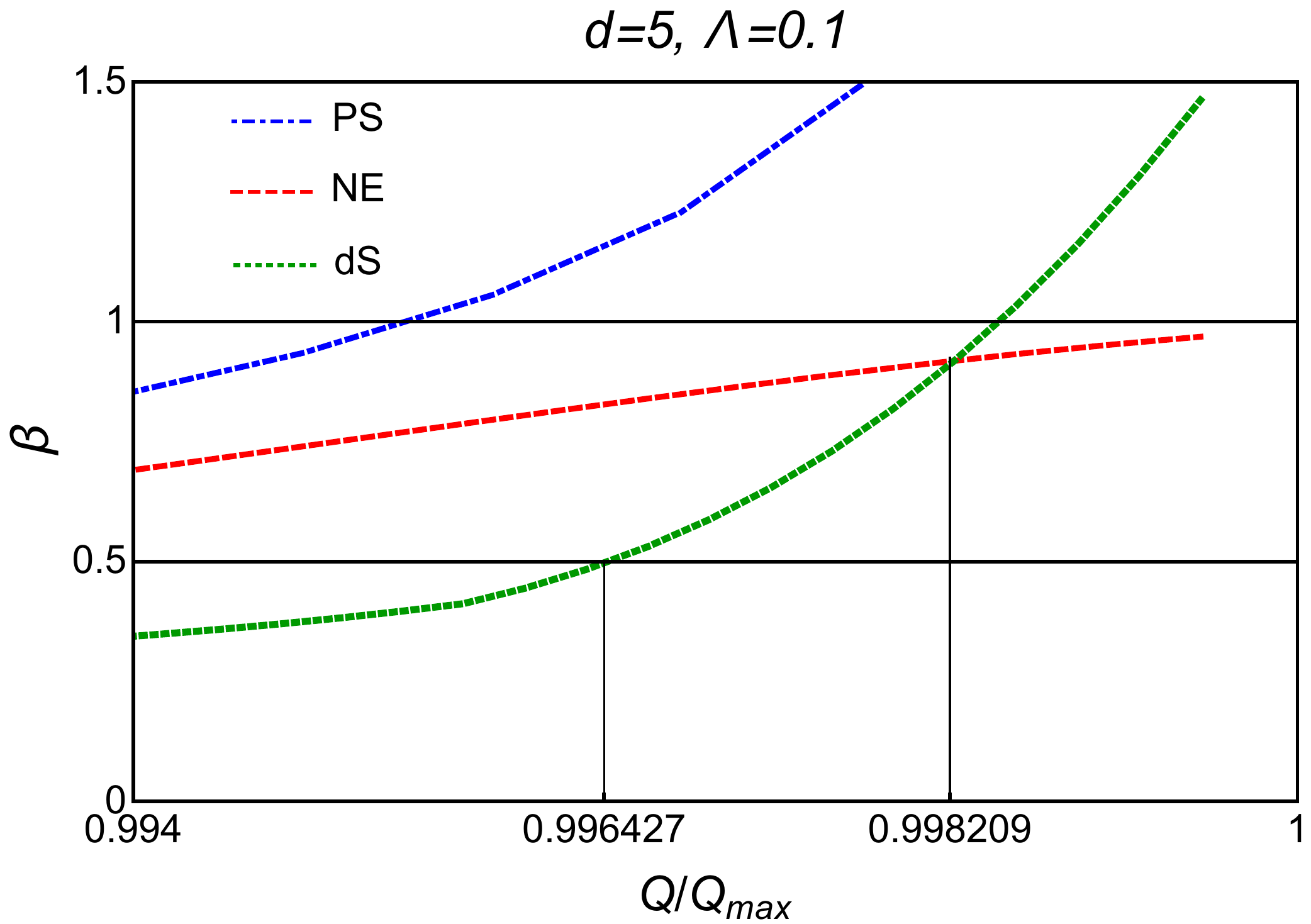}
\endminipage\hfill
\minipage{0.32\textwidth}
  \includegraphics[width=\linewidth]{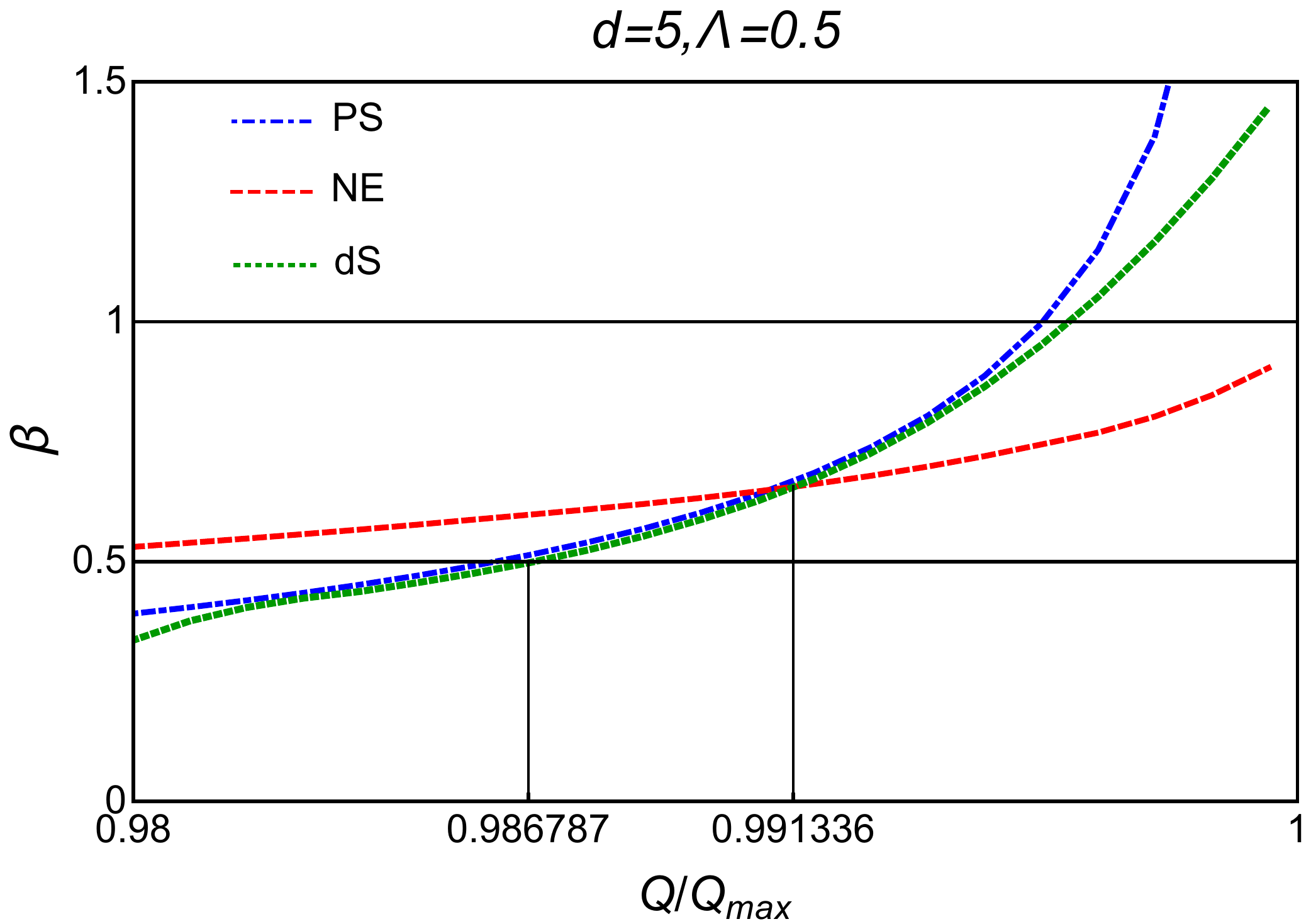}
\endminipage\hfill
\minipage{0.32\textwidth}%
  \includegraphics[width=\linewidth]{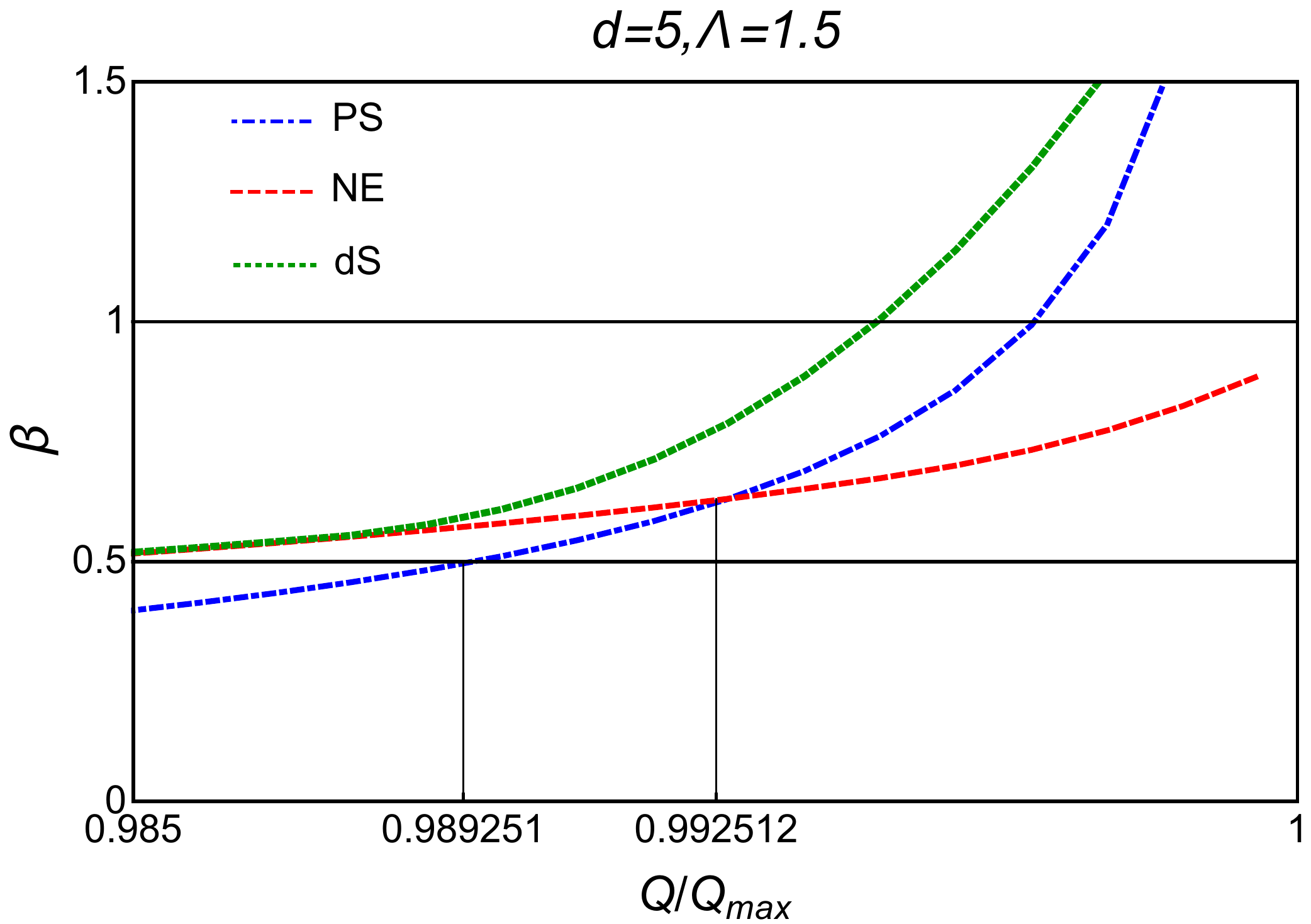}
\endminipage\hfill
\minipage{0.32\textwidth}
  \includegraphics[width=\linewidth]{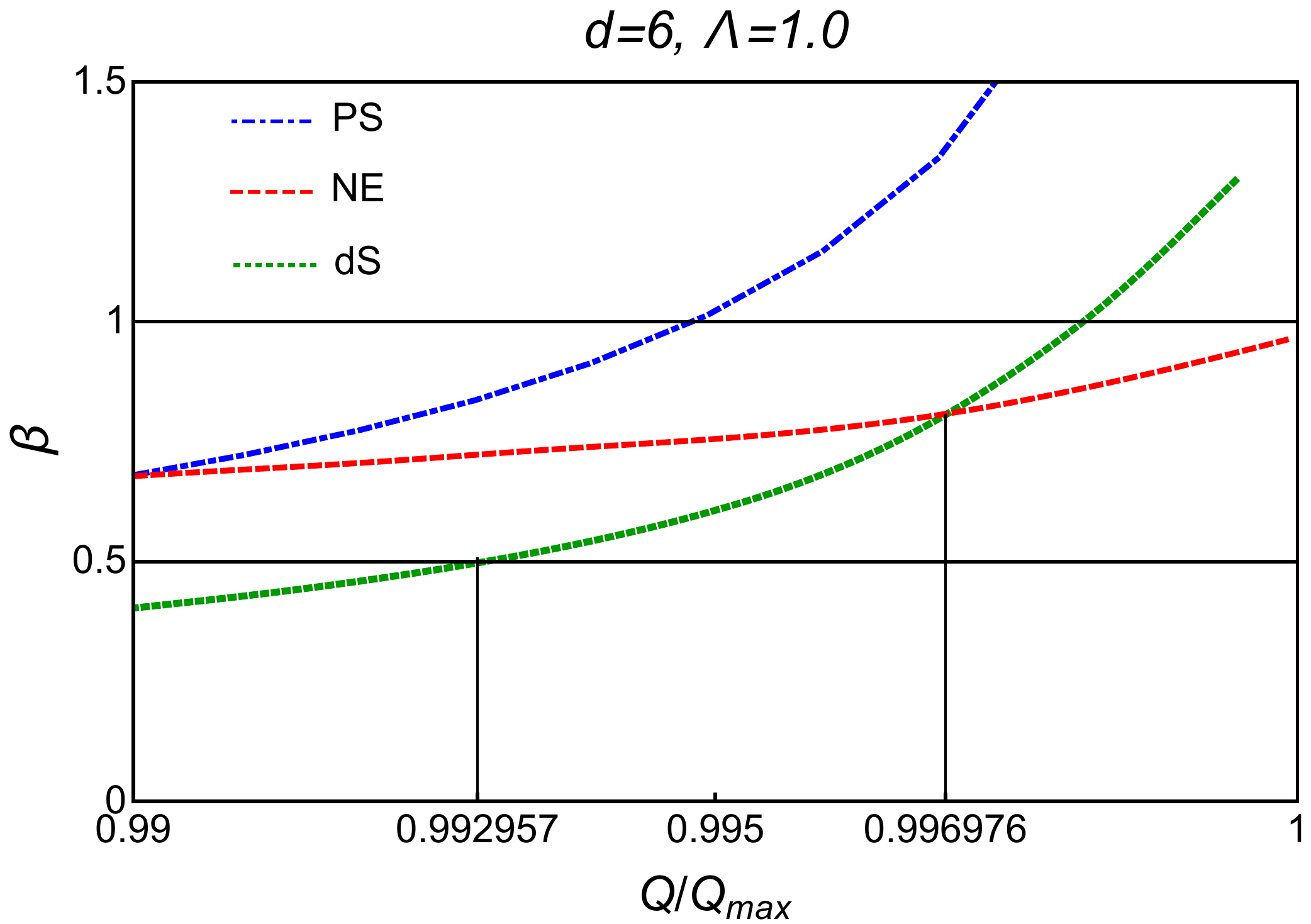}
\endminipage\hfill
\minipage{0.32\textwidth}
  \includegraphics[width=\linewidth]{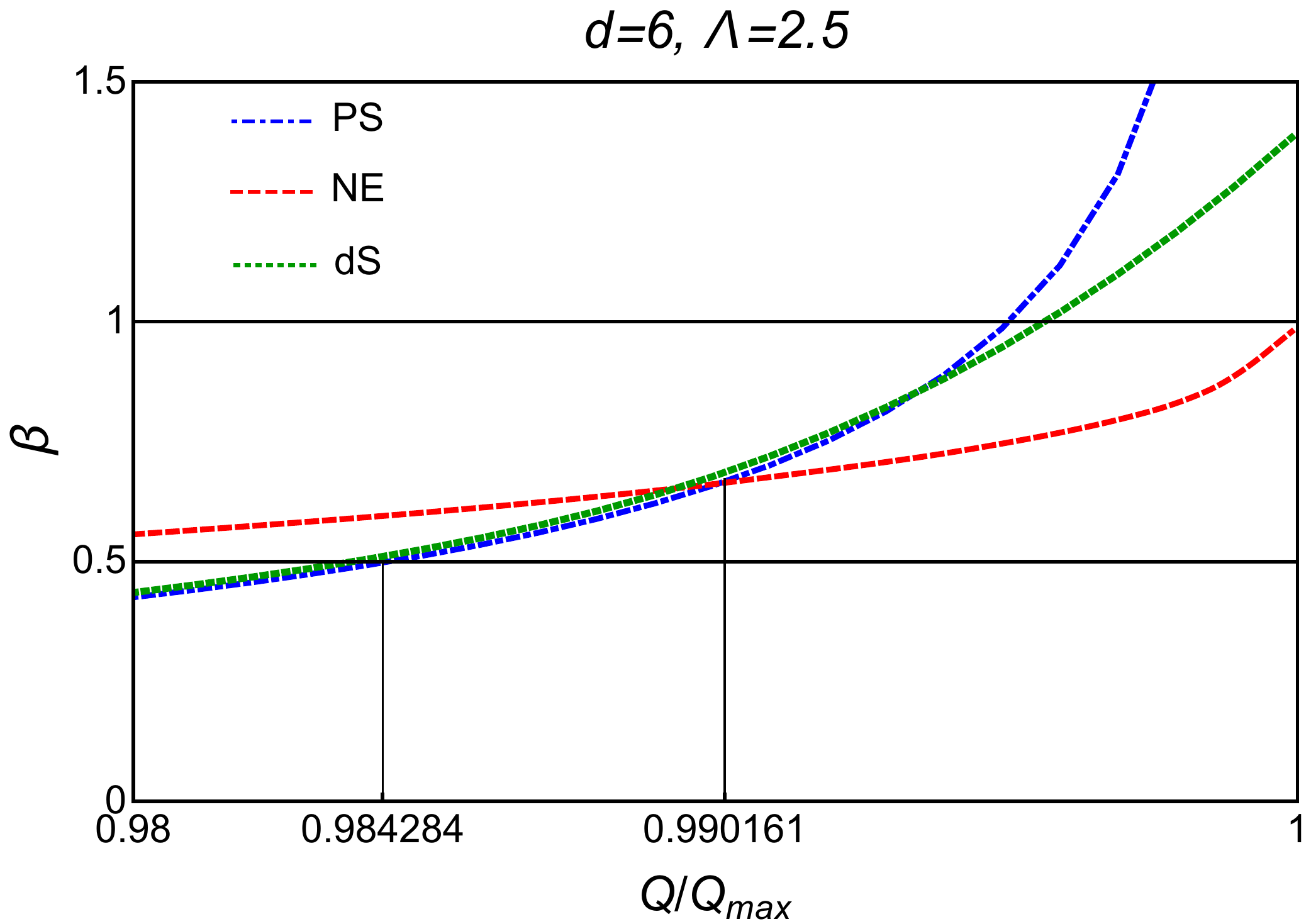}
\endminipage\hfill
\minipage{0.32\textwidth}%
  \includegraphics[width=\linewidth]{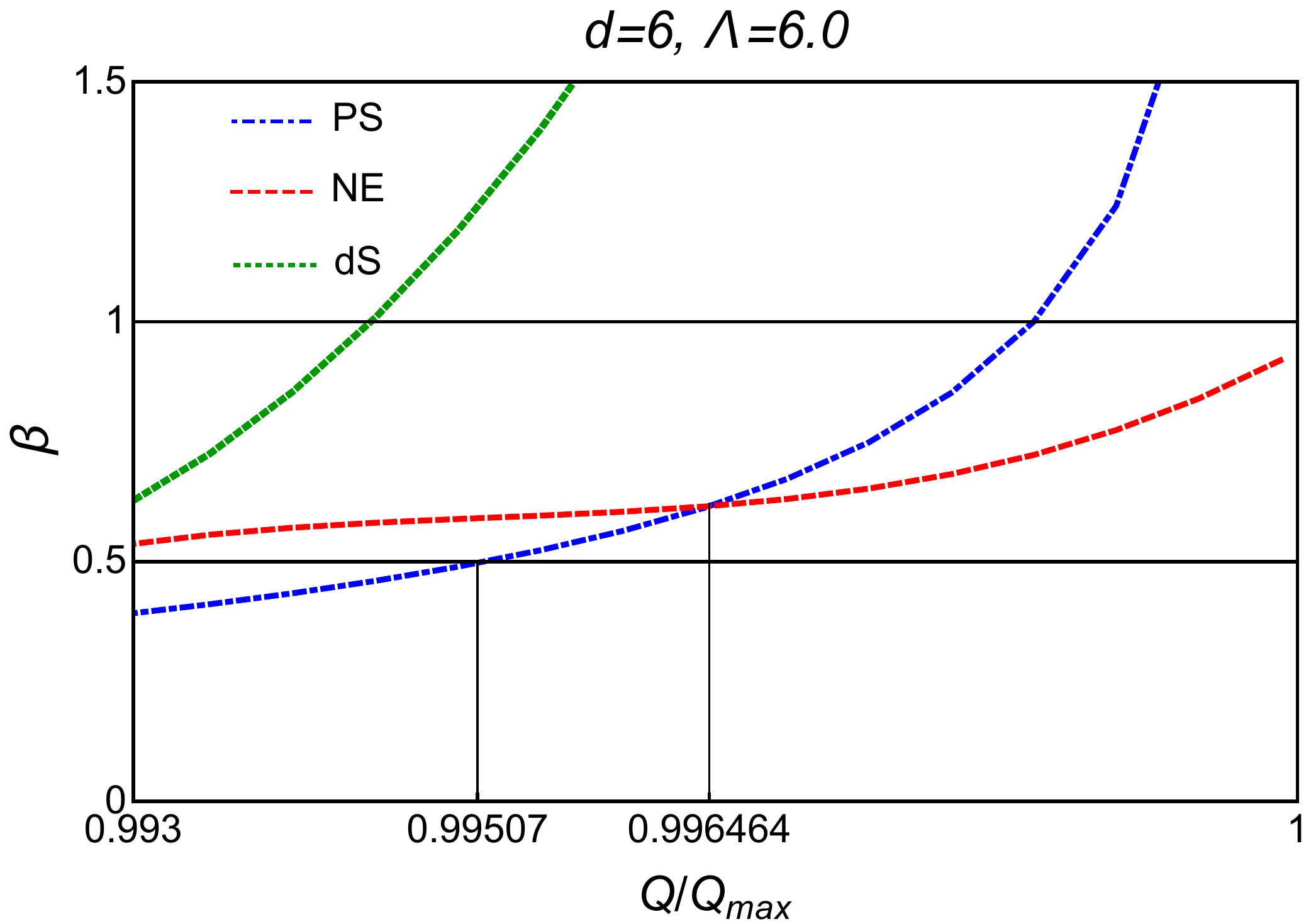}
\endminipage
\caption{The variation of $\beta$, related to the imaginary part of the quasi-normal mode frequency, with $(Q/Q_{\rm max})$ has been presented for different spacetime dimension $d$ and different choices of the cosmological constant $\Lambda$. The plots for $\beta \equiv -(\textrm{Im}~\omega/\kappa _{-})$ explicitly demonstrate that as the spacetime dimension increases, the value of $(Q/Q_{\rm max})$, where the quasi-normal modes cross $\beta=(1/2)$, becomes smaller. Thus the parameter space associated with violation of strong cosmic censorship conjecture becomes larger. Note that ultimately, the near extremal modes (red dashed line) starts dominating over the photon sphere modes (blue dashed line) or the de Sitter modes (green dashed line) before reaching $\beta=1$. Each rows of these plots are for a fixed spacetime dimension but for different choices of the cosmological constant. The first black vertical line corresponds to the value of $(Q/Q_{\rm max})$ where strong cosmic censorship conjecture is violated and the second vertical line presents the location of $(Q/Q_{\rm max})$, where near extremal modes (drawn for $\ell=0$) starts to dominate over and above the photon sphere modes (drawn for $\ell=10$).}
\label{RNDSNE}
\end{figure}
For curiosity we also write down the expression for the Lyapunov exponent in four spacetime dimensions, which takes the following form
\begin{equation}\label{lyaRNdSin4}
\lambda_{\rm RNdS}^{(d=4)}=\sqrt{\frac{f(r_{\rm ph})}{r_{\rm ph}^{2}}\left(1-\frac{2Q^{2}}{r_{\rm ph}^{2}}\right)}~.
\end{equation}
Note that in four dimensions the Lyapunov exponent does not depend on the Mass parameter of the black hole, while in higher dimension it does. This has to do with the fact that the mass parameter in the higher dimensional Lyapunov exponent presented in \ref{lyaRNdSind} appears with a pre-factor of $(d-4)$. This gives us a hint that in presence of higher dimensions the Lyapunov exponent behaves differently. Proceeding further, since we have both the surface gravity and the Lyapunov exponent at our hand, we can immediately compute the parameter $\beta_{\rm ph}$ for static and spherically symmetric situation using \ref{betastatic}, which becomes
\begin{equation}\label{betaRNdS}
\beta^{\rm RNdS}_{\rm ph}=\frac{\sqrt{\frac{f(r_{\rm ph})}{r_{\rm ph}^{2}}\left[1+\frac{\varpi_{d-2}M}{r_{\rm ph}^{d-3}}\left(\frac{(d-1)(d-4)}{2}\right)-\frac{(d-2)\varpi_{d-2}^{2}}{8(d-3)}\frac{Q^{2}}{r_{\rm ph}^{2d-6}}\left((d-3)(2d-5)-1\right)\right]}}{\frac{(d-3)\varpi_{d-2}M}{r_{-}^{d-2}}-\frac{(d-2)\varpi_{d-2}^{2}}{4}\frac{Q^{2}}{r_{-}^{2d-5}}-\frac{4\Lambda}{(d-1)(d-2)}r_{-}}~.
\end{equation} 
Here as well, $r_{\rm ph}$ stands for the photon circular orbit and $r_{-}$ represents the location of the Cauchy horizon. Thus the parameter $\beta_{\rm ph}^{\rm RNdS}$ is dependent on the Mass $M$, Charge $Q$ and the Cosmological constant $\Lambda$ through both explicit presence of these terms in \ref{betaRNdS}, as well as through implicit dependence of these parameters on the photon circular orbit $r_{\rm ph}$ and the Cauchy horizon $r_{-}$. Thus if for any range of the above parameters, before the black hole turns extremal, the value of $\beta^{\rm RNdS}_{\rm ph}$ becomes larger than half it follows that violation of cosmic censorship conjecture will take place, as far as the photon sphere modes are considered. 

As pointed out in \cite{Cardoso:2017soq}, in four dimensions there indeed exists a certain region of parameter space where $\beta\equiv -\{\textrm{Im}(\omega)\}_{\rm min}/\kappa_{-}$ is larger than half and hence the four dimensional \RN-\dS\ black holes indeed violates cosmic censorship conjecture. To see the effect of higher spatial dimensions on violation of cosmic censorship conjecture, we have plotted $\beta$ as a function of $(Q/Q_{\rm max})$ and it is clear from \ref{RNDSNE} that strong cosmic censorship conjecture gets violated near the extremal limit. This result is independent of the number of spacetime dimensions. Thus the presence of higher dimensions do \emph{not} save the doomsday. Rather, as evident from \ref{RNDSNE}, violation of strong cosmic censorship conjecture can be more severe for higher dimensional charged black holes, depending upon the value of the cosmological constant $\Lambda$. This is because the value of $Q/Q_{\rm max}$, where $\beta$ becomes greater than $(1/2)$, is smaller for higher dimensional black holes, implying a larger parameter space where deterministic nature of general relativity breaks down (see, e.g., the middle column of \ref{RNDSNE}). In other words, for certain choices of the cosmological constant, the deterministic nature of higher dimensional Einstein's equations are more of a concern than the usual four dimensional ones. 

Using numerical techniques, besides the photon sphere modes, we have also presented the near extremal modes as well. Alike the case of brane world black hole, for higher dimensional \RN-\dS\ black hole as well the near extremal modes start to dominate over and above the photon sphere or de Sitter modes as the black hole reaches near extremality and keeps $\beta$ less than unity. Further, as evident from \ref{RNDSNE}, for small values of the cosmological constant the de Sitter modes govern the decay rate of scalar perturbation, while for higher values of the cosmological constant the photon sphere modes start to dominate. This behaviour appears to be generic and holds true for both five and six dimensional charged de Sitter black hole. Further \ref{RNDSNE} also demonstrates that for higher dimensions the lowest lying quasi-normal modes crosses the $\beta\equiv -\textrm{Im}~\omega/\kappa _{-}=(1/2)$ line, leading to violation of strong cosmic censorship conjecture, for smaller values of the charge parameter $(Q/Q_{\rm max})$ in higher dimensions (for certain choices of the cosmological constant). This is in exact agreement with our analytical estimations as well. This further shows that the analytical results derived earlier in this section are in consonance with our numerical estimations. This helps one to conclude that violation of strong cosmic censorship conjecture is a generic feature of \RN-\dS\ black holes irrespective of spacetime dimensions.
\section{Strong Cosmic Censorship Conjecture for Rotating Black Holes: General Analysis}\label{SCC_Rotating}

In the previous two sections we have been discussing the cosmic censorship conjecture for static and spherically symmetric black holes. However in astrophysical scenarios all the black holes are rotating and hence they must be represented by Kerr-like solutions. Interestingly, in the context of cosmic censorship conjecture it has been found that even though the \RN-\dS\ black hole indeed violates the censorship conjecture, the Kerr-de Sitter black holes do not \cite{PhysRevD.97.104060}. Since in the previous section, we have shown that presence of higher dimensions lead to a stronger violation of the cosmic censorship conjecture for charged black holes, it is legitimate to ask what happens for a higher dimensional rotating black holes. In particular, whether presence of higher dimensions can lead to a violation of cosmic censorship conjecture even for rotating black holes is the question we would like to answer in this section.    

For generality and wider applicability, we will first determine the Lyapunov exponent in a general rotating spacetime and hence the quantity $\beta$. This will enable us to apply the formalism to any rotating metric that becomes available in the future. Since the trajectory we are interested in lies in the four-dimensional spacetime, specifically on the equatorial plane, we can write down the metric elements using the following metric ansatz
\begin{equation}\label{generalrotating}
ds^{2}=-(e^{2\nu}-\omega^{2}e^{2\psi}) dt^{2}-2\omega e^{2\psi} dtd\phi+e^{2\mu_{2}}dr^{2}+e^{2\psi}d\phi^{2}~,
\end{equation}
where, $\nu$, $\psi$, $\omega$ and $\mu_{2}$ are arbitrary functions of the radial coordinate alone, since we are working in the equatorial plane. For static and spherically symmetric spacetime discussed earlier, one has the following correspondence: $\omega=0$, $e^{2\psi}=r^{2}$ along with $e^{2\nu}=f(r)$ and $e^{-2\mu_{2}}=g(r)$. Hence whether we can reproduce the result for Lyapunov exponent presented in \ref{lyanospin} starting from the result in the context of rotating black hole will show the correctness of the derived result. 

In order to determine the Lyapunov exponent for the general rotating black hole spacetime, we need to determine the potential associated with the radial null geodesics and hence obtain the equatorial photon circular orbits in this rotating black hole spacetime. Since the metric is independent of both time and azimuthal coordinate $\phi$, it follows that there exists conserved quantities, e.g., the energy $E$ and the angular momentum $L$. These conserved quantities can be expressed in terms of the metric elements as,
\begin{equation}\label{energymomentum}
\begin{aligned}
E&=(e^{2\nu}-\omega^{2}e^{2\psi})\dot{t}+\omega e^{2\psi}\dot{\phi}~;
\\
L&= e^{2\psi}\dot{\phi}-\omega e^{2\psi}\dot{t}~.
\end{aligned}
\end{equation}
Since we are working on the equatorial plane characterized by $\theta=\pi/2$, the radial equation of motion of a particle can be easily determined with the help of the relation: $p^{\alpha}p_{\alpha}=m^{2}\delta_{1}$. For spacelike or timelike trajectory $\delta _{1}=\pm 1$, while for null trajectory $\delta _{1}=0$. Even though we have kept $m$ in the above relation, it is beneficial to work with particles having unit mass. This leads to the following radial equation of motion \cite{Gyulchev:2006zg}
\begin{equation}\label{radialgeodesics}
\dot{r}^{2}=e^{-2\mu_{2}}\left[\delta_{1}+e^{-2\nu}E^{2}-2\omega e^{-2\nu}EL-(e^{-2\psi}-\omega^{2}e^{-2\nu})L^{2}\right]~.
\end{equation}
Since the determination of the Lyapunov exponent depends on the photon circular orbit, in this work we will exclusively consider null  geodesics corresponding to $\delta_{1}=0$. Thus for such null trajectories, the radial geodesic equation presented above in \ref{radialgeodesics}, can be casted as
\begin{equation}\label{nullradialgeodesics}
\dot{r}^{2}\equiv V_{\rm eff}(r)=e^{-2\mu_{2}}e^{-2\nu}E^{2}e^{-2\psi}\Big[e^{2\psi}-2\omega e^{2\psi}\ell+\left(\omega ^{2}e^{2\psi}- e^{2\nu}\right)\ell^{2}\Big]~,
\end{equation}
where $\ell=L/E$ is the specific angular momentum associated with the motion along null geodesic depicted above, which sometimes is also refereed to as the impact parameter. However, in this work, we will refer it to the specific angular momentum. As an immediate verification of this result one can explicitly check that in the context of static and spherically symmetric spacetime (with $\omega=0$, $e^{2\nu}=f(r)$, $e^{-2\mu_{2}}=g(r)$ and $e^{2\psi}=r^{2}$) this equation reduces to the one presented in \ref{potentialnospin}.

As evident from \ref{nullradialgeodesics} it immediately follows that it would be advantageous to introduce the following definitions, instead of working with $e^{2\nu}$, $e^{2\psi}$ and other metric components explicitly,
\begin{equation}\label{redefination}
\mathcal{A}(r)\equiv -e^{2\nu}+\omega^{2}e^{2\psi}~;
\qquad
\mathcal{B}(r)\equiv \omega e^{2\psi};
\qquad
\mathcal{C}(r)\equiv e^{2\psi}~.
\end{equation}
Thus, given the above definitions, one can compute the following quantity, namely $\mathcal{B}^{2}-\mathcal{A}\mathcal{C}=e^{2\nu}e^{2\psi}$. Hence the radial null geodesic equation presented in \ref{nullradialgeodesics} takes the following form,
\begin{equation}\label{redefinedV}
\dot{r}^{2}=V_{\rm eff}(r)=e^{-2\mu_{2}}E^{2}\left\{\frac{\mathcal{C}-2\mathcal{B}\ell+\mathcal{A}\ell^{2}}{\mathcal{B}^{2}-\mathcal{A}\mathcal{C}}\right\}~.
\end{equation}
This facilitates the computation of the photon sphere, which requires setting both the potential and its derivative to zero. As immediate from \ref{redefinedV}, setting the potential to zero results into a quadratic expression for the specific angular momentum $\ell$, which can be immediately solved resulting into,
\begin{equation}\label{impactfactor}
\ell=\frac{\mathcal{B}}{\mathcal{A}}\pm \sqrt{\left(\frac{\mathcal{B}}{\mathcal{A}}\right)^{2}-\frac{\mathcal{C}}{\mathcal{A}}}~.
\end{equation}
On the other hand, the other condition necessary for having a photon circular orbit corresponds to setting the derivative of the potential to zero, i.e., we must set $V_{\rm eff}'(r)=0$. Using the fact that $V_{\rm eff}(r)$ also vanishes at that location, the above condition provides yet another quadratic equation for the specific angular momentum $\ell$, such that $\mathcal{C}'-2\mathcal{B}'\ell+\mathcal{A}'\ell^{2}=0$. Here, as mentioned earlier, `prime' denotes derivative with respect to the radial coordinate. This can also be solved for an expression of the specific angular momentum $\ell$. Equating it to the expression derived in \ref{impactfactor}, we get the following condition for determination of the photon circular orbit,
\begin{equation}\label{photonorbit}
\frac{\mathcal{B}}{\mathcal{A}}\pm \sqrt{\left(\frac{\mathcal{B}}{\mathcal{A}}\right)^{2}-\frac{\mathcal{C}}{\mathcal{A}}}
=\frac{\mathcal{B}'}{\mathcal{A}'}\pm \sqrt{\left(\frac{\mathcal{B}'}{\mathcal{A}'}\right)^{2}-\frac{\mathcal{C}'}{\mathcal{A}'}}~.
\end{equation}
The above relation can be simplified further along the following lines, first of all one multiplies both the sides by $\mathcal{A}\mathcal{A}'$ and subsequently taking square of them. Finally manipulating the resulting expression appropriately, we are left with a single square root term, which subsequently is again squared to get rid of the same. This results into the following equation among the metric components determining the photon sphere on the equatorial plane \cite{Gyulchev:2006zg, Rahman:2018fgy},
\begin{equation}\label{SpinPhoton}
\left(\mathcal{A}\mathcal{C}'-\mathcal{A}'\mathcal{C}\right)^{2}=4\left(\mathcal{A}\mathcal{B}'-\mathcal{A}'\mathcal{B}\right)
\left(\mathcal{B}\mathcal{C}'-\mathcal{B}'\mathcal{C}\right)~.
\end{equation}
Note that in the context of static and spherically symmetric spacetime we have $\mathcal{B}=0$ along with $\mathcal{A}(r)=f(r)$, and $\mathcal{C}(r)=r^{2}$. Thus the above expression in \ref{SpinPhoton} reduces to the following expression, $2f=rf'$, the equation determining the photon sphere in static, spherically symmetric spacetime. Thus our expression presented in \ref{SpinPhoton} indeed reduces to the respective expression in static and spherically symmetric spacetime, as it should. 

Having determined the necessary and sufficient condition for determination of the photon circular orbit, we now concentrate on the calculation of the Lyapunov exponent $\lambda _{\rm R}$, which is intimately tied with the infinitesimal fluctuations around the photon orbit and given by \ref{lyapunov}. For this purpose, we need to determine the second derivative of the potential presented in \ref{redefinedV} on the photon circular orbit, which yields,
\begin{equation}\label{ddpot}
V_{\rm eff}''(r_{\rm ph})=e^{-2\mu_{2}}E^{2}\left\{\frac{\mathcal{C}''-2\mathcal{B}''\ell+\mathcal{A}''\ell^{2}}{\mathcal{B}^{2}-\mathcal{A}\mathcal{C}}\right\}~,
\end{equation}
where we have used the fact that on the photon circular orbit both the potential and its first derivative identically vanishes. In order to find the Lyapunov exponent, we need to determine the temporal component of the four velocity as well. Thus using \ref{energymomentum} yields, $\dot{t}=\{\mathcal{C}E-\mathcal{B}L\}(\mathcal{B}^{2}-\mathcal{A}\mathcal{C})^{-1}$. Further, we can substitute for $\ell ^{2}=(1/\mathcal{A})\{2\mathcal{B}\ell-\mathcal{C}\}$ in \ref{ddpot} and then use the above expression for $\dot{t}$, yielding the Lyapunov exponent to be
\begin{equation}\label{lyaspin}
\begin{aligned}
\lambda_{\rm R}&=\left[e^{-2\mu_{2}}E^{2}\left\{\frac{\left(\mathcal{A}\mathcal{C}''-\mathcal{A}''\mathcal{C}\right)+2\ell \left(\mathcal{A}''\mathcal{B}-\mathcal{A}\mathcal{B}''\right)}{2\mathcal{A}\left(\mathcal{B}^{2}-\mathcal{A}\mathcal{C}\right)\dot{t}^{2}}\right\}\right]^{\frac{1}{2}}
\\
&=\left[e^{-2\mu_{2}}\left(\mathcal{B}^{2}-\mathcal{A}\mathcal{C}\right)\left\{\frac{\left(\mathcal{A}\mathcal{C}''-\mathcal{A}''\mathcal{C}\right)+2\ell \left(\mathcal{A}''\mathcal{B}-\mathcal{A}\mathcal{B}''\right)}{2\mathcal{A}\left(\mathcal{C}-\mathcal{B}\ell\right)^{2}}\right\}\right]^{\frac{1}{2}}~,
\end{aligned}
\end{equation}
where in the last line we have used the expression for $\dot{t}$ introduced earlier. Here as well the subscript `R' stands for rotating black hole. Besides the Lyapunov exponent, one also requires to determine the Cauchy horizon, which corresponds to the smallest root of the equation $e^{-2\mu_{2}}=0$. Hence one can also find out the surface gravity of the Cauchy horizon by computing the acceleration associated with the Killing vector field determining the Killing horizon. This enables one to determine the parameter $\beta _{\rm R}$ for a rotating black hole using \ref{sccviolation}. 
\section{Application: Rotating Black Hole in Higher Dimensions}\label{SCC_Rot_App}

In this section, we will apply the formalism presented above in the context of rotating black holes in presence of higher dimensions. There can be two possibilities, firstly the black hole itself could be higher dimensional, otherwise one may consider black hole solution originating from effective gravitational field equations on the four dimensional brane. In what follows, we will discuss both these scenarios. In the context of higher dimensional black holes, we will try to understand the status of cosmic censorship conjecture for Kerr-de Sitter black holes. On the other hand, for effective gravitational theory, we will analyze a black hole in four dimensional spacetime inheriting rotation, cosmological constant as well as tidal charge due to presence of higher dimensions. We elaborate on them in the subsequent sections.

\subsection{Cosmic Censorship Conjecture for Higher Dimensional Kerr-de Sitter Black Hole}

As we have witnessed in the context of higher dimensional \RN-\dS\ black hole, the presence of higher dimension leads to a stronger violation of the cosmic censorship conjecture. This leads to the natural question, what happens for higher dimensional Kerr-de Sitter black holes? Since in four dimensions the cosmic censorship conjecture is respected for Kerr-de Sitter black holes, it is important to understand whether presence of higher dimensions can lead to a possible violation of the same. The general formalism necessary for this purpose has already been laid down in the previous section, here we apply the same demonstrating validity/violation of cosmic censorship conjecture in the context of Kerr-de Sitter black hole. For this purpose, we write down the Kerr-de Sitter black hole solution in $d$-spacetime dimensions in the Boyer-Lindquist co-ordinate, which takes the following form \cite{Gibbons:2004uw, Gibbons:2004js}
\begin{align}\label{spinmetric}
ds^{2}&=-W\left(1-\frac{\Lambda}{(d-1)} r^{2}\right)dt^{2}+\dfrac{U dr^{2}}{F-2M}+\frac{2M}{U}\left(dt-\sum_{i=1}^{N}\dfrac{a_{i}\mu_{i}^{2}d\phi_{i}}{1+\frac{\Lambda}{(d-1)} a_{i}^{2}}\right)^{2}+ \sum_{i=1}^{N+\epsilon}\dfrac{r^{2}+a_{i}^{2}}{1+\frac{\Lambda}{(d-1)} a_{i}^{2}}d\mu_{i}^{2}
\nonumber
\\
&+ \sum_{i=1}^{N}\dfrac{r^{2}+a_{i}^{2}}{1+\frac{\Lambda}{(d-1)} a_{i}^{2}}\mu_{i}^{2}\left(d\phi_{i}-\frac{\Lambda}{(d-1)} a_{i}dt\right)^{2}+\dfrac{\frac{\Lambda}{(d-1)}}{W\left(1-\frac{\Lambda}{(d-1)} r^{2}\right)}\left(\sum_{i=1}^{N+\epsilon}\dfrac{r^{2}+a_{i}^{2}}{1+\frac{\Lambda}{(d-1)} a_{i}^{2}}\mu_{i}d\mu_{i}\right)^{2}~.
\end{align}
In order to simplify the line element for the rotating black hole spacetime in presence of higher dimensions, we have introduced several short hand definitions in \ref{spinmetric}, these quantities can be expressed as, 
\begin{align}\label{rotatingcoefficient}
F&=\dfrac{\left(1-\frac{\Lambda}{(d-1)} r^{2}\right)}{r^{2-\epsilon}}\prod_{i=1}^{N}(r^{2}+a_{i}^{2})~; 
\quad
W=\sum_{i=1}^{N+\epsilon}\dfrac{\mu_{i}^{2}}{1+\frac{\Lambda}{(d-1)} a_{i}^{2}}~;
\nonumber
\\
U&=r^{\epsilon}\sum_{i=1}^{N+\epsilon}\dfrac{\mu_{i}^{2}}{r^{2}+a_{i}^{2}}\prod_{j=1}^{N}(r^{2}+a_{j}^{2})~;
\quad
\sum_{i=1}^{N+\epsilon}\mu_{i}^{2}=1~.
\end{align}
In the above expressions $N$ stands for the number of azimuthal co-ordinates and hence the number of independent orthogonal planes, $a_{i}$ represents the rotation parameter in each of these planes and $\epsilon$ is $0$ for odd dimensional black holes ($d=2n+1$) whereas it is unity for even dimensional black holes ($d=2n$). The parameter $n$ associated with spacetime dimensions is related to $N$ and $\epsilon$ by the relation $n=N+\epsilon$. Further the quantities $\mu_{i}$ are the direction cosines associated with each of these planes. The location of the horizons in this black hole spacetime can be found by solving the equation
\begin{equation}\label{horizons}
F(r_{*})=2M~,
\end{equation}
where $r_{*}=r_{-},r_{+}~\textrm{and}~r_{c}$ corresponds to the Cauchy, Event and Cosmological horizons respectively. Since the Kerr-de Sitter spacetime depicts a stationary black hole, it has a Killing horizon, defined by the following Killing vector field   
\begin{equation}\label{killing_vectors}
\xi^{\alpha}=t^{\alpha}+\sum_{i=1}^{N}\Omega_{i}\phi_{i}^{\alpha}~.
\end{equation}
As one can explicitly check, the location of the Killing horizon coincides with the location of the event horizon. This is essentially due to the Killing vector field $\xi^{\mu}$  becoming null at the event horizon. Further, $t^{\alpha}=(\partial/\partial t)^{\alpha}$ appearing in the expression for $\xi^{\alpha}$ is the Killing vector associated with the time translation and the quantities $\Omega_{i}$ appearing in \ref{killing_vectors} are the angular velocities of the black hole horizon with respect to the various planes introduced earlier \cite{Gibbons:2004uw},
\begin{equation}\label{angularvelocity}
\Omega_{i}(r_{*})=\frac{a_{i}\left(1-\frac{\Lambda}{d-1} r_{*}^{2}\right)}{r_{*}^{2}+a^{2}_{i}}~.
\end{equation}
Note that for $a_{i}=0$, the angular velocities also vanish as it should. Finally, given the Killing field, we would like to find out the expression for surface gravity as well, which is essentially the non-affinity parameter $\kappa$ appearing in the geodesic equation $\xi^{\alpha}\nabla_{\alpha}\xi^{\beta}=\kappa \xi^{\beta}$. The surface gravity $\kappa$ associated with the Killing horizon at $r=r_{*}$ can be determined using derivatives of the metric elements appearing in \ref{spinmetric}, which can be given by the following expression \cite{Gibbons:2004uw}
\begin{equation}\label{surfacegravitykerr}
\kappa_{*}=r_{*}\left(1-\frac{\Lambda}{(d-1)} r_{*}^{2}\right)\left(\sum_{i=1}^{N}\frac{1}{r_{*}^{2}+a^{2}_{i}}+\frac{\epsilon}{2r_{*}^{2}}\right)-\frac{1}{r_{*}}~.
\end{equation}
This expression is absolutely essential in determining the quantity $\beta$ governing the validity of the censorship conjecture. However it is difficult to handle the most general rotating metric depicted in \ref{spinmetric} and hence for simplicity and for our purpose, it will suffice to consider a situation in which all the rotation parameters are identical, i.e., $a_{i}=a$. Further we will consider motion in a plane, such that only one direction cosine among all the $\mu_{i}$'s is $1$ (say $\mu_{1}=\mu=1$), while rest of them are identically vanishing. Thus \ref{spinmetric} becomes
\begin{align}\label{simplespinmetric}
ds^{2}=-W\left(1-\frac{\Lambda}{(d-1)} r^{2}\right)dt^{2}&+\frac{U dr^{2}}{F-2M}+\frac{2M}{U}\left(dt-\frac{a d\phi}{1+\frac{\Lambda}{(d-1)} a^{2}}\right)
\nonumber
\\
&+\frac{r^{2}+a^{2}}{1+\frac{\Lambda}{(d-1)} a^{2}}\left(d\phi-\frac{\Lambda}{d-1} a dt\right)^{2}~.
\end{align}
Here the metric coefficients, namely $W$, $U$ and $F$ appearing in \ref{simplespinmetric} takes the following simplified form,
\begin{eqnarray}\label{simplerotcoefficient}
F=\dfrac{\left(1-\frac{\Lambda}{(d-1)} r^{2}\right)}{r^{2-\epsilon}}\left(r^{2}+a^{2}\right)^{N}~; 
\quad
W=\dfrac{1}{1+\frac{\Lambda}{(d-1)} a^{2}}~;
\quad
U=\dfrac{r^{\epsilon}}{r^{2}+a^{2}}\left(r^{2}+a^{2}\right)^{N}~.
\end{eqnarray}
Since $\epsilon$ behaves differently for even and odd spacetime dimensions, it follows that one needs to compute the coefficients $\mathcal{A}$, $\mathcal{B}$ and $\mathcal{C}$ separately for even and odd spacetime dimensions respectively. This is what we work out explicitly below.
\begin{figure}
\centering 
\begin{minipage}[b]{0.475\textwidth}
\includegraphics[width=\textwidth]{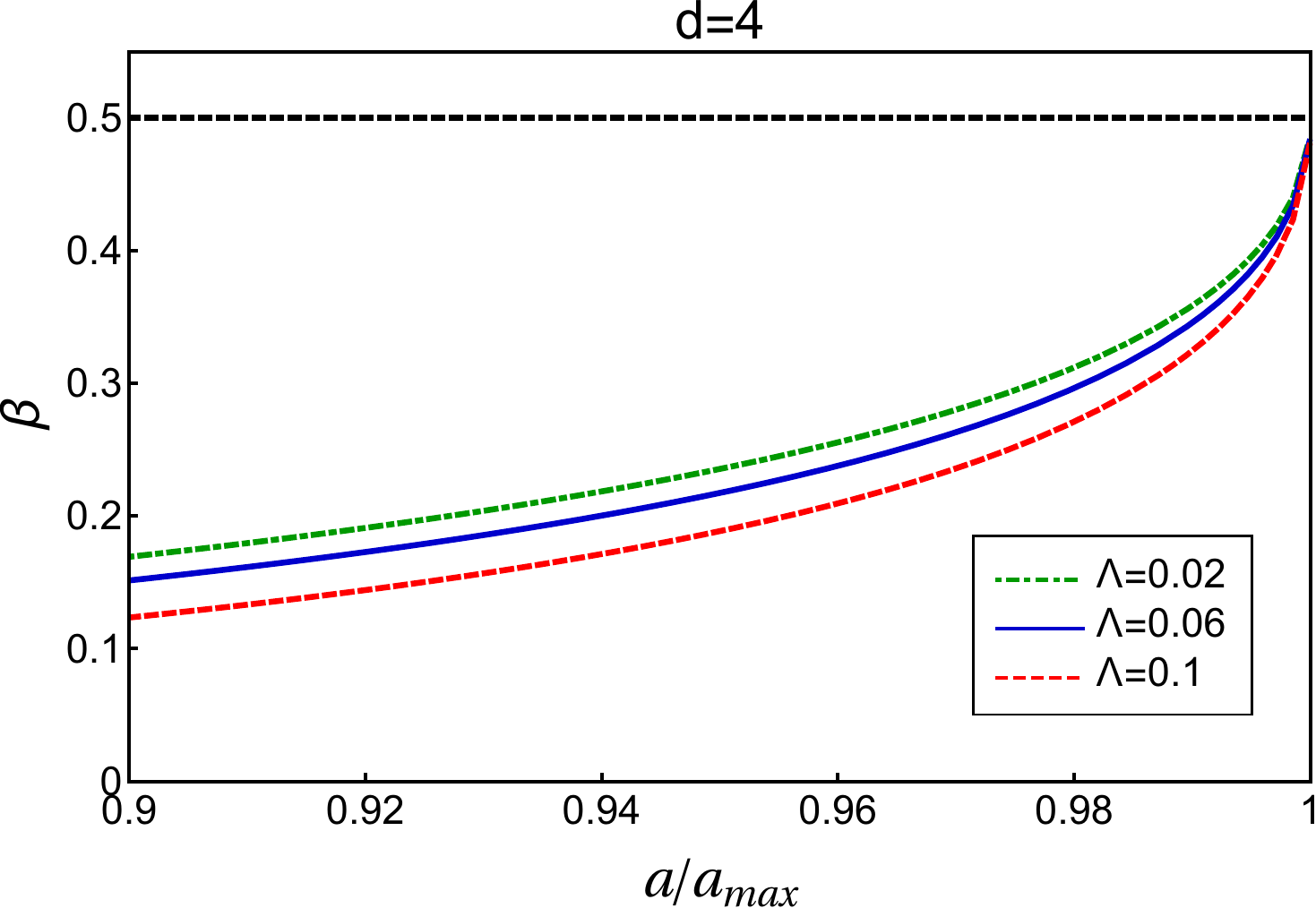}
\end{minipage}
\hfill
\begin{minipage}[b]{0.475\textwidth}
\includegraphics[width=\textwidth]{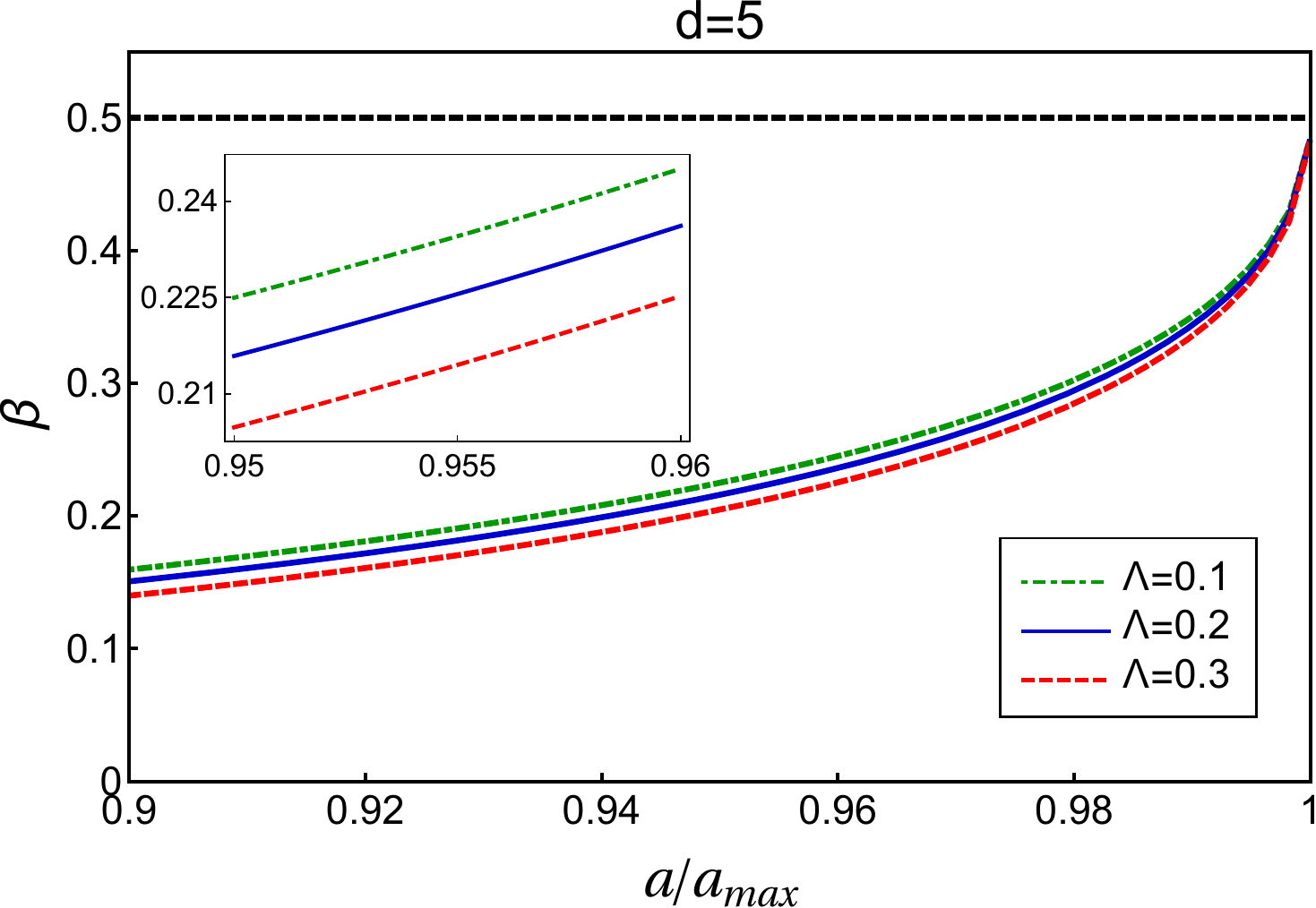}
\end{minipage}
\hfill
\begin{minipage}[b]{0.475\textwidth}
\includegraphics[width=\textwidth]{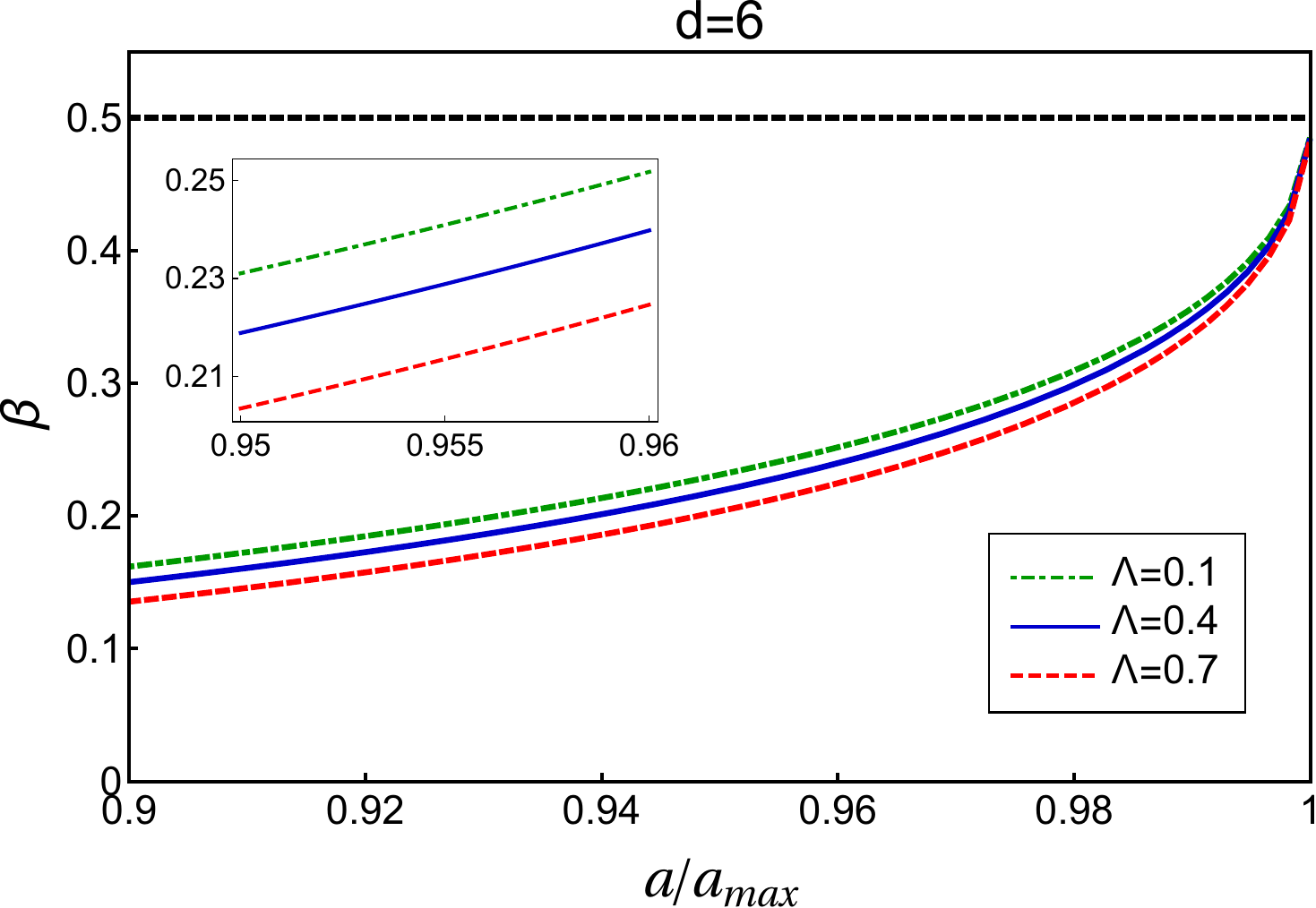}
\end{minipage}
\caption{The variation of the parameter $\beta$ with $a/a_{\rm max}$ (with $a_{\rm max}$ denoting the value of the maximal rotation parameter) for different values of $\Lambda$ in a d dimensional Kerr-\dS\ black hole has been presented. The plots associated with four and five dimensional Kerr-\dS\ black hole have been presented at the upper-left and upper-right corners respectively, while that for six dimensional Kerr-\dS\ black hole is presented at the bottom. In each of these plot of the parameter $\beta$, the black dashed line is associated with the curve corresponding to $\beta=1/2$. As evident these plots indicate that cosmic censorship conjecture is respected in any d-dimensional Kerr-\dS\ black holes. See text for more discussions.}
\label{KDSlambda}
\end{figure}
\begin{figure}[ht]
\centering
\begin{minipage}[b]{0.60\textwidth}
\includegraphics[width=\textwidth]{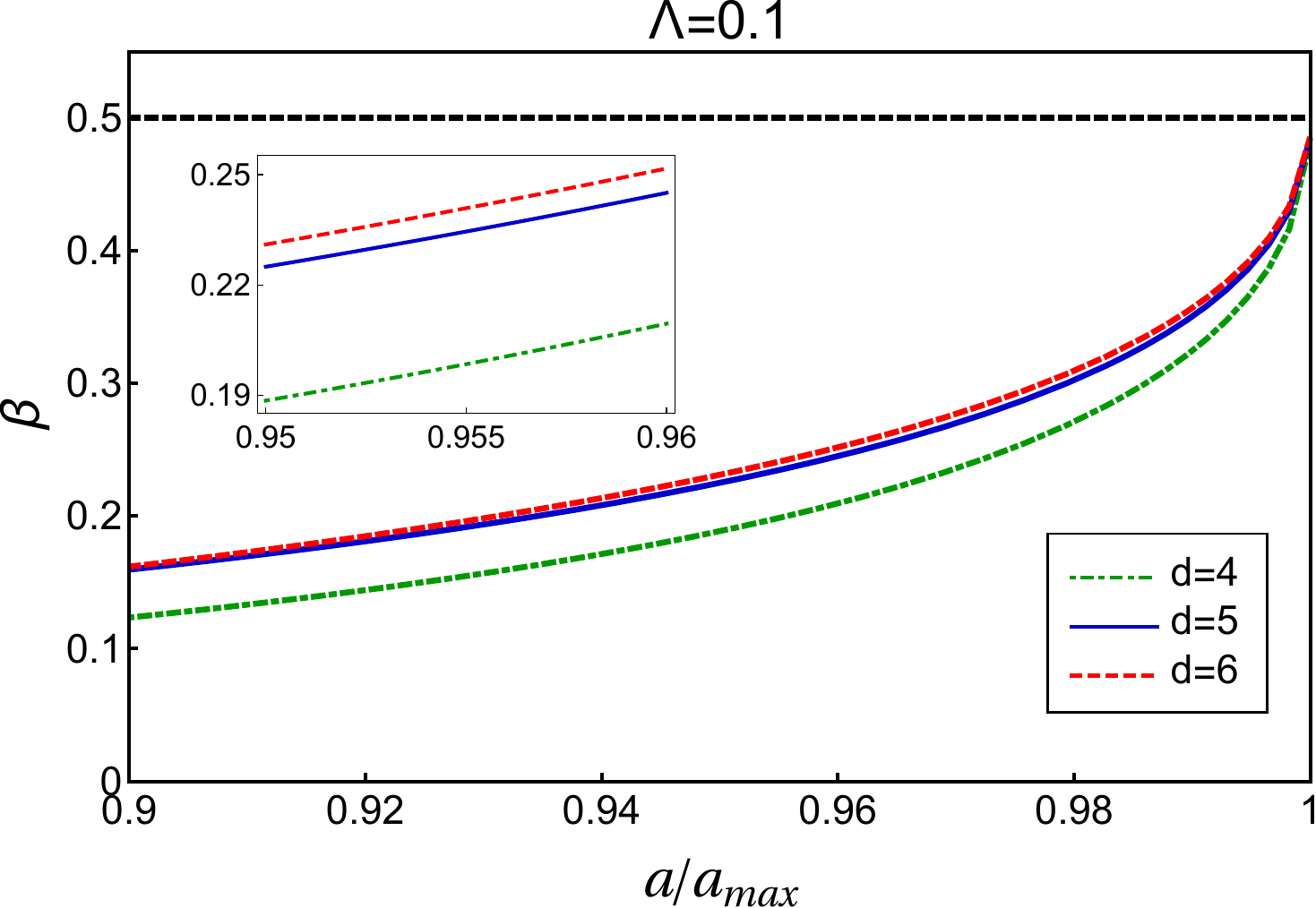}
\end{minipage}
\caption{A comparative study of the behaviour of the parameter $\beta$ with spacetime dimensions has been presented. For this purpose we have drawn $\beta$ as a function of $(a/a_{\rm max})$ at a fixed value of the cosmological constant ($\Lambda=0.1$) in four (green dot-dashed line), five (blue solid line) and six (red dashed line) dimensional spacetimes for Kerr-\dS\ black holes. As the  plots clearly depict, none of them really crosses the black dashed line corresponds to $\beta=1/2$. Thus cosmic censorship conjecture is being uphold for Kerr-\dS\ black holes even in presence of higher dimensions.}
\label{KDSdimension}
\end{figure}
\paragraph*{Metric Coefficients in Even Spacetime Dimensions} For an even dimensional Kerr-de Sitter black hole, we have $d=2n$ and $\epsilon=1$, such that the parameter $N$ appearing in \ref{spinmetric} is $N-1$. Thus one can determine, the metric coefficients necessary for our computation, i.e., the quantities $\mathcal{A}$, $\mathcal{B}$ and $\mathcal{C}$ respectively  for an even dimensional Kerr-de Sitter black hole as,
\begin{equation}\label{even}
\begin{aligned}
\mathcal{A}(r)&=-1+\frac{2M}{r\left(r^2+a^2\right)^{\frac{\text{d}-4}{2}}}+\frac{\Lambda}{d-1}\left(r^2+a^2\right);\\
\mathcal{B}(r)&=\frac{a}{1+\frac{\Lambda}{d-1}a^{2}}\left[\frac{2M}{r\left(r^2+a^2\right)^{\frac{\text{d}-4}{2}}}+\frac{\Lambda}{d-1}  \left(r^2+a^2\right)\right];\\
\mathcal{C}(r)&=\frac{r^2+a^2}{1+\frac{\Lambda}{d-1}a^2}+\frac{2a^{2}M}{\left(1+\frac{\Lambda}{d-1}a^2\right)^{2}r\left(r^2+a^2\right)^{\frac{\text{d}-4}{2}}}~.
\end{aligned}
\end{equation}
Note that for $d=4$, the above metric components coincides with the metric of four dimensional Kerr-de Sitter black hole. Finally in order to determine the horizon location, we need to know the zeros of the $g^{rr}$ component, which in the present context reads, $g^{rr}=\Delta(r)r^{-2}(r^2+a^2)^{\frac{4-d}{2}}$. Thus locations of the horizon can be determined by solving the algebraic equation, $\Delta(r)=0$, where, $\Delta(r)=[1-\{\Lambda/(d-1)\}r^2](r^2+a^2)^{(d-2)/2}-2Mr$. The smallest root corresponds to the Cauchy horizon and the largest root corresponds to the cosmological horizon. Thus the above provides all the relevant ingredients using which one can immediately compute the parameter $\beta$. 
\paragraph*{Metric Coefficient for Odd Spacetime Dimensions} For odd dimensional Kerr-de Sitter black hole, it follows that $d=2n+1$, with $\epsilon=0$ and hence $N=n$. Thus in this case the relevant quantities, namely $\mathcal{A}$, $\mathcal{B}$ and $\mathcal{C}$, constructed out of the metric elements correspond to,
\begin{equation}\label{odd}
\begin{aligned}
\mathcal{A}(r)&=-1+\frac{2M}{\left(r^2+a^2\right)^{\frac{\text{d}-3}{2}}}+\frac{\Lambda}{d-1}\left(r^2+a^2\right);\\
\mathcal{B}(r)&=\frac{a}{1+\frac{\Lambda}{d-1}a^{2}}\left[\frac{2M}{\left(r^2+a^2\right)^{\frac{\text{d}-3}{2}}}+\frac{\Lambda}{d-1}  \left(r^2+a^2\right)\right];\\
\mathcal{C}(r)&=\frac{r^2+a^2}{1+\frac{\Lambda}{d-1}a^{2}}+\frac{2 a^2 M}{\left(1+\frac{\Lambda}{d-1}a^{2}\right)^{2} \left(r^2+a^2\right)^{\frac{\text{d}-3}{2}}}~.
\end{aligned}
\end{equation}
The only additional information necessary for our purpose correspond to the location of the horizon. This can be determined by computing zeros of the $g^{rr}$ component which reads, $\Delta(r)r^{-2}(r^2+a^2)^{(3-d)/2}$, where $\Delta(r)=[1-\{\Lambda/
(d-1)\}r^2](r^2+a^2)^{(d-1)/2}-2Mr^{2}$. Thus the horizons are located at the zeros of the algebraic equation $\Delta(r)=0$. The lowest root of the same being the Cauchy horizon, while the largest one depicts the cosmological event horizon. Using these informations, we can easily compute the Lyapunov exponent and hence the parameter $\beta$ associated with the Kerr-de Sitter black hole. 

Using all these results, we have plotted the parameter $\beta$ as a function of $a/a_{\rm max}$ for various choices of the cosmological constant $\Lambda$ and spacetime dimension $d$ in \ref{KDSlambda} and \ref{KDSdimension} respectively. In particular, as evident from \ref{KDSlambda}, where the parameter $\beta$ has been plotted against the rotation parameter for various choices of the cosmological constant $\Lambda$ and spacetime dimension, the cosmic censorship conjecture is never violated. A similar behaviour of $\beta$ is more explicitly from \ref{KDSdimension} and hence one can safely conclude that for rotating Kerr-\dS\ black holes, the parameter $\beta$ never crosses $(1/2)$ and hence strong cosmic censorship conjecture is respected. 

Further, we would like to emphasize that in the context of rotating black holes in higher dimensions, the photon sphere modes are sufficiently accurate in providing the associated quasi-normal modes. This can be ascertained from the fact that the relative error between the numerical estimations and the analytical expressions using photon sphere modes is $\mathcal{O}(10^{-4})$ even in the extremal limit. This is a direct generalization of the result presented in \cite{PhysRevD.97.104060} in presence of higher dimensions. This being the primary reason for not presenting the de Sitter and near-extremal modes in the context of rotating black holes.
\subsection{Cosmic Censorship Conjecture for Rotating Black Hole on the Brane}

Another possibility to incorporate the effect of higher dimensions on the cosmic censorship conjecture is to consider a black hole on the four dimensional brane. This can be derived by solving the effective gravitational field equations involving bulk Weyl tensor projected on the brane. The effective field equations can also involve the presence of a brane cosmological constant inherited from the bulk spacetime. Thus in this case as well it is possible to arrive at a rotating solution to the gravitational field equations on the brane which will be asymptotically de-Sitter. Interestingly, besides the rotation parameter and the brane cosmological constant it will also involve a tidal charge parameter inherited from the bulk Weyl tensor \cite{Modgil:2001hm,Frolov:2004wy,Larranaga:2013aoa}. This essentially correspond to the four dimensional Kerr-Newman-de Sitter spacetime, with the sign of the charge parameter reversed. Thus the associated line element takes the following form,
\begin{align}
ds^{2}=-\frac{\Delta_{r}}{\rho^{2}}\left(dt-\frac{a}{\Upsilon}\sin ^{2}\theta d\phi \right)^{2}+\frac{\rho^{2}}{\Delta_{r}}dr^{2}+\frac{\rho^{2}}{\Delta _{\theta}}d\theta ^{2}+\frac{\Delta _{\theta}\sin^{2}\theta}{\rho^{2}}\left\{-\frac{\left(r^{2}+a^{2}\right)}{\Upsilon}d\phi+adt \right\}^{2}
\end{align}
where, $\Delta_{r}=\left(r^{2}+a^{2}\right)\left\{1-(\Lambda/3) r^{2}\right\}-2Mr-q$, $\Delta _{\theta}=1+(\Lambda/3) a^{2}\cos ^{2}\theta$, $\rho^{2}=r^{2}+a^{2}\cos^{2}\theta$ and finally $\Upsilon = 1+(\Lambda/3) a^{2}$. Here the quantity $q$ is appearing from the presence of higher dimensions through the bulk Weyl tensor. One can expand out the above expression for the line element, thus obtaining the metric components, relevant for our computation. In particular, we need to compute the quantities introduced in \ref{redefination}. These are somewhat simplified, since we need them on the equatorial plane in order to determine the coefficient $\beta$. These quantities in the present context read, 
\begin{align}
\mathcal{A}&=-\frac{\Delta _{r}-a^{2}}{r^{2}} 
=\frac{2M}{r}+\frac{q}{r^{2}}-1+\frac{\Lambda}{3} \left(r^{2}+a^{2}\right)
\\
\mathcal{B}&=-\frac{\Delta _{r}}{r^{2}}\frac{a}{\Upsilon}+\frac{a}{r^{2}}\frac{r^{2}+a^{2}}{\Upsilon}
=\frac{a}{r^{2}\Upsilon}\left\{\left(r^{2}+a^{2}\right)-\Delta _{r} \right\}
=\frac{a}{\Upsilon} \left\{\frac{2M}{r}+\frac{q}{r^{2}}+\frac{\Lambda}{3}\left(r^{2}+a^{2} \right)\right\}
\\
\mathcal{C}&=-\frac{\Delta _{r}}{r^{2}}\frac{a^{2}}{\Upsilon^{2}}+\frac{1}{r^{2}}\frac{\left(r^{2}+a^{2}\right)^{2}}{\Upsilon^{2}}
=\frac{1}{\Upsilon ^{2}}\left\{a^{2}\left(\frac{2M}{r}+\frac{q}{r^{2}}\right)+\left(r^{2}+a^{2}\right)r^{2}\Upsilon \right\}
\end{align}
Given the above, the location of the photon circular orbit as well as the double derivative of the potential can be determined using \ref{SpinPhoton} and \ref{ddpot} respectively. Subsequently one can use the expression for $\dot{t}$ along with the above results to determine the Lyapunov exponent as in \ref{lyaspin}. The other necessary ingredient corresponds to the location of the Cauchy horizon, which can be determined by solving for the lowest root of the algebraic relation, $\Delta _{r}=0$. Thus one can compute the expression for the surface gravity on the Cauchy horizon, which can be used along with the Lyapunov exponent in order to determine the quantity $\beta$.
\begin{figure}
\centering 
\begin{minipage}[b]{0.475\textwidth}
\includegraphics[width=\textwidth]{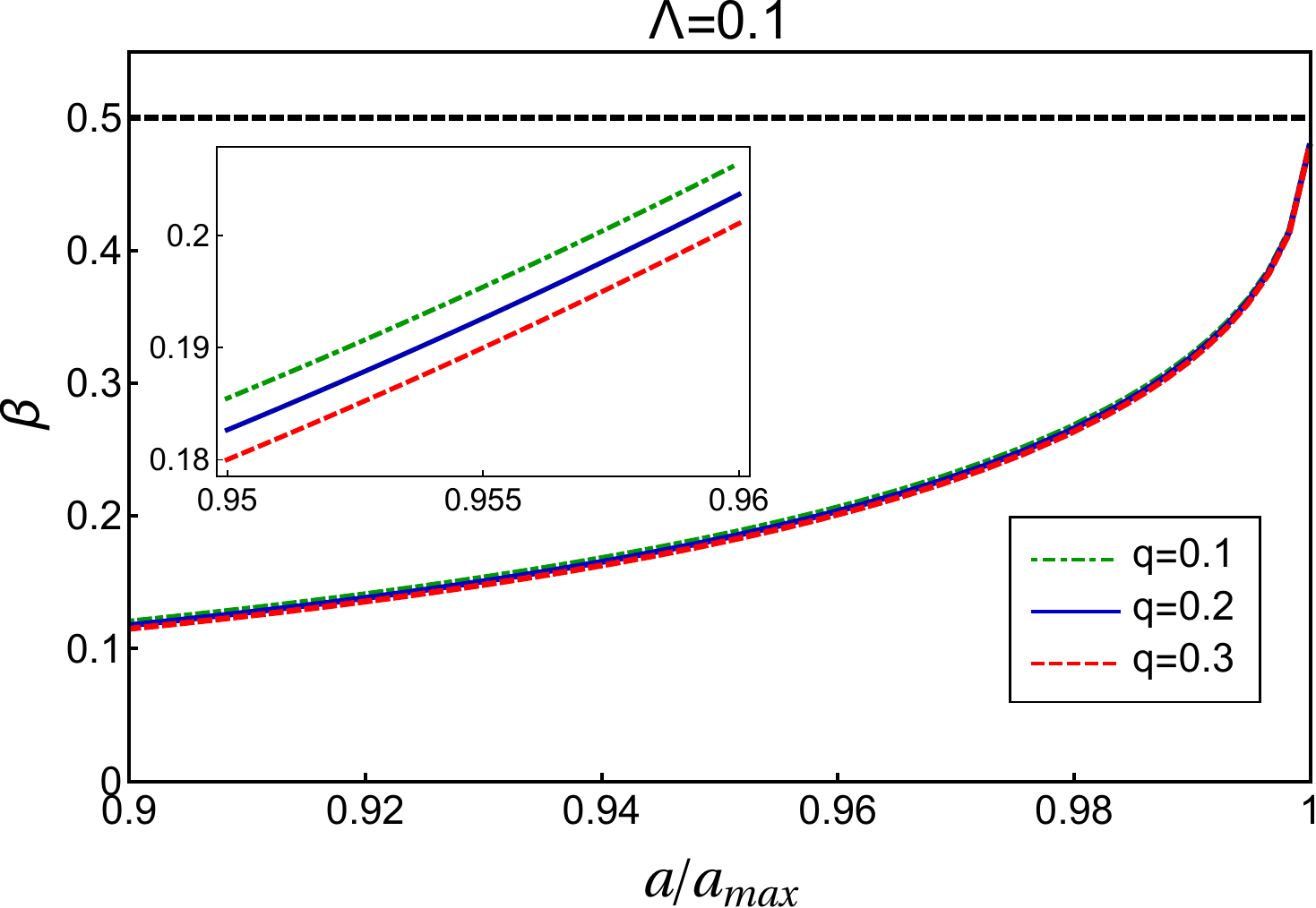}
\end{minipage}
\hfill
\begin{minipage}[b]{0.475\textwidth}
\includegraphics[width=\textwidth]{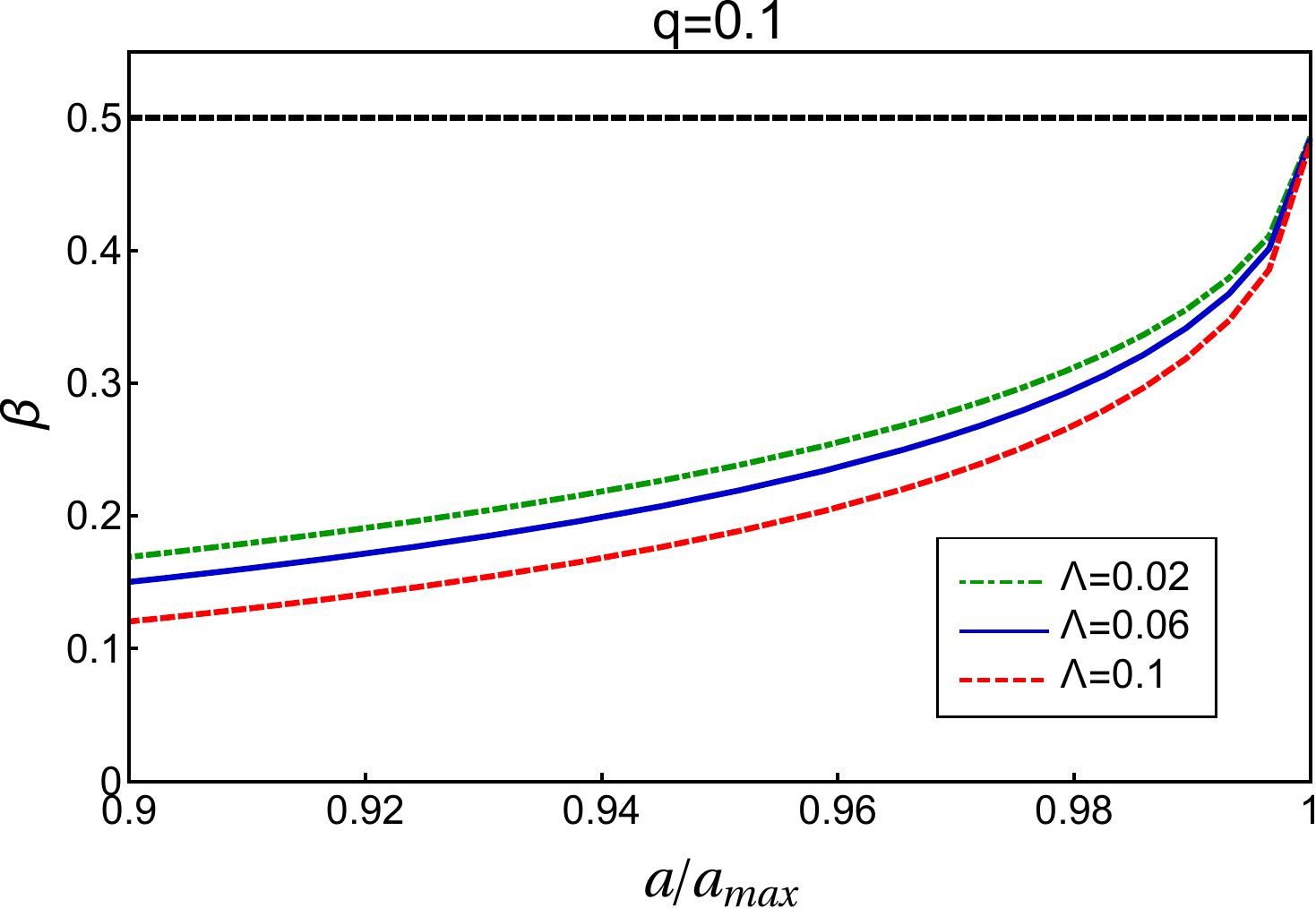}
\end{minipage}
\caption{The variation of the parameter $\beta$ against the normalized rotation parameter $(a/a_{\rm max})$, with $a_{\rm max}$ denoting the maximal rotation, for rotating black hole in four dimensional brane for different values of $q$ (left corner) and $\Lambda$ (right corner) have been presented. In each plot, the black dashed line corresponds to the limiting case $\beta=1/2$. As evident both of these plots indicate that cosmic censorship conjecture is respected in a rotating black hole spacetime in the four dimensional brane.}
\label{RNDSeffective}
\end{figure}

\begin{figure}
\centering 
\begin{minipage}[b]{0.51\textwidth}
\includegraphics[width=\textwidth]{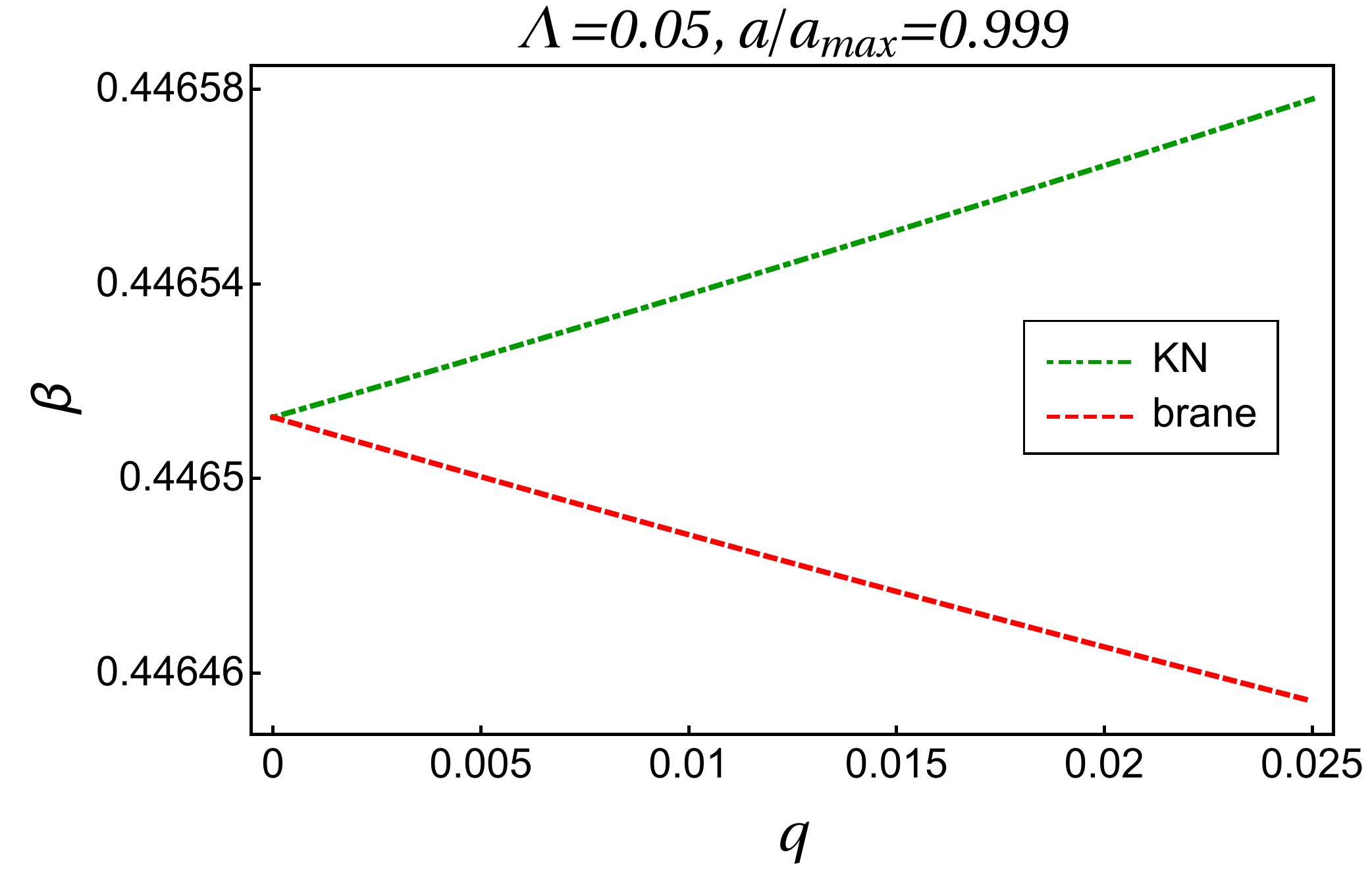}
\end{minipage}
\hfill
\begin{minipage}[b]{0.475\textwidth}
\includegraphics[width=\textwidth]{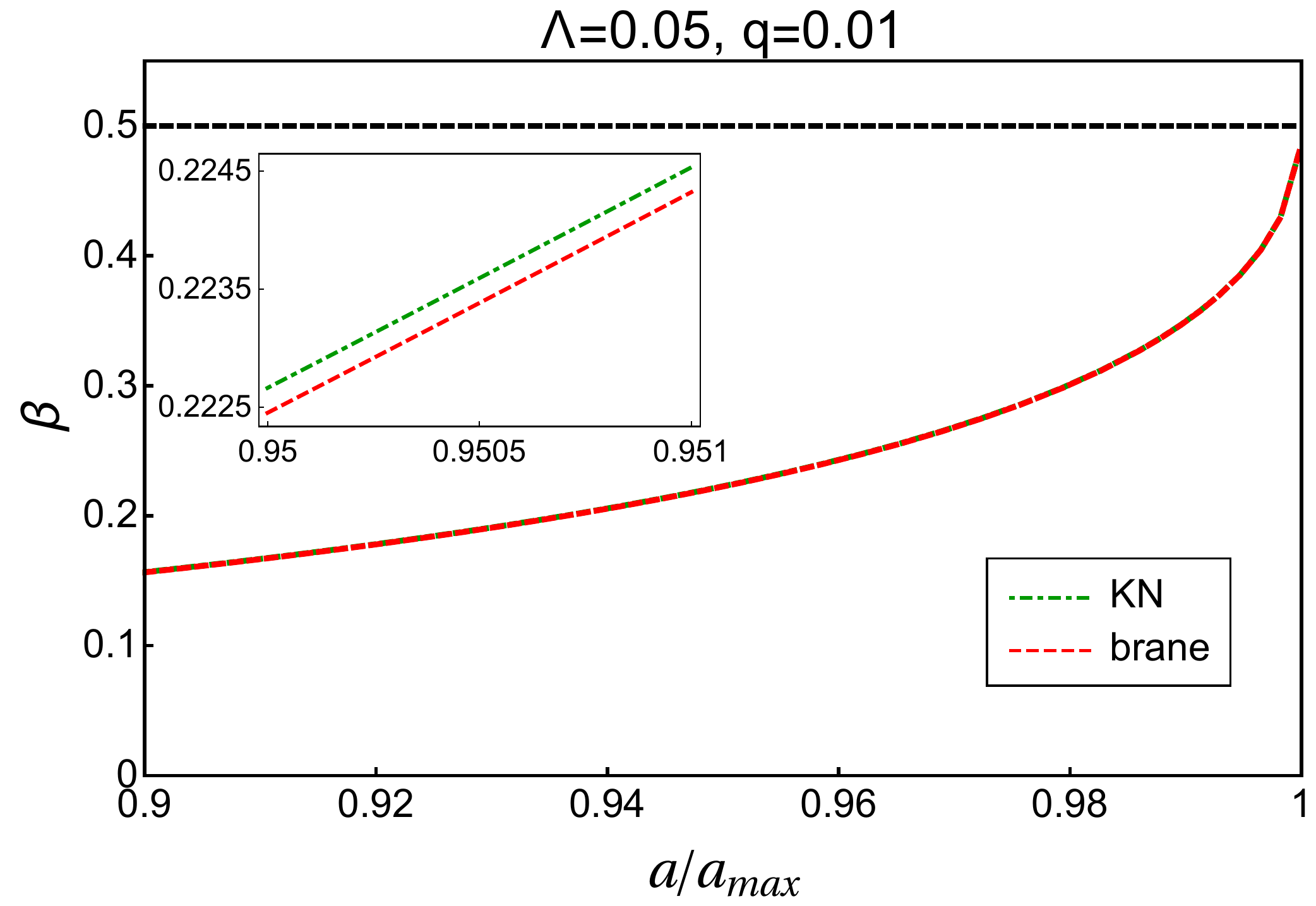}
\end{minipage}
\caption{A comparison between four dimensional Kerr-Newman-de Sitter black hole and the Kerr-de Sitter black hole in effective gravitational theory on the brane has been presented. On the left panel, the variation of the parameter $\beta$ against the ``charge" in the respective contexts has been plotted with a fixed cosmological constant $\Lambda$ and normalized rotation parameter $(a/a_{\rm max})$. While the plot on the right depicts variation of $\beta$ with $(a/a_{\rm max})$ for a fixed ``charge" and cosmological constant. As evident both of these plots indicate that the parameter $\beta$ is always smaller for the rotating black hole spacetime in the four dimensional brane compared to the four dimensional Kerr-Newman-\dS\ spacetime.}
\label{RNDScomparison}
\end{figure}

The consequence of such a computation has been presented in \ref{RNDSeffective}, where the parameter $\beta$ has been presented against the rotation parameter, normalized to extremal values. This has been done for several different choices of the charge parameter inherited from the extra dimensions and the cosmological constant. All of them depicts the parameter $\beta$ being less than $(1/2)$. Thus we can safely conclude that in the context of brane world as well, where the gravitational field equations receive corrections due to presence of higher dimensions, for rotating black holes the cosmic censorship conjecture is respected. Thus rotation acts as a key ingredient to uphold the cosmic censorship conjecture in different higher dimensional scenarios, including brane world.

At this stage, it is also an interesting idea to compare the rotating brane world black hole with the four dimensional Kerr-Newman-\dS\ black hole. In particular, for a given value of the charge parameter $q$, one can compare the parameter $\beta$ among these two black holes, one in which the charge is providing a negative contribution (the brane world black hole) and the Kerr-Newman-\dS\ black hole where the charge has a positive impact on the metric. This situation has been depicted in \ref{RNDScomparison}. As evident from the figures, the parameter $\beta$ for a given value of the charge parameter is always less in the brane world black hole when compared to the Kerr-Newman-\dS\ black hole. Note that in the context of rotating black hole on the brane as well the photon sphere modes are accurate enough to provide correct estimation of the black hole quasi-normal modes. This holds true even when the black hole is in the near-extremal zone. Thus the analytical estimation presented above suffices and it suggests that the cosmic censorship conjecture is respected in the context of brane world black holes in a more stringent fashion than the Kerr-Newman-\dS\ black hole. 
\section{Concluding Remarks}

Strong cosmic censorship conjecture asserts that the dynamics of gravity can be formulated in a deterministic manner. It basically predicts doom for an observer who is curious enough to cross the Cauchy horizon, the end of the deterministic world for any gravity theory. Intriguingly, in presence of a positive cosmological constant, it turns out, the metric can be safely extended beyond the Cauchy horizon with locally square integrable connections, if the decay rate of any perturbing scalar field living in the spacetime becomes comparable to the blueshift at the Cauchy horizon. This implies possible breakdown of predictability and hence violation of strong cosmic censorship conjecture. The above result can be presented in a quantitative manner by considering the following quantity $\beta=-\{\textrm{Im}(\omega)\}/\kappa_{-}$. An analytical estimation can be obtained in the context of photon sphere modes, which can be derived using eikonal approximation. In the context of photon sphere modes the ratio of the Lyapunov exponent associated with instability of the photon circular orbit and the surface gravity at the Cauchy horizon, provides the desired expression for $\beta_{\rm ph}$. The violation of the strong cosmic censorship conjecture is arrived at whenever the parameter $\beta$ becomes larger than half. So far, validity of the strong cosmic censorship conjecture is tested for four dimensional black holes either having electric charge or rotation or both. In this paper, we have investigated the validity of strong cosmic censorship in presence of higher spatial  dimensions. In our study, we have discussed two possibilities which may arise in the context of higher dimensions, namely --- (a) The black hole itself exists in a higher dimensional spacetime or, (b) The black hole is living on the four dimensional brane hypersurface, while the spacetime is intrinsically higher dimensional.

By calculating the lowest lying quasinormal modes in both the geometrical optics limit, as well as for de Sitter and near extremal modes we have explicitly demonstrated that the violation of strong cosmic censorship is a generic feature in \RN-\dS\ black holes irrespective of the spacetime dimension they live in. In fact, it is evident from our study (see the middle column of \ref{RNDSNE}) that violation of strong cosmic censorship is more severe for higher dimensional black holes, for certain choices of the cosmolgical constant as the value of properly normalized electric charge, where $\beta$ becomes greater than $(1/2)$, is smaller. Thus for higher dimensional \RN-\dS\ black holes, this implies a larger parameter space where deterministic nature of general relativity breaks down. This result is further confirmed using numerical analysis through the method of continued fraction as well. Even when we consider a charged black hole that lives on a four dimensional brane, the cosmic censorship conjecture is still violated as long as effect from extra dimension is subdominant compared to its Maxwell charge. However, in this scenario, the parameter space that leads to the violation of the conjecture, becomes smaller. As a result, violation happens at larger values of the Maxwell charge (see \ref{dS_NE_PS_Modes}). Thus for a brane world black hole, which may seem more relevant from a physical point of view, the effect of extra dimension is to protect it from violation of cosmic censorship conjecture. This is due to the fact that the effective charge of the black hole, denoted as $Q^{2}-q$ gets reduced in the presence of extra dimensions. In particular, it should be emphasized that for large enough $q$, it is entirely possible to completely change the nature of the spacetime as Cauchy horizon may cease to exist. Thus in the context of a four dimensional black hole, embedded in a higher dimensional spacetime, the cosmic censorship conjecture is only weakly violated or, not violated at all, depending on the ``charge'' $q$ inherited from the bulk Weyl tensor. 

The other arena, where violation of strong cosmic censorship conjecture may have significant implications, corresponds to the rotating black holes. However, it was demonstrated that for four dimensional rotating black holes the parameter $\beta$ never becomes larger than the critical value $1/2$. It merely reaches the critical value, but in the extremal limit. Taking a cue from our previous discussion regarding violation of cosmic censorship conjecture for static and spherically symmetric black holes in presence of higher dimension, we consider the same for rotating black holes. We have clearly depicted that, for a general $d$-dimensional ($d\geq 4$) rotating Kerr-\dS\ black hole, the parameter $\beta$ never crosses the value half, while for a given rotation parameter the value of $\beta$ is higher for a higher dimensional black hole (see \ref{KDSdimension}). This, in turn implies, that in the higher dimensional rotating black hole spacetimes, the decay rate of perturbation along the event horizon is slow enough that it is overwhelmed by its exponential growth at Cauchy horizon. As a result, \SCC\ is respected. Moreover, our result is in accord with Ref. \cite{PhysRevD.97.104060} for the four dimensional scenario. The same conclusion can also be drawn from the analysis of a rotating black hole that lives on a four-dimensional brane. In this case, for a given rotation parameter, the estimation of $\beta$ is smaller for the brane world black hole compared to its four dimensional counter part (see \ref{RNDScomparison}). Thus one can conclude that rotation acts as a key ingredient to uphold the cosmic censorship conjecture even in presence of higher dimensions. The cosmic censorship conjecture is respected for both higher dimensional as well as brane world black holes. 

We would like to emphasize that in this work we have derived expressions for the quantity $\beta$ in an arbitrary static and spherically symmetric spacetime as well in an axisymmetric spacetime. Both these expressions can be used to assess the validity of cosmic censorship conjecture in other black hole spacetimes as well. Further, it will be interesting to understand the fate of strong cosmic censorship if one chooses to trade off the smoothness of the initial data \cite{Dafermos:2018tha,Dias:2018etb}. In particular, whether non-smooth, but physically motivated initial data can rescue the strong cosmic censorship conjecture even in presence of higher dimension will be another avenue to explore, which we leave for the future. 
\section*{Acknowledgements}

M.R. thanks INSPIRE-DST, Government of India for a Junior Research Fellowship. Research of S.C. is funded by the INSPIRE Faculty Fellowship (Reg. No. DST/INSPIRE/04/2018/000893) from Department of Science and Technology, Government of India. The research of S.S.G is supported by the Science and Engineering Research Board-Extra Mural Research Grant (No. EMR/2017/001372), Government of India. The authors thank the anonymous referee for his/her comments in improving this manuscript.
\appendix
\labelformat{section}{Appendix #1}
\labelformat{subsection}{Appendix #1}
\labelformat{subsubsection}{Appendix #1}

\section{Near-Extremal Modes of Higher Dimensional Black Hole}\label{App_NE}

The line element for a \RN\ black hole in $d$-dimension is given by \ref{staticspherically} where the metric co-efficients $f(r)$ and $g(r)$ takes the form
\begin{equation}\label{RN}
f(r)=g(r)=1-\frac{\varpi_{d-2}M}{r^{d-3}}+\frac{(d-2)\varpi_{d-2}^{2}}{8(d-3)}\frac{Q^{2}}{r^{2d-6}}~.
\end{equation}
The position of the horizons correspond to the real positive roots of the equation $f(r)=0$. It turns out that the corresponding equation has two real, positive roots, denoted by $r_{+}$ and $r_{-}$, which obeys the following inequality, $r_{+}\geq r_{-}$.  This in turn allows us to identify $r_{+}$ and $r_{-}$ as the position of the Event horizon and Cauchy horizon respectively. In the limit ${\varpi_{d-2}M}\to 2~Q_{d}$, these two horizons coincide with each other, which corresponds to a extremal black hole. Here, for the sake of notational simplicity, we have taken $Q_{d}^{2}=(d-2)\varpi_{d-2}^{2}~Q^{2}/8(d-3)$. Since we are interested in the near-extremal black hole scenario, we take the following co-ordinate transformation
\begin{eqnarray}{\label{NEcondition}}
& &{r}^{d-3}\to Q_{d}+\epsilon \rho\,,\qquad 
{\varpi_{d-2}M}\to 2~\sqrt{Q_{d}^{2}+\epsilon^{2}B^{2}}\,, \qquad 
{t}\to \frac{\tau}{\epsilon}.
\end{eqnarray}
where, the parameter $B$ denotes deviation from extremity. Under this transformation, the line element for the \RN\ black hole in the near extremal limit can be written as follows
\begin{equation}\label{RNNE}
ds^{2}=-f(\rho)~d\tau^{2}+\frac{p_{1}}{f(\rho)}~d\rho^{2}+p_{2}~d\Omega_{d-2}^{2}
\end{equation}
where,
\begin{eqnarray}{\label{NEmetric}}
& &{f(\rho)}=\frac{\rho^{2}-B^{2}}{Q_{d}^{2}}\,,\qquad 
{p_{2}}=Q_{d}^{2}~p_{1}=Q^{\frac{2}{d-3}}_{d}.
\end{eqnarray}
For the metric given in \ref{RNNE}, the surface gravity at the Cauchy horizon becomes
\begin{equation}\label{RNNEkappa}
\begin{aligned}
\kappa_{-}\approx\kappa_{+}&=\frac{1}{2}\frac{1}{\sqrt{-g_{\tau\tau}g_{\rho\rho}}}~\bigg{|}\frac{dg_{\tau\tau}}{d\rho}\bigg{|}_{\rho=B}\\&=\frac{B}{\sqrt{p_{2}}~Q_{d}}
\end{aligned}
\end{equation}
Let us consider a situation where this black hole gets perturbed by a massless scalar field which satisfies the Klein-Gordon equation, $\Box\Phi=0$. Since the spacetime admits both time translational and  spherical symmetry, the scalar field can be decomposed as $\Phi(\tau,\rho,\theta,\phi)=e^{-i\omega\tau}~R(\rho)~Y_{lm}(\theta,\phi)$, where $Y_{lm}(\theta,\phi)$ are Spherical harmonics associated with $(d-2)$ dimensional sphere and $R(\rho)$ satisfies a second order differential equation which can be written as
\begin{equation}\label{NEeq}
\dfrac{d}{d\rho}\left[(\rho^{2}-B^{2})\dfrac{dR(\rho)}{d\rho}\right]+\left[\frac{\omega^{2}p_{2}Q_{d}^{2}}{\rho^{2}-B^{2}}-l(l+d-3)\right]R(\rho)=0
\end{equation}
The solution of this equation can be expressed in terms of hyper-geometric functions $F_{1}(a,b,c;z)$ as follows
\begin{equation}
\begin{aligned}
R(x)&=C_{1}~(x^{2}-1)^{\frac{i\mu}{2}}~F_{1}(1+i\mu+\sigma,i\mu-\sigma,i\mu+1;\frac{1-x}{2})\\&+C_{2}~\frac{(x+1)^{\frac{i\mu}{2}}}{(x-1)^{\frac{i\mu}{2}}}~F_{1}(-\sigma,\sigma+1,1-i\mu;\frac{1-x}{2})
\end{aligned}
\end{equation}
where, $C_{1}$ and $C_{2}$ are two arbitrary constants which can be fixed from the boundary conditions, $x=\rho/B$ , $\mu=\sqrt{p_{2}}~\omega~Q_{d}/B=\omega/\kappa_{-}$ and $\sigma$ satisfies the equation, $\sigma(\sigma+1)=l(l+d-3)$. Since we are interested in finding the quasi-normal modes of black hole under perturbation, the boundary condition can be fixed as follows: there are only ingoing modes at the event horizon and only outgoing modes on the boundary. Near the horizon where $x=1$, the solution takes the form
\begin{equation}
\begin{aligned}
R(x)&=C_{H}^{in}~\frac{(x+1)^{\frac{i\mu}{2}}}{(x-1)^{\frac{i\mu}{2}}}+C_{H}^{out}~(x^2-1)^{\frac{i\mu}{2}}\\
&\approx C_{H}^{in}~2^{\frac{i\mu}{2}}(x-1)^{-\frac{i\mu}{2}}+C_{H}^{out}~2^{\frac{i\mu}{2}}(x-1)^{\frac{i\mu}{2}}
\end{aligned}
\end{equation}
where, $C_{H}^{in}=C_{2}$ and $C_{H}^{out}=C_{1}$. Here we have used the property of hyper-geometric function, $F_{1}(a,b,c;0)=1$. Since quasi-normal mode demands the presence of only ingoing modes at the horizon, the constant $C_{1}$ vanishes.
Using the property of hyper-geometric function given by 
\begin{equation}
\begin{aligned}
F_{1}(a,b,c;z)&=\frac{\Gamma(c)~\Gamma(b-a)}{\Gamma(b)~\Gamma(c-a)}z^{-a}~F_{1}(a,1-c+b,1-b+a~;\frac{1}{z})\\
&+~\frac{\Gamma(c)~\Gamma(a-b)}{\Gamma(b)~\Gamma(c-b)}(-z)^{-b}~F_{1}(b,1-c+b,1-a+b~;\frac{1}{z})
\end{aligned}
\end{equation}
and putting $F_{1}(a,b,c;0)=1$, we can easily obtain solution near the boundary ($x\to \infty$) as
\begin{equation}
\begin{aligned}
R(x)&=C_{B}^{in}~\left(\frac{x-1}{2}\right)^{\sigma}+C_{B}^{out}~\left(\frac{x-1}{2}\right)^{-\sigma-1}
\end{aligned}
\end{equation}
where
\begin{equation}
\begin{aligned}
C_{B}^{in}&=C_{2}~\frac{\Gamma(1-i\mu)~\Gamma(2\sigma+1)}{\Gamma(1-i\mu+\sigma)~\Gamma(\sigma+1)}\\
C_{B}^{out}&=C_{2}~\frac{\Gamma(1-i\mu)~\Gamma(-2\sigma-1)}{\Gamma(-i\mu-\sigma)~\Gamma(\sigma+1)}
\end{aligned}
\end{equation}
Demanding there will be only outgoing mode near the boundary, the only non-trivial choice leads to the condition
\begin{equation}\label{NEqnm}
\frac{1}{\Gamma(1-i\mu+\sigma)}=0
\end{equation}
This in turn implies $1-i\mu+\sigma=-n$, where $n$ is a positive integer. After rearranging and substituting $\mu=\omega/\kappa_{-}$ in this equation, we obtain the quasi-normal frequencies of a near extremal black hole as
\begin{equation}\label{NEqnf}
\omega_{NE}=-i(n+\sigma+1)\kappa_{+}\approx -i(n+\sigma+1)\kappa_{-}
\end{equation}
As evident from the above equation these near extremal modes are purely imaginary.  Note that for four dimensional spacetime $\sigma=\ell$ and thus \ref{NEqnf} matches with the result presented in \cite{Kim:2012mh, Cardoso:2017soq}.

\bibliography{reference}

\providecommand{\href}[2]{#2}\begingroup\raggedright\begin{thebibliography}{10}

\bibitem{PhysRevLett.14.57}
R.~Penrose, ``Gravitational collapse and space-time singularities,''
  \href{http://dx.doi.org/10.1103/PhysRevLett.14.57}{{\em Phys. Rev. Lett.}
  {\bfseries 14} (Jan, 1965) 57--59}.
  \url{https://link.aps.org/doi/10.1103/PhysRevLett.14.57}.

\bibitem{Wald:106274}
R.~M. Wald, {\em {General relativity}}.
\newblock Chicago Univ. Press, Chicago, IL, 1984.
\newblock \url{https://cds.cern.ch/record/106274}.

\bibitem{nla.cat-vn3002454}
P.~T. Chrusciel and A.~N. University., {\em On uniqueness in the large of
  solutions of Einstein's equations : "strong cosmic censorship' / Piotr T.
  Chrusciel}.
\newblock Centre for Mathematics and its Applications, ANU [Canberra], 1991.

\bibitem{Costa:2017tjc}
J.~L. Costa, P.~M. Girão, J.~Natário, and J.~D. Silva, ``{On the Occurrence
  of Mass Inflation for the Einstein–Maxwell-Scalar Field System with a
  Cosmological Constant and an Exponential Price Law},''
  \href{http://dx.doi.org/10.1007/s00220-018-3122-z}{{\em Commun. Math. Phys.}
  {\bfseries 361} no.~1, (2018) 289--341},
\href{http://arxiv.org/abs/1707.08975}{{\ttfamily arXiv:1707.08975 [gr-qc]}}.

\bibitem{Costa:2014yha}
J.~L. Costa, P.~M. Girão, J.~Natário, and J.~D. Silva, ``{On the global
  uniqueness for the Einstein-Maxwell-scalar field system with a cosmological
  constant: I. Well posedness and breakdown criterion},''
  \href{http://dx.doi.org/10.1088/0264-9381/32/1/015017}{{\em Class. Quant.
  Grav.} {\bfseries 32} no.~1, (2015) 015017},
\href{http://arxiv.org/abs/1406.7245}{{\ttfamily arXiv:1406.7245 [gr-qc]}}.

\bibitem{Chandrasekhar:579245}
S.~Chandrasekhar, {\em {The mathematical theory of black holes}}.
\newblock Oxford classic texts in the physical sciences. Oxford Univ. Press,
  Oxford, 2002.
\newblock \url{https://cds.cern.ch/record/579245}.

\bibitem{poisson_2004}
E.~Poisson, \href{http://dx.doi.org/10.1017/CBO9780511606601}{{\em A
  Relativist's Toolkit: The Mathematics of Black-Hole Mechanics}}.
\newblock Cambridge University Press, 2004.

\bibitem{1973IJTP....7..183S}
M.~{Simpson} and R.~{Penrose}, ``{Internal Instability in a
  Reissner-Nordstr{\"o}m Black Hole},''
  \href{http://dx.doi.org/10.1007/BF00792069}{{\em International Journal of
  Theoretical Physics} {\bfseries 7} (Apr., 1973) 183--197}.

\bibitem{PhysRevD.41.1796}
E.~Poisson and W.~Israel, ``Internal structure of black holes,''
  \href{http://dx.doi.org/10.1103/PhysRevD.41.1796}{{\em Phys. Rev. D}
  {\bfseries 41} (Mar, 1990) 1796--1809}.
  \url{https://link.aps.org/doi/10.1103/PhysRevD.41.1796}.

\bibitem{Dafermos:2003wr}
M.~Dafermos, ``{The Interior of charged black holes and the problem of
  uniqueness in general relativity},'' {\em Commun. Pure Appl. Math.}
  {\bfseries 58} (2005) 0445--0504,
\href{http://arxiv.org/abs/gr-qc/0307013}{{\ttfamily arXiv:gr-qc/0307013
  [gr-qc]}}.

\bibitem{Dafermos2014}
M.~Dafermos, ``Black holes without spacelike singularities,''
  \href{http://dx.doi.org/10.1007/s00220-014-2063-4}{{\em Communications in
  Mathematical Physics} {\bfseries 332} no.~2, (Dec, 2014) 729--757}.
  \url{https://doi.org/10.1007/s00220-014-2063-4}.

\bibitem{PhysRevLett.67.789}
A.~Ori, ``Inner structure of a charged black hole: An exact mass-inflation
  solution,'' \href{http://dx.doi.org/10.1103/PhysRevLett.67.789}{{\em Phys.
  Rev. Lett.} {\bfseries 67} (Aug, 1991) 789--792}.
  \url{https://link.aps.org/doi/10.1103/PhysRevLett.67.789}.

\bibitem{0264-9381-16-12A-302}
D.~Christodoulou, ``On the global initial value problem and the issue of
  singularities,'' {\em Classical and Quantum Gravity} {\bfseries 16} no.~12A,
  (1999) A23. \url{http://stacks.iop.org/0264-9381/16/i=12A/a=302}.

\bibitem{Cardoso:2017soq}
V.~Cardoso, J.~L. Costa, K.~Destounis, P.~Hintz, and A.~Jansen, ``{Quasinormal
  modes and Strong Cosmic Censorship},''
  \href{http://dx.doi.org/10.1103/PhysRevLett.120.031103}{{\em Phys. Rev.
  Lett.} {\bfseries 120} no.~3, (2018) 031103},
\href{http://arxiv.org/abs/1711.10502}{{\ttfamily arXiv:1711.10502 [gr-qc]}}.

\bibitem{Christodoulou:2008nj}
D.~Christodoulou, \href{http://dx.doi.org/10.1142/9789814374552_0002}{``{The
  Formation of Black Holes in General Relativity},''} in {\em {On recent
  developments in theoretical and experimental general relativity, astrophysics
  and relativistic field theories. Proceedings, 12th Marcel Grossmann Meeting
  on General Relativity, Paris, France, July 12-18, 2009. Vol. 1-3}},
  pp.~24--34.
\newblock 2008.
\newblock
\href{http://arxiv.org/abs/0805.3880}{{\ttfamily arXiv:0805.3880 [gr-qc]}}.
\newblock

\bibitem{Chambers:1997ef}
C.~M. Chambers, ``{The Cauchy horizon in black hole de sitter space-times},''
  {\em Annals Israel Phys. Soc.} {\bfseries 13} (1997) 33,
  \href{http://arxiv.org/abs/gr-qc/9709025}{{\ttfamily arXiv:gr-qc/9709025
  [gr-qc]}}.

\bibitem{PhysRevD.5.2419}
R.~H. Price, ``Nonspherical perturbations of relativistic gravitational
  collapse. i. scalar and gravitational perturbations,''
  \href{http://dx.doi.org/10.1103/PhysRevD.5.2419}{{\em Phys. Rev. D}
  {\bfseries 5} (May, 1972) 2419--2438}.
  \url{https://link.aps.org/doi/10.1103/PhysRevD.5.2419}.

\bibitem{Dafermos:2014cua}
M.~Dafermos, I.~Rodnianski, and Y.~Shlapentokh-Rothman, ``{Decay for solutions
  of the wave equation on Kerr exterior spacetimes III: The full subextremal
  case |a| < M},''
\href{http://arxiv.org/abs/1402.7034}{{\ttfamily arXiv:1402.7034 [gr-qc]}}.

\bibitem{Angelopoulos:2016wcv}
Y.~Angelopoulos, S.~Aretakis, and D.~Gajic, ``{Late-time asymptotics for the
  wave equation on spherically symmetric, stationary spacetimes},''
  \href{http://dx.doi.org/10.1016/j.aim.2017.10.027}{{\em Adv. Math.}
  {\bfseries 323} (2018) 529--621},
\href{http://arxiv.org/abs/1612.01566}{{\ttfamily arXiv:1612.01566 [math.AP]}}.

\bibitem{PhysRevD.19.2821}
R.~A. Matzner, N.~Zamorano, and V.~D. Sandberg, ``Instability of the cauchy
  horizon of reissner-nordstr\"om black holes,''
  \href{http://dx.doi.org/10.1103/PhysRevD.19.2821}{{\em Phys. Rev. D}
  {\bfseries 19} (May, 1979) 2821--2826}.
  \url{https://link.aps.org/doi/10.1103/PhysRevD.19.2821}.

\bibitem{HISCOCK1981110}
W.~A. Hiscock, ``Evolution of the interior of a charged black hole,''
  \href{http://dx.doi.org/https://doi.org/10.1016/0375-9601(81)90508-9}{{\em
  Physics Letters A} {\bfseries 83} no.~3, (1981) 110 -- 112}.
  \url{http://www.sciencedirect.com/science/article/pii/0375960181905089}.

\bibitem{PhysRevLett.80.3432}
P.~R. Brady, I.~G. Moss, and R.~C. Myers, ``Cosmic censorship: As strong as
  ever,'' \href{http://dx.doi.org/10.1103/PhysRevLett.80.3432}{{\em Phys. Rev.
  Lett.} {\bfseries 80} (Apr, 1998) 3432--3435}.
  \url{https://link.aps.org/doi/10.1103/PhysRevLett.80.3432}.

\bibitem{Brady:1996za}
P.~R. Brady, C.~M. Chambers, W.~Krivan, and P.~Laguna, ``{Telling tails in the
  presence of a cosmological constant},''
  \href{http://dx.doi.org/10.1103/PhysRevD.55.7538}{{\em Phys. Rev.} {\bfseries
  D55} (1997) 7538--7545},
\href{http://arxiv.org/abs/gr-qc/9611056}{{\ttfamily arXiv:gr-qc/9611056
  [gr-qc]}}.

\bibitem{Dyatlov:2013hba}
S.~Dyatlov, ``{Asymptotics of linear waves and resonances with applications to
  black holes},'' \href{http://dx.doi.org/10.1007/s00220-014-2255-y}{{\em
  Commun. Math. Phys.} {\bfseries 335} no.~3, (2015) 1445--1485},
\href{http://arxiv.org/abs/1305.1723}{{\ttfamily arXiv:1305.1723 [gr-qc]}}.

\bibitem{Bony2008}
J.-F. Bony and D.~H{\"a}fner, ``Decay and non-decay of the local energy for the
  wave equation on the de sitter--schwarzschild metric,''
  \href{http://dx.doi.org/10.1007/s00220-008-0553-y}{{\em Communications in
  Mathematical Physics} {\bfseries 282} no.~3, (Sep, 2008) 697--719}.
  \url{https://doi.org/10.1007/s00220-008-0553-y}.

\bibitem{Dyatlov2012}
S.~Dyatlov, ``Asymptotic distribution of quasi-normal modes for kerr--de sitter
  black holes,'' \href{http://dx.doi.org/10.1007/s00023-012-0159-y}{{\em
  Annales Henri Poincar{\'e}} {\bfseries 13} no.~5, (Jul, 2012) 1101--1166}.
  \url{https://doi.org/10.1007/s00023-012-0159-y}.

\bibitem{PhysRevD.61.064016}
A.~Ori, ``Strength of curvature singularities,''
  \href{http://dx.doi.org/10.1103/PhysRevD.61.064016}{{\em Phys. Rev. D}
  {\bfseries 61} (Feb, 2000) 064016}.
  \url{https://link.aps.org/doi/10.1103/PhysRevD.61.064016}.

\bibitem{Costa:2014aia}
J.~L. Costa, P.~M. Girão, J.~Natário, and J.~D. Silva, ``{On the global
  uniqueness for the Einstein-Maxwell-scalar field system with a cosmological
  constant. Part 3: Mass inflation and extendibility of the solutions},''
\href{http://arxiv.org/abs/1406.7261}{{\ttfamily arXiv:1406.7261 [gr-qc]}}.

\bibitem{Hintz:2015jkj}
P.~Hintz and A.~Vasy, ``{Analysis of linear waves near the Cauchy horizon of
  cosmological black holes},'' \href{http://dx.doi.org/10.1063/1.4996575}{{\em
  J. Math. Phys.} {\bfseries 58} no.~8, (2017) 081509},
\href{http://arxiv.org/abs/1512.08004}{{\ttfamily arXiv:1512.08004 [math.AP]}}.

\bibitem{Dafermos:2017dbw}
M.~Dafermos and J.~Luk, ``{The interior of dynamical vacuum black holes I: The
  $C^0$-stability of the Kerr Cauchy horizon},''
\href{http://arxiv.org/abs/1710.01722}{{\ttfamily arXiv:1710.01722 [gr-qc]}}.

\bibitem{PhysRevD.97.104060}
O.~J.~C. Dias, F.~C. Eperon, H.~S. Reall, and J.~E. Santos, ``Strong cosmic
  censorship in de sitter space,''
  \href{http://dx.doi.org/10.1103/PhysRevD.97.104060}{{\em Phys. Rev. D}
  {\bfseries 97} (May, 2018) 104060}.
  \url{https://link.aps.org/doi/10.1103/PhysRevD.97.104060}.

\bibitem{Emparan:2008eg}
R.~Emparan and H.~S. Reall, ``{Black Holes in Higher Dimensions},''
  \href{http://dx.doi.org/10.12942/lrr-2008-6}{{\em Living Rev. Rel.}
  {\bfseries 11} (2008) 6},
\href{http://arxiv.org/abs/0801.3471}{{\ttfamily arXiv:0801.3471 [hep-th]}}.

\bibitem{Reall:2015esa}
H.~S. Reall, ``{Higher dimensional black holes},''
  \href{http://dx.doi.org/10.1142/S0218271812300017}{{\em Int. J. Mod. Phys.}
  {\bfseries D21} (2012) 1230001},
  \href{http://arxiv.org/abs/1210.1402}{{\ttfamily arXiv:1210.1402 [gr-qc]}}.
[,105(2015)].

\bibitem{Emparan:2001wn}
R.~Emparan and H.~S. Reall, ``{A Rotating black ring solution in
  five-dimensions},''
  \href{http://dx.doi.org/10.1103/PhysRevLett.88.101101}{{\em Phys. Rev. Lett.}
  {\bfseries 88} (2002) 101101},
\href{http://arxiv.org/abs/hep-th/0110260}{{\ttfamily arXiv:hep-th/0110260
  [hep-th]}}.

\bibitem{Emparan:2007wm}
R.~Emparan, T.~Harmark, V.~Niarchos, N.~A. Obers, and M.~J. Rodriguez, ``{The
  Phase Structure of Higher-Dimensional Black Rings and Black Holes},''
  \href{http://dx.doi.org/10.1088/1126-6708/2007/10/110}{{\em JHEP} {\bfseries
  10} (2007) 110},
\href{http://arxiv.org/abs/0708.2181}{{\ttfamily arXiv:0708.2181 [hep-th]}}.

\bibitem{Arcioni:2004ww}
G.~Arcioni and E.~Lozano-Tellechea, ``{Stability and critical phenomena of
  black holes and black rings},''
  \href{http://dx.doi.org/10.1103/PhysRevD.72.104021}{{\em Phys. Rev.}
  {\bfseries D72} (2005) 104021},
\href{http://arxiv.org/abs/hep-th/0412118}{{\ttfamily arXiv:hep-th/0412118
  [hep-th]}}.

\bibitem{Gibbons:2002av}
G.~W. Gibbons, D.~Ida, and T.~Shiromizu, ``{Uniqueness and nonuniqueness of
  static black holes in higher dimensions},''
  \href{http://dx.doi.org/10.1103/PhysRevLett.89.041101}{{\em Phys. Rev. Lett.}
  {\bfseries 89} (2002) 041101},
\href{http://arxiv.org/abs/hep-th/0206049}{{\ttfamily arXiv:hep-th/0206049
  [hep-th]}}.

\bibitem{Gregory:1993vy}
R.~Gregory and R.~Laflamme, ``{Black strings and p-branes are unstable},''
  \href{http://dx.doi.org/10.1103/PhysRevLett.70.2837}{{\em Phys. Rev. Lett.}
  {\bfseries 70} (1993) 2837--2840},
\href{http://arxiv.org/abs/hep-th/9301052}{{\ttfamily arXiv:hep-th/9301052
  [hep-th]}}.

\bibitem{Horowitz:2012nnc}
G.~T. Horowitz, ed., {\em {Black holes in higher dimensions}}.
\newblock Cambridge Univ. Pr., Cambridge, UK, 2012.
\newblock
\url{http://www.cambridge.org/de/knowledge/isbn/item6633780}.
\newblock

\bibitem{Myers:1986un}
R.~C. Myers and M.~J. Perry, ``{Black Holes in Higher Dimensional
  Space-Times},''
\href{http://dx.doi.org/10.1016/0003-4916(86)90186-7}{{\em Annals Phys.}
  {\bfseries 172} (1986) 304}.

\bibitem{Shiromizu:1999wj}
T.~Shiromizu, K.-i. Maeda, and M.~Sasaki, ``{The Einstein equation on the
  3-brane world},'' \href{http://dx.doi.org/10.1103/PhysRevD.62.024012}{{\em
  Phys. Rev.} {\bfseries D62} (2000) 024012},
\href{http://arxiv.org/abs/gr-qc/9910076}{{\ttfamily arXiv:gr-qc/9910076
  [gr-qc]}}.

\bibitem{Maartens:2001jx}
R.~Maartens, \href{http://dx.doi.org/10.1142/9789812810021_0008}{``{Geometry
  and dynamics of the brane world},''} in {\em {Spanish Relativity Meeting on
  Reference Frames and Gravitomagnetism (EREs2000) Valladolid, Spain, September
  6-9, 2000}}.
\newblock 2001.
\newblock
\href{http://arxiv.org/abs/gr-qc/0101059}{{\ttfamily arXiv:gr-qc/0101059
  [gr-qc]}}.
\newblock

\bibitem{Dadhich:2000am}
N.~Dadhich, R.~Maartens, P.~Papadopoulos, and V.~Rezania, ``{Black holes on the
  brane},'' \href{http://dx.doi.org/10.1016/S0370-2693(00)00798-X}{{\em Phys.
  Lett.} {\bfseries B487} (2000) 1--6},
\href{http://arxiv.org/abs/hep-th/0003061}{{\ttfamily arXiv:hep-th/0003061
  [hep-th]}}.

\bibitem{Germani:2001du}
C.~Germani and R.~Maartens, ``{Stars in the brane world},''
  \href{http://dx.doi.org/10.1103/PhysRevD.64.124010}{{\em Phys. Rev.}
  {\bfseries D64} (2001) 124010},
\href{http://arxiv.org/abs/hep-th/0107011}{{\ttfamily arXiv:hep-th/0107011
  [hep-th]}}.

\bibitem{Casadio:2012pu}
R.~Casadio and J.~Ovalle, ``{Brane-world stars and (microscopic) black
  holes},'' \href{http://dx.doi.org/10.1016/j.physletb.2012.07.041}{{\em Phys.
  Lett.} {\bfseries B715} (2012) 251--255},
\href{http://arxiv.org/abs/1201.6145}{{\ttfamily arXiv:1201.6145 [gr-qc]}}.

\bibitem{Harko:2004ui}
T.~Harko and M.~K. Mak, ``{Vacuum solutions of the gravitational field
  equations in the brane world model},''
  \href{http://dx.doi.org/10.1103/PhysRevD.69.064020}{{\em Phys. Rev.}
  {\bfseries D69} (2004) 064020},
\href{http://arxiv.org/abs/gr-qc/0401049}{{\ttfamily arXiv:gr-qc/0401049
  [gr-qc]}}.

\bibitem{Chakraborty:2014xla}
S.~Chakraborty and S.~SenGupta, ``{Spherically symmetric brane spacetime with
  bulk $f(\mathcal {R})$ gravity},''
  \href{http://dx.doi.org/10.1140/epjc/s10052-014-3234-3}{{\em Eur. Phys. J.}
  {\bfseries C75} no.~1, (2015) 11},
\href{http://arxiv.org/abs/1409.4115}{{\ttfamily arXiv:1409.4115 [gr-qc]}}.

\bibitem{Chakraborty:2015bja}
S.~Chakraborty and S.~SenGupta, ``{Effective gravitational field equations on
  $m$-brane embedded in n-dimensional bulk of Einstein and $f(\mathcal {R})$
  gravity},'' \href{http://dx.doi.org/10.1140/epjc/s10052-015-3768-z}{{\em Eur.
  Phys. J.} {\bfseries C75} no.~11, (2015) 538},
\href{http://arxiv.org/abs/1504.07519}{{\ttfamily arXiv:1504.07519 [gr-qc]}}.

\bibitem{Chakraborty:2015taq}
S.~Chakraborty and S.~SenGupta, ``{Spherically symmetric brane in a bulk of
  $f(R)$ and Gauss–Bonnet gravity},''
  \href{http://dx.doi.org/10.1088/0264-9381/33/22/225001}{{\em Class. Quant.
  Grav.} {\bfseries 33} no.~22, (2016) 225001},
\href{http://arxiv.org/abs/1510.01953}{{\ttfamily arXiv:1510.01953 [gr-qc]}}.

\bibitem{Chakraborty:2017qve}
S.~Chakraborty, K.~Chakravarti, S.~Bose, and S.~SenGupta, ``{Signatures of
  extra dimensions in gravitational waves from black hole quasinormal modes},''
  \href{http://dx.doi.org/10.1103/PhysRevD.97.104053}{{\em Phys. Rev.}
  {\bfseries D97} no.~10, (2018) 104053},
\href{http://arxiv.org/abs/1710.05188}{{\ttfamily arXiv:1710.05188 [gr-qc]}}.

\bibitem{Mukherjee:2017fqz}
S.~Mukherjee and S.~Chakraborty, ``{Horndeski theories confront the Gravity
  Probe B experiment},''
  \href{http://dx.doi.org/10.1103/PhysRevD.97.124007}{{\em Phys. Rev.}
  {\bfseries D97} no.~12, (2018) 124007},
\href{http://arxiv.org/abs/1712.00562}{{\ttfamily arXiv:1712.00562 [gr-qc]}}.

\bibitem{Banerjee:2017hzw}
I.~Banerjee, S.~Chakraborty, and S.~SenGupta, ``{Excavating black hole
  continuum spectrum: Possible signatures of scalar hairs and of higher
  dimensions},'' \href{http://dx.doi.org/10.1103/PhysRevD.96.084035}{{\em Phys.
  Rev.} {\bfseries D96} no.~8, (2017) 084035},
\href{http://arxiv.org/abs/1707.04494}{{\ttfamily arXiv:1707.04494 [gr-qc]}}.

\bibitem{Chakraborty:2016lxo}
S.~Chakraborty and S.~SenGupta, ``{Strong gravitational lensing --- A probe for
  extra dimensions and Kalb-Ramond field},''
  \href{http://dx.doi.org/10.1088/1475-7516/2017/07/045}{{\em JCAP} {\bfseries
  1707} no.~07, (2017) 045},
\href{http://arxiv.org/abs/1611.06936}{{\ttfamily arXiv:1611.06936 [gr-qc]}}.

\bibitem{Du:2004jt}
D.-P. Du, B.~Wang, and R.-K. Su, ``{Quasinormal modes in pure de Sitter
  space-times},'' \href{http://dx.doi.org/10.1103/PhysRevD.70.064024}{{\em
  Phys. Rev.} {\bfseries D70} (2004) 064024},
\href{http://arxiv.org/abs/hep-th/0404047}{{\ttfamily arXiv:hep-th/0404047
  [hep-th]}}.

\bibitem{Berti:2009kk}
E.~Berti, V.~Cardoso, and A.~O. Starinets, ``{Quasinormal modes of black holes
  and black branes},''
  \href{http://dx.doi.org/10.1088/0264-9381/26/16/163001}{{\em Class. Quant.
  Grav.} {\bfseries 26} (2009) 163001},
\href{http://arxiv.org/abs/0905.2975}{{\ttfamily arXiv:0905.2975 [gr-qc]}}.

\bibitem{Cardoso:2004cj}
V.~Cardoso, G.~Siopsis, and S.~Yoshida, ``{Scalar perturbations of higher
  dimensional rotating and ultra-spinning black holes},''
  \href{http://dx.doi.org/10.1103/PhysRevD.71.024019}{{\em Phys. Rev.}
  {\bfseries D71} (2005) 024019},
\href{http://arxiv.org/abs/hep-th/0412138}{{\ttfamily arXiv:hep-th/0412138
  [hep-th]}}.

\bibitem{Ida:2002zk}
D.~Ida, Y.~Uchida, and Y.~Morisawa, ``{The Scalar perturbation of the higher
  dimensional rotating black holes},''
  \href{http://dx.doi.org/10.1103/PhysRevD.67.084019}{{\em Phys. Rev.}
  {\bfseries D67} (2003) 084019},
\href{http://arxiv.org/abs/gr-qc/0212035}{{\ttfamily arXiv:gr-qc/0212035
  [gr-qc]}}.

\bibitem{PhysRevD.31.290}
B.~Mashhoon, ``Stability of charged rotating black holes in the eikonal
  approximation,'' \href{http://dx.doi.org/10.1103/PhysRevD.31.290}{{\em Phys.
  Rev. D} {\bfseries 31} (Jan, 1985) 290--293}.
  \url{https://link.aps.org/doi/10.1103/PhysRevD.31.290}.

\bibitem{Cornish:2003ig}
N.~J. Cornish and J.~J. Levin, ``{Lyapunov timescales and black hole
  binaries},'' \href{http://dx.doi.org/10.1088/0264-9381/20/9/304}{{\em Class.
  Quant. Grav.} {\bfseries 20} (2003) 1649--1660},
\href{http://arxiv.org/abs/gr-qc/0304056}{{\ttfamily arXiv:gr-qc/0304056
  [gr-qc]}}.

\bibitem{Cardoso:2008bp}
V.~Cardoso, A.~S. Miranda, E.~Berti, H.~Witek, and V.~T. Zanchin, ``{Geodesic
  stability, Lyapunov exponents and quasinormal modes},''
  \href{http://dx.doi.org/10.1103/PhysRevD.79.064016}{{\em Phys. Rev.}
  {\bfseries D79} (2009) 064016},
\href{http://arxiv.org/abs/0812.1806}{{\ttfamily arXiv:0812.1806 [hep-th]}}.

\bibitem{0264-9381-9-12-004}
L.~Bombelli and E.~Calzetta, ``Chaos around a black hole,'' {\em Classical and
  Quantum Gravity} {\bfseries 9} no.~12, (1992) 2573.
  \url{http://stacks.iop.org/0264-9381/9/i=12/a=004}.

\bibitem{PhysRevLett.52.1361}
V.~Ferrari and B.~Mashhoon, ``Oscillations of a black hole,''
  \href{http://dx.doi.org/10.1103/PhysRevLett.52.1361}{{\em Phys. Rev. Lett.}
  {\bfseries 52} (Apr, 1984) 1361--1364}.
  \url{https://link.aps.org/doi/10.1103/PhysRevLett.52.1361}.

\bibitem{Konoplya:2017wot}
R.~A. Konoplya and Z.~Stuchlík, ``{Are eikonal quasinormal modes linked to the
  unstable circular null geodesics?},''
  \href{http://dx.doi.org/10.1016/j.physletb.2017.06.015}{{\em Phys. Lett.}
  {\bfseries B771} (2017) 597--602},
\href{http://arxiv.org/abs/1705.05928}{{\ttfamily arXiv:1705.05928 [gr-qc]}}.

\bibitem{Hod:2009td}
S.~Hod, ``{Black-hole quasinormal resonances: Wave analysis versus a
  geometric-optics approximation},''
  \href{http://dx.doi.org/10.1103/PhysRevD.80.064004}{{\em Phys. Rev.}
  {\bfseries D80} (2009) 064004},
\href{http://arxiv.org/abs/0909.0314}{{\ttfamily arXiv:0909.0314 [gr-qc]}}.

\bibitem{Hod:2018dpx}
S.~Hod, ``{Strong cosmic censorship in charged black-hole spacetimes: As strong
  as ever},''
\href{http://arxiv.org/abs/1801.07261}{{\ttfamily arXiv:1801.07261 [gr-qc]}}.

\bibitem{Cardoso:2018nvb}
V.~Cardoso, J.~L. Costa, K.~Destounis, P.~Hintz, and A.~Jansen, ``{Strong
  cosmic censorship in charged black-hole spacetimes: still subtle},''
\href{http://arxiv.org/abs/1808.03631}{{\ttfamily arXiv:1808.03631 [gr-qc]}}.

\bibitem{Ge:2018vjq}
B.~Ge, J.~Jiang, B.~Wang, H.~Zhang, and Z.~Zhong, ``{Strong cosmic censorship
  for the massless Dirac field in the Reissner-Nordstrom-de Sitter
  spacetime},''
\href{http://arxiv.org/abs/1810.12128}{{\ttfamily arXiv:1810.12128 [gr-qc]}}.

\bibitem{Mo:2018nnu}
Y.~Mo, Y.~Tian, B.~Wang, H.~Zhang, and Z.~Zhong, ``{Strong cosmic censorship
  for the massless charged scalar field in the Reissner-Nordstrom-de Sitter
  spacetime},''
\href{http://arxiv.org/abs/1808.03635}{{\ttfamily arXiv:1808.03635 [gr-qc]}}.

\bibitem{LopezOrtega:2012vi}
A.~Lopez-Ortega, ``{On the quasinormal modes of the de Sitter spacetime},''
  \href{http://dx.doi.org/10.1007/s10714-012-1398-4}{{\em Gen. Rel. Grav.}
  {\bfseries 44} (2012) 2387--2400},
\href{http://arxiv.org/abs/1207.6791}{{\ttfamily arXiv:1207.6791 [gr-qc]}}.

\bibitem{Abdalla:2002hg}
E.~Abdalla, K.~H.~C. Castello-Branco, and A.~Lima-Santos, ``{Support of dS /
  CFT correspondence from space-time perturbations},''
  \href{http://dx.doi.org/10.1103/PhysRevD.66.104018}{{\em Phys. Rev.}
  {\bfseries D66} (2002) 104018},
\href{http://arxiv.org/abs/hep-th/0208065}{{\ttfamily arXiv:hep-th/0208065
  [hep-th]}}.

\bibitem{Kim:2012mh}
Y.-W. Kim, Y.~S. Myung, and Y.-J. Park, ``{Quasinormal modes and hidden
  conformal symmetry in the Reissner-Nordstr\'om black hole},''
  \href{http://dx.doi.org/10.1140/epjc/s10052-013-2440-8}{{\em Eur. Phys. J.}
  {\bfseries C73} (2013) 2440},
\href{http://arxiv.org/abs/1205.3701}{{\ttfamily arXiv:1205.3701 [hep-th]}}.

\bibitem{Chen:2012zn}
C.-M. Chen, S.~P. Kim, I.-C. Lin, J.-R. Sun, and M.-F. Wu, ``{Spontaneous Pair
  Production in Reissner-Nordstrom Black Holes},''
  \href{http://dx.doi.org/10.1103/PhysRevD.85.124041}{{\em Phys. Rev.}
  {\bfseries D85} (2012) 124041},
\href{http://arxiv.org/abs/1202.3224}{{\ttfamily arXiv:1202.3224 [hep-th]}}.

\bibitem{Chamblin:2000ra}
A.~Chamblin, H.~S. Reall, H.-a. Shinkai, and T.~Shiromizu, ``{Charged brane
  world black holes},''
  \href{http://dx.doi.org/10.1103/PhysRevD.63.064015}{{\em Phys. Rev.}
  {\bfseries D63} (2001) 064015},
\href{http://arxiv.org/abs/hep-th/0008177}{{\ttfamily arXiv:hep-th/0008177
  [hep-th]}}.

\bibitem{Cai:2001tv}
R.-G. Cai, ``{Cardy-Verlinde formula and thermodynamics of black holes in de
  Sitter spaces},'' \href{http://dx.doi.org/10.1016/S0550-3213(02)00064-0}{{\em
  Nucl. Phys.} {\bfseries B628} (2002) 375--386},
\href{http://arxiv.org/abs/hep-th/0112253}{{\ttfamily arXiv:hep-th/0112253
  [hep-th]}}.

\bibitem{refId0}
{Zhang, Li-Chun}, {Ma, Meng-Sen}, {Zhao, Hui-Hua}, and {Zhao, Ren},
  ``Thermodynamics of phase transition in higher-dimensional
  reissner-nordstr\"om-de sitter black hole,''
  \href{http://dx.doi.org/10.1140/epjc/s10052-014-3052-7}{{\em Eur. Phys. J. C}
  {\bfseries 74} no.~9, (2014) 3052}.
  \url{https://doi.org/10.1140/epjc/s10052-014-3052-7}.

\bibitem{Gyulchev:2006zg}
G.~N. Gyulchev and S.~S. Yazadjiev, ``{Kerr-Sen dilaton-axion black hole
  lensing in the strong deflection limit},''
  \href{http://dx.doi.org/10.1103/PhysRevD.75.023006}{{\em Phys. Rev.}
  {\bfseries D75} (2007) 023006},
\href{http://arxiv.org/abs/gr-qc/0611110}{{\ttfamily arXiv:gr-qc/0611110
  [gr-qc]}}.

\bibitem{Rahman:2018fgy}
M.~Rahman and A.~A. Sen, ``{Astrophysical Signatures of Black holes in
  Generalized Proca Theories},''
\href{http://arxiv.org/abs/1810.09200}{{\ttfamily arXiv:1810.09200 [gr-qc]}}.

\bibitem{Gibbons:2004uw}
G.~W. Gibbons, H.~Lu, D.~N. Page, and C.~N. Pope, ``{The General Kerr-de Sitter
  metrics in all dimensions},''
  \href{http://dx.doi.org/10.1016/j.geomphys.2004.05.001}{{\em J. Geom. Phys.}
  {\bfseries 53} (2005) 49--73},
\href{http://arxiv.org/abs/hep-th/0404008}{{\ttfamily arXiv:hep-th/0404008
  [hep-th]}}.

\bibitem{Gibbons:2004js}
G.~W. Gibbons, H.~Lu, D.~N. Page, and C.~N. Pope, ``{Rotating black holes in
  higher dimensions with a cosmological constant},''
  \href{http://dx.doi.org/10.1103/PhysRevLett.93.171102}{{\em Phys. Rev. Lett.}
  {\bfseries 93} (2004) 171102},
\href{http://arxiv.org/abs/hep-th/0409155}{{\ttfamily arXiv:hep-th/0409155
  [hep-th]}}.

\bibitem{Modgil:2001hm}
M.~S. Modgil, S.~Panda, and G.~Sengupta, ``{Rotating brane world black
  holes},'' \href{http://dx.doi.org/10.1142/S0217732302007442}{{\em Mod. Phys.
  Lett.} {\bfseries A17} (2002) 1479--1488},
\href{http://arxiv.org/abs/hep-th/0104122}{{\ttfamily arXiv:hep-th/0104122
  [hep-th]}}.

\bibitem{Frolov:2004wy}
V.~P. Frolov, D.~V. Fursaev, and D.~Stojkovic, ``{Rotating black holes in brane
  worlds},'' \href{http://dx.doi.org/10.1088/1126-6708/2004/06/057}{{\em JHEP}
  {\bfseries 06} (2004) 057},
\href{http://arxiv.org/abs/gr-qc/0403002}{{\ttfamily arXiv:gr-qc/0403002
  [gr-qc]}}.

\bibitem{Larranaga:2013aoa}
A.~Larrañaga, C.~Grisales, and M.~Londoño, ``{A Topologically Charged
  Rotating Black Hole in the Brane},''
\href{http://dx.doi.org/10.1155/2013/727294}{{\em Adv. High Energy Phys.}
  {\bfseries 2013} (2013) 727294}.

\bibitem{Dafermos:2018tha}
M.~Dafermos and Y.~Shlapentokh-Rothman, ``{Rough initial data and the strength
  of the blue-shift instability on cosmological black holes with $\Lambda >
  0$},'' \href{http://dx.doi.org/10.1088/1361-6382/aadbcf}{{\em Class. Quant.
  Grav.} {\bfseries 35} no.~19, (2018) 195010},
\href{http://arxiv.org/abs/1805.08764}{{\ttfamily arXiv:1805.08764 [gr-qc]}}.

\bibitem{Dias:2018etb}
O.~J.~C. Dias, H.~S. Reall, and J.~E. Santos, ``{Strong cosmic censorship:
  taking the rough with the smooth},''
  \href{http://dx.doi.org/10.1007/JHEP10(2018)001}{{\em JHEP} {\bfseries 10}
  (2018) 001},
\href{http://arxiv.org/abs/1808.02895}{{\ttfamily arXiv:1808.02895 [gr-qc]}}.

\end{thebibliography}\endgroup

\bibliographystyle{utphys1}
\end{document}